\documentclass[aps,prl,twocolumn,showpacs,psfig,superscriptaddress,longbibliography]{revtex4-1}

\usepackage{textcomp}
\usepackage{times}
\usepackage{graphicx}
\usepackage{float}
\usepackage{latexsym,amsmath,amssymb,bm,euscript}
\usepackage{color}
\usepackage{subfigure}
\usepackage{epstopdf}
\usepackage[colorlinks=true,linkcolor=blue,citecolor=blue]{hyperref}
\usepackage{hyperref}
\usepackage{soul}
\usepackage[normalem]{ulem}
\usepackage{mathrsfs}
\usepackage{amsmath}
\usepackage{lettrine}
\usepackage{xspace}
\usepackage{textcomp}

\def\para{\ensuremath{/\kern -0.8em /}\xspace}

\renewcommand{\i}{\text{i}}

\graphicspath{{../Figs/}}

\begin{document}

\title{Monte Carlo Study of Critical Fermi Surface with Spatially Disordered Interactions}

\author{Tu~Hong}
\affiliation{Key Laboratory of Artificial Structures and Quantum Control (Ministry of Education)\\
 School of Physics and Astronomy, Shanghai Jiao Tong University, Shanghai 200240, China}

\author{Xiao~Yan~Xu}
\email{xiaoyanxu@sjtu.edu.cn}
\affiliation{Key Laboratory of Artificial Structures and Quantum Control (Ministry of Education)\\
 School of Physics and Astronomy, Shanghai Jiao Tong University, Shanghai 200240, China}
\affiliation{Hefei National Laboratory, Hefei 230088, China
}

\begin{abstract}
\end{abstract}

\date{\today}
\maketitle

\noindent {\bf{Abstract}}\\
\textbf{
Non-Fermi liquids are an important topic in condensed matter physics, as their characteristics challenge the framework of traditional Fermi liquid theory and reveal the complex behavior of electrons in strongly interacting systems. 
Both the experimentally observed smeared region and the theoretically predicted marginal Fermi liquid suggest that spatial disorder seems to be an important driver of these phenomena~\cite{patelUniversalTheoryStrange2023}. By performing large-scale determinant quantum Monte Carlo (DQMC) simulations in the ferromagnetic spin-fermion model at finite $N$, beyond the large-$N$ used in previous theoretical work, we investigated the role of spatial disorder in the critical Fermi surface (FS) of this model. We proposed a corrected theory of our system, which is based on a modified Eliashberg theory~\cite{kleinNormalStateProperties2020} and a universal theory of strange metals~\cite{patelUniversalTheoryStrange2023}. This theory agrees well with the data obtained from DQMC, particularly in capturing the $\omega \ln \omega$ type self-energy characteristic of marginal Fermi liquid behavior, and observing the linear-in-temperature resistivity. Our findings offer strong and unbiased validation of the universal theory of strange metals, broaden the applicability of the modified Eliashberg theory, and provide insights for numerically searching for marginal Fermi liquid and linear-in-temperature resistivity.
\\
}

\noindent {\bf{Introduction}}\\
In recent decades, numerous behaviors that go beyond traditional Landau Fermi liquid theory have been discovered in quantum materials like Cu- and Fe-based superconductors, heavy-fermion metals, transition-metal alloys, and even recently widely studied Moir\'e systems and Ni-based superconductors \cite{lohneysenFermiliquidInstabilitiesMagnetic2007a,leeRecentDevelopmentsNonFermi2018, liSuperconductivityInfinitelayerNickelate2019, sunSignaturesSuperconductivity802023, zhuSuperconductivityPressurizedTrilayer2024, cao2018unconventional, cao2020strange}. The emergence of these novel behaviors leads to the study of non-Fermi liquid (nFL) \cite{stewart2001non, lohneysenFermiliquidInstabilitiesMagnetic2007a, hartnollHolographicQuantumMatter2016,keimerQuantumMatterHightemperature2015} in itinerant electron systems. Nowadays, understanding nFL has become one of the central issues in condensed matter physics.

The theoretical paradigm concerning itinerant electron systems can date back to the famous Hertz-Mills-Moriya (HMM) framework \cite{hertzQuantumCriticalPhenomena1976,millisEffectNonzeroTemperature1993,moriyaSpinFluctuationsItinerant1985}, which is based on calculations of the one-loop diagram and predicts mean-field scaling. However, subsequent research indicates that the original HMM framework is not complete, propelling the development of advanced perturbative computational techniques such as the large-$N$ expansion and perturbative renormalization group within this general paradigm. These methods have found the existence of anomalous dimensions by computing higher-order loop diagrams \cite{polchinskiLowEnergyDynamics1994, polchinskiLowEnergyDynamics1994, kimGaugeinvariantResponseFunctions1994, metlitskiQuantumPhaseTransitions2010, abanovQuantumcriticalTheorySpinfermion2001, mrossControlledExpansionCertain2010}. While our work remains firmly within the general HMM framework, {it is worth noting that alternative theoretical approaches have been developed for nFL, particularly in heavy fermion compounds. The Kondo breakdown scenario \cite{si2001locally, coleman2001fermiliquids, senthil2004weak} represents one such alternative framework that addresses phenomena in heavy fermion materials through different physical mechanisms than those we explore here \cite{schroder2000onset, custers2003break, paschen2004hall, friedemann2010fermi}.}

Returning to developments within the general HMM framework, with the advancements in computing power, large-scale simulations have become feasible, providing a non-perturbative approach to studying itinerant electron systems. Numerically, research continues the theoretical spirit. Following the groundbreaking work of Berg, Metlitski, and Sachdev~\cite{bergSignproblemfreeQuantumMonte2012a}, an array of models involving FS coupling to critical bosonic fields, such as nematic-Ising \cite{schattnerIsingNematicQuantum2016, ledererSuperconductivityNonFermiLiquid2017, grossmanSpecificHeatQuantum2021}, ferromagnetic~\cite{xuNonFermiLiquid+12017,jiangMonteCarloStudy2022} and antiferromagnetic~\cite{schattnerCompetingOrdersNearly2016,liNatureEffectiveInteraction2017a,liuItinerantQuantumCritical2018, liuItinerantQuantumCritical2019,lunts2023non}, have been proposed. These models have unveiled nFL signals in the critical region.

Despite both theoretical and numerical aspects having made significant advancements, the puzzle of nFL remains incompletely resolved. Sung-Sik Lee discovered a class of Feynman diagrams that had been overlooked by previous researchers, indicating that perturbative calculations are not convergent even in the large-$N$ limit~\cite{leeLowenergyEffectiveTheory2009}. Due to the high computational complexity of fermionic systems, the current sizes of simulated systems are not substantial, and the conclusions drawn for the thermodynamic limit are also not very reliable. Within the general HMM framework itself, many challenges remain unresolved, particularly in reconciling analytical predictions with numerical simulations and experimental observations~\cite{michonReconcilingScalingOptical2023a}. 

Experimentally, linear-$T$ resistivity is frequently observed within a widening fan region emerging from the quantum critical point (QCP) \cite{cooper2009anomalous,greeneStrangeMetalState2020a}, leading researchers to argue for a significant role of spatial disorder therein, and a universal theory of strange metals (abbreviated as UT) is proposed where a FS couples to a $2$D critical bosonic field with spatially random interactions~\cite{guoLarge$N$Theory2022,esterlisLarge$N$Theory2021,patelUniversalTheoryStrange2023}. {Note that disorder effects on nFL are also studied in other groups~\cite{nosov2020,foster2022}.} Inspired by the infinite-range Sachdev-Ye-Kitaev (SYK) model \cite{sachdevGaplessSpinFluidGround1993,kitaev2015talks, maldacena2016remarks,sachdevBekensteinHawkingEntropyStrange2015}, UT treats the spatial randomness in a self-average manner, resulting in a spatially independent effective action. The theory predicts that the introduction of spatially disordered interactions can lead to the observation of several properties of strange metals, including linear-$T$ resistivity, in the critical region. This behavior can be attributable to the emerging marginal Fermi liquid (MFL) term in self-energy \cite{patelUniversalTheoryStrange2023,patelLocalizationOverdampedBosonic2024}. However, UT is developed within the framework of large-$N$ approximation where the bosonic fields and fermions have $N$ flavors, and random couplings are also random in flavor space. Considering that both the bosonic fields and the fermions have finite flavors in realistic materials and the couplings between them may not be random in flavor space, it may make UT less precise. Although some conclusions have been supported by mean field calculations (for $N \ge 2$) and the Hybrid Monte Carlo method (for $N = 1$) \cite{patelLocalizationOverdampedBosonic2024}, these numerical methods are based on the effective action obtained after integrating out the fermions. Therefore, an unbiased numerical validation is necessary. 

While determinant quantum Monte Carlo (DQMC) \cite{blankenbeclerMonteCarloCalculations1981, scalapinoMonteCarloCalculations1981, assaadWorldlineDeterminantalQuantum2008, lohStableNumericalSimulations1992, sorella1988numerical,sugiyama1986auxiliary} can help us perform unbiased simulations on lattice systems by starting from the complete Hamiltonian, it is still very challenging to verify UT. UT assumes that the disorder strength is much greater than the Matsubara frequency, which means we need to be at extremely low temperatures. However, it has been studied that above various types of quantum phase transition points, there is commonly observed a superconducting dome~\cite{xuNonFermiLiquid+12017,ledererSuperconductivityNonFermiLiquid2017}. Even in a spin-fermion bilayer model where the superconducting transition temperature is suppressed to a very low level \cite{xuRevealingFermionicQuantum2019}, this condition is still difficult to satisfy. Interestingly, the spin-fermion bilayer model in Ref.~\cite{xuNonFermiLiquid+12017} can be described by modified Eliashberg theory (MET) \cite{kleinNormalStateProperties2020,xuIdentificationNonFermiLiquid2020} at finite temperature above $T_c$, which is based on perturbation expansion and is equivalent to the case in UT when no spatial disorder is present. The combination of these approaches appears promising for verifying UT at finite temperatures.

{In this work, we study the effect of spatially disordered interactions in the spin-fermion bilayer model proposed in Ref.~\cite{xuNonFermiLiquid+12017}. While the original model was defined on a square lattice, we extend it to a triangular lattice due to its potential for weaker anisotropy.} By adjusting the parameters of the bosonic component, a critical region can be achieved and characterized. Especially, this kind of model can be free of the infamous sign-problem in DQMC~\cite{bergSignproblemfreeQuantumMonte2012a,wuSufficientConditionAbsence2005,xuNonFermiLiquid+12017}, enabling large-scale simulations to become possible. To avoid the influence of the superconducting dome in measurements, we employ finite-temperature DQMC above $T_c$ \cite{assaadWorldlineDeterminantalQuantum2008}. In the absence of spatial disorder, nFL behaviors emerge in the quantum critical region, and the self-energy of this model can be described well by MET analytically. By applying the MET, we can exclude the contribution of thermal fluctuations and predict that the quasiparticle weight will vanish at zero temperature. When spatial disorder is applied, the form of MET is no longer established. We fix the MET by absorbing the spirits in Ref.~\cite{patelUniversalTheoryStrange2023} and propose the UT of our case, which is equivalent to adding two contributions from disorder to the fermionic self-energy in MET. 
{When a random transverse field is applied, our numerical results indicate that the self-energy data align well with the derived formula. More interestingly, when a completely random Yukawa coupling is applied, we successfully observe an $\omega \ln \omega$ type self-energy characteristic of MFL behavior, and most importantly, observing the linear-in-$T$ resistivity predicted by the universal theory. Our results provide important validation of the universal theory of strange metals beyond the large-$N$ approximation, and inspire the search for linear-$T$ resistivity numerically.}

\bigskip \noindent {\bf Model and phase diagram in the clean limit}\\
In this work, we focus on investigating the role of spatially disordered interactions. Before that, we first consider the clean limit in this section. 
The clean model is based on the spatially uniform Hamiltonian proposed in Ref.~\cite{xuNonFermiLiquid+12017},
\begin{gather}
  \hat{H} = \hat{H}_f + \hat{H}_s + \hat{H}_{sf}, \label{eq:eq1}\\
  \hat{H}_f = -t \sum_{\langle ij \rangle \lambda \sigma} \hat{c}^{\dagger}_{i \lambda \sigma} \hat{c}_{j \lambda \sigma} 
  + \text{H.c.} - \mu \sum_{i \lambda \sigma} \hat{n}_{i \lambda \sigma}, \\
  \hat{H}_{s} = -J \sum_{\langle ij \rangle} \hat{Z}_i \hat{Z}_j - h \sum_{i} \hat{X}_i, \\
  \hat{H}_{sf} = -\xi \sum_i \hat{Z}_i(\hat{\sigma}^z_{i1} + \hat{\sigma}^z_{i2}).
\end{gather}

\begin{figure*}[t]
	\includegraphics[width=16cm]{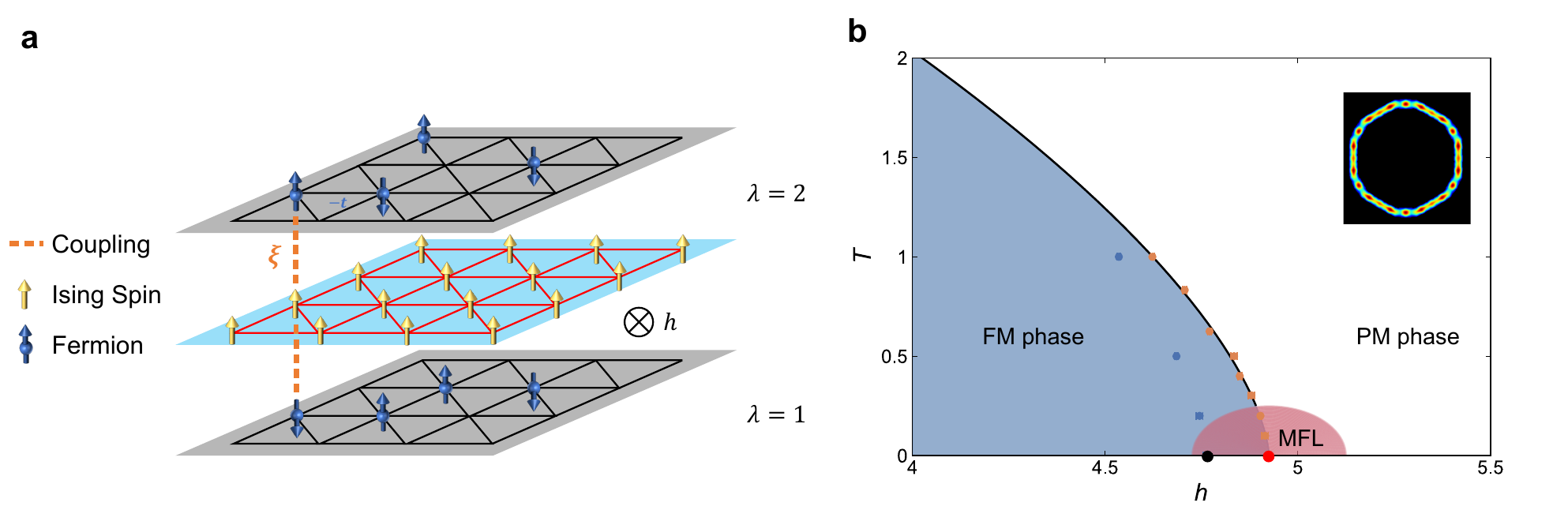}
    \centering
	\caption{\label{fig:fig1}
	{Model and Phase Diagram.}
	(a) Graph of the spin-fermion bilayer model defined in~\eqref{eq:eq1}. The top and bottom layers are fermion layers($\lambda=1,2$), and in each fermion layer, spin-1/2 fermions live on a triangular lattice. Between the fermion layers is an Ising spin layer which mediates the fermion-fermion interaction. The Ising spins have two tunable parameters (ferromagnetic interaction $J$ and transverse magnetic field $h$) to control the FM-PM phase transition. 
	(b) $h-T$ phase diagram of model in~\eqref{eq:eq1}. In this study, we use coupling strength $\xi = 1$ and $\mu = 0.8348$ to realize a hall-filling case ($\langle n_{i\lambda} \rangle = 1.0$). The orange data points are FM-PM phase transition critical points at finite temperatures. By using the formula $T_N(h) \sim |h - h_c|^{c}$, we have fitted a phase boundary (the black line), with quantum critical point $h_c = 4.9256(3)$ and $c = 0.635(2)$. For comparison, we also plot the phase boundary of the transverse field Ising model (blue points), which corresponds to the no coupling ($\xi = 0$) case. This boundary is consistent with $(2 + 1)$D Ising universality class and can be described by $T_N(h) \sim |h - h_c|^{z \nu}$ with $\nu z \approx 0.63$ and $h_c = 4.76(7)$. The position of QCP (the black dot and the red dot on the $h$-axis) is changed due to the coupling term $\hat{H}_{sf}$.
    The inset is the low-energy spectral weight at QCP, which is obtained from $G(\mathbf{k}, \beta/2)$. Twisted boundary conditions are used to increase the resolution. The light red MFL region corresponds to the case when the coupling term is completely spatially random.}
\end{figure*}

This model has two fermion layers and a transverse field Ising model (TFIM) in between, as shown in Fig.~\ref{fig:fig1}{\bf a}. The top and bottom are fermion layers, which are governed by the first part of Hamiltonian $\hat{H}_f$ with $\lambda=1,2$ the fermion layer index. On each fermion layer live spin-$1/2$ fermions, and we use $\sigma = \uparrow, \downarrow$ to represent different spins. The grand-canonical ensemble is used with a chemical potential term to control the average number of fermions. Here we choose $\mu/t = 0.8348$ to approach a half-filling case, e.g., $\langle \hat{n}_{i\lambda\sigma} \rangle \approx 0.5$.
The in between TFIM ($\hat{H}_s$) has nearest neighbor ferromagnetic (FM) interaction. When reducing transverse field $h$ and the temperature $T$, the Ising part $\hat{H}_s$ will undergo a paramagnetic (PM)-FM phase transition, which corresponds to the spontaneous breaking of $Z_2$ symmetry. The Ising spins also mediate the fermion-fermion interactions by the coupling term $\hat{H}_{sf}$ and $\hat{\sigma}_{i \lambda} = \frac{1}{2}(\hat{n}_{i\lambda\uparrow} - \hat{n}_{i\lambda\downarrow})$ is the net magnetic moment of fermions on site $i$ of layer  $\lambda$.
The details of DQMC implementation can be seen in SI Appendix 1.

The properties of the above model on a square lattice have already been explored in Ref.~\cite{xuNonFermiLiquid+12017}. We further implement the model on a triangular lattice instead of the original square one. This modification is motivated by two key considerations. First, it reduces anisotropy, thus provides a more ideal platform for validating theoretical frameworks which assumes isotropy. 
Second, using another lattice geometry also offers additional validation for employing MET to analyze spin-fermion models. In Fig.~\ref{fig:fig1}{\bf b}, we show the phase diagram of the model with $T$ and $h$ expressed in units of $t$ and $J$, respectively. When the coupling strength is zero ($\xi = 0$), this model is a 2D Ising model with a transverse field, which can be mapped to a $(2+1)$D classical Ising model. Near the quantum critical point (QCP), the phase boundary can be described by $T_N(h) \sim |h - h_c|^{\nu z}$. Our numerical results give out a QCP $h_c = 4.763(8)$ at $T=0$ with critical exponent $\nu z \approx 0.63$ that is in accordance with the $(2+1)$D Ising universality class~\cite{pfeuty1971ising}. When the coupling is opened ($\xi = 1$), the fermions and spins are correlated strongly near the critical point, and the PM-FM transition happens at higher temperatures and transverse fields. Similar to the square lattice case in Ref.~\cite{xuNonFermiLiquid+12017}, the exponent no longer obeys the relation $c = \nu z$. But unlike results on the square lattice where the $c = 0.77$ is much larger than the no coupling one, here we get a QCP $h_c = 4.9256(3)$ with $c = 0.635(2)$. It is interesting to mention that a recent theory predicts the absence of a unique dynamical critical exponent in nFL with critical FS~\cite{kukreja2024space}.

To determine the phase boundary, we fix the temperature and scan different transverse fields $h$. We apply the data collapse method \cite{bhattacharjee2001measure} and finite-size analysis \cite{shao2016quantum,qinDualityDeconfinedQuantumCritical2017}.
We find that both Ising spin and fermion spin susceptibility data yield the same critical point value in the thermodynamic limit, indicating that they belong to the same universality class.  
Further details and results regarding data collapse and finite-size analysis are included in SI Appendix 3.

By lowering the transverse field and going across the PM-FM phase boundary, the $Z_2$ symmetry of Ising spins is spontaneously broken. Hence, the degeneracy of fermions with different spins is also broken, and two distinct FS are formed. In the inset of Fig.~\ref{fig:fig1} we demonstrate the fermionic spectral weight extract from the imaginary time Green's function $G(\mathbf{k}, \tau = \beta / 2) \sim N(\mathbf{k}, \omega = 0)$ which can be used to locate the FS~\cite{schattnerIsingNematicQuantum2016}. The nearly circular FS indicates that the triangular lattice has weaker anisotropy, which is also supported by the subsequent self-energy data.

The interaction that causes nFL behaviors also gives rise to superconductivity, and one can improve the transition temperature $T_c$ by enhancing the interaction. As discussed in SI Appendix 4, many superconducting order parameters are allowed in our model, and the strongest channel is the layer-singlet, spin-triplet one. Until the lowest temperature $T = 0.025$ that we can access, no signature of superconductivity is found, as shown in SI Appendix Fig. S5. In fact, the extremely low transition temperature has been seen in other spin-fermion bilayer models, though the mechanism is not yet fully understood~\cite{xuRevealingFermionicQuantum2019}, which makes the spin-fermion bilayer models an ideal platform to study critical quantum fluctuations. Based on the calculation of fermionic Green's function and the self-energy, we find the clean model on the triangle lattice can be well described by the MET, similar to the square lattice case~\cite{xuNonFermiLiquid+12017}, as shown in SI Appendix 6 and 7.

\begin{figure*}[t]
  \includegraphics[scale=0.60]{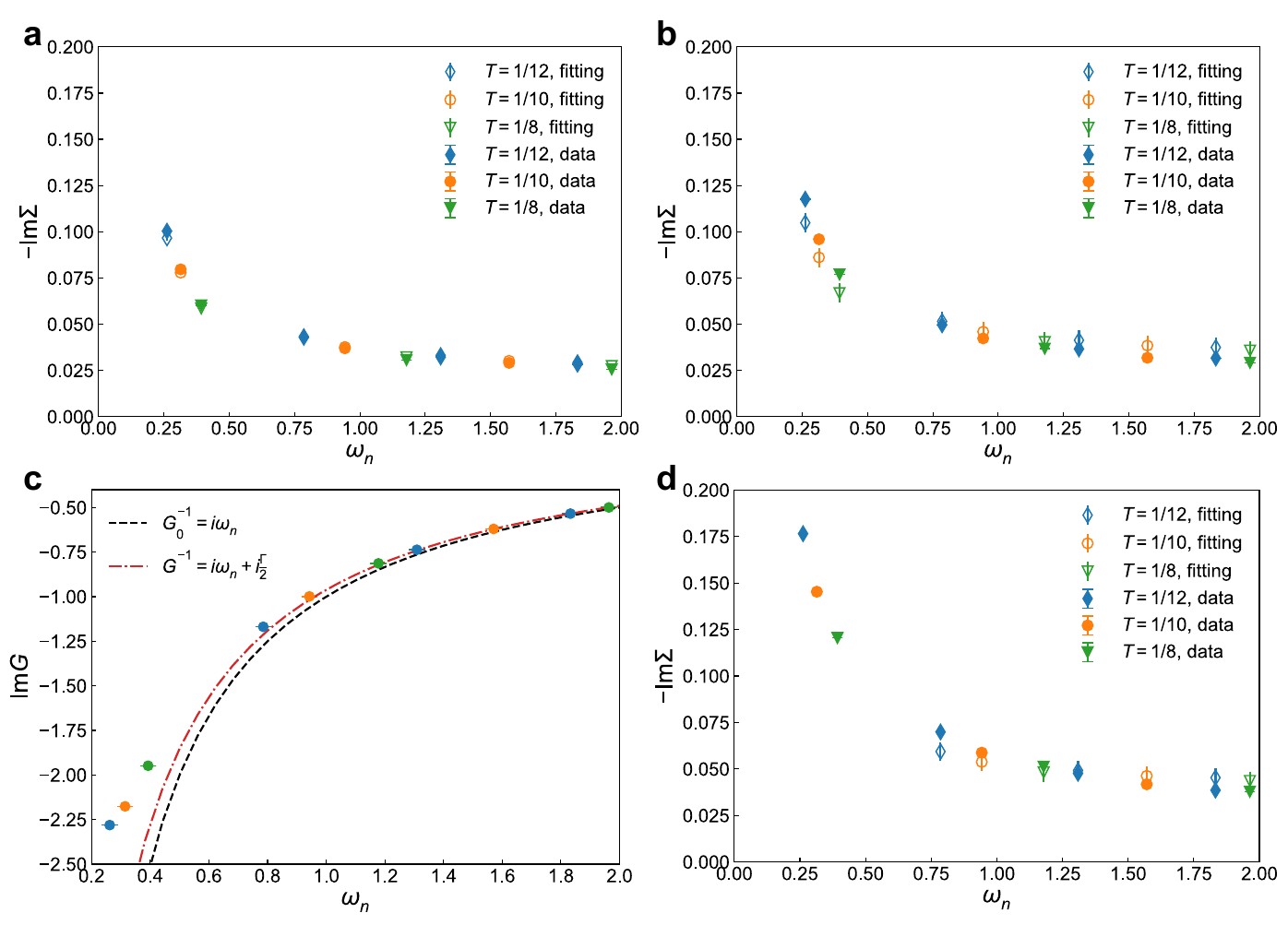}
  \centering
  \caption{\label{fig:hsigma}{Fitting DQMC data by appling UT.}
  {(a)} Fitting of self-energy when $\sigma_h = 0.2$.
  {(b)} Fitting of self-energy when $\sigma_h = 0.5$.
  {(c)} Fitting of Green's funtion when $\sigma_h = 1.0$.
  {(d)} Fitting of self-energy after dropping the first Matsubara frequency when $\sigma_h = 1.0$.}
\end{figure*}

\bigskip\noindent
\textbf{Fermionic properties for the spatially disordered model} \\
In this part, we study the effects of spatially disordered interactions by combining MET and UT. We first introduce a disorder field term into the Ising part of the original Hamiltonian,
\begin{eqnarray}
	\hat{H}_{s} = -J \sum_{\langle ij \rangle} \hat{Z}_i \hat{Z}_j - \sum_{i} (h+h'_i) \hat{X}_i,
\end{eqnarray}
where $h$ is the original spatially uniform part, and $h'_i$ is the spatially disordered part and conforms to the Gaussian distribution
\begin{eqnarray}
\overline{h'_i h'_j} = \sigma_h^2 \delta_{ij}, \quad
\overline{h'_i} = 0.	
\end{eqnarray}
In field theory, the disorder field acts as a ``scalar mass disorder", which is equivalent to a spatially random Yukawa coupling term ($g'$ term in the following) and a spatially random potential term ($v$ term in the following)~\cite{patelUniversalTheoryStrange2023}.

When the spatial disorder is applied, the translation symmetry is broken. After the disorder average procedure similar to the Yukawa SYK model, a statistical spatial translational symmetry is restored, and a similar effective action is derived~\cite{patelUniversalTheoryStrange2023}. We can thus propose the universal theory of our model, which is actually equivalent to a correction of MET when the spatial disorder is applied.
\begin{align}
	\Pi(\mathbf{{r}},\tau) =& -D_0 N_f g^{2}G(-\mathbf{{r}}, -\tau)G(\mathbf{{r}},\tau) \nonumber  \\
  &-D_0 g'^{2}G(\mathbf{{r}}=0,-\tau)G(\mathbf{{r}}=0,\tau)\delta^{2}(\mathbf{{r}}),\\
	\text{\ensuremath{\Sigma(\mathbf{{r}}, \tau)}}=&g^{2}D(\mathbf{{r}},\tau)G(\mathbf{{r}},\tau)
  +v^{2}G(\mathbf{{r}}=\mathbf{{0}},\tau)\delta^{2}(\mathbf{{r}}) \nonumber \\
  &+g'^{2}G(\mathbf{{r}}=\mathbf{{0}},\tau)D(\mathbf{{r}}=\mathbf{{0}},\tau)\delta^{2}(\mathbf{{r}}),\\
	G(\mathbf{{k}},\omega_n)=&\frac{{1}}{\i\omega_n-\varepsilon_{{k}}+\mu-\Sigma(\mathbf{{k}},\omega_n)}, \\
	D(\mathbf{{q}},\Omega_m)=&\frac{{D_0}}{c^{-2}\Omega_m^{2}+q^{2}+M^{2}-\Pi(\mathbf{{q}},\Omega_m)}.
\end{align}

Unlike the case in Yukawa SYK model, the contributions from random Yukawa coupling in our model turn out to be weak. Thus, to study the effect of disorder, we include the explicit leading term $\i\frac{\Gamma}{2} \mathrm{sgn}(\omega_n) = \i\frac{v^2 k_F}{2 v_F} \mathrm{sgn}(\omega_n)$ in Green's function which comes from random potential $v^2 G(\mathbf{r}=\mathbf{0},\tau)\delta^2(\mathbf{r})$ in $\Sigma(\mathbf{r},\tau)$. Following the procedures in MET, we first calculate the bosonic self-energy, which has the form
\begin{eqnarray}
  \Pi(\mathbf{q},\Omega_m)
  = - \mathcal{N} \overline{g} \frac{|\Omega_m|}{\sqrt{(\Gamma+\Omega_m)^2 + v_F^2 q^2}}
  - \frac{\pi}{2}\mathcal{N}^2 \overline{g}'|\Omega_m| \nonumber \\
\end{eqnarray}
with $\mathcal{N} = \frac{k_F}{2\pi v_F}$, $\overline{g}' = D_0 g'^2$ and $\overline{g}=D_0 g^2 = D_0 (\xi/2)^2$. Due to $\overline{g}'$ and $\mathcal{N}$ are small numbers, the second term can be neglected when calculating the fermionic self-energy.

The thermal part $\Sigma_T(\mathbf{q},\omega_n)$ is 
  \begin{align}
    &\Sigma_{T}(\mathbf{{k}},\omega_n)
    =\Sigma_{T,g}(\mathbf{{k}},\omega_n)+\Sigma_{T,g'}(\omega_n),\\
    &\Sigma_{T,g}(\mathbf{{k}},\omega_n) \nonumber \\
    &=-\i\overline{g}T\mathrm{{sgn}}(\omega_n)
    \int_{0}^{+\infty}\frac{{qdq}}{2\pi}
    \frac{{1}}{\sqrt{(\omega_n+\frac{{\Gamma}}{2})^{2}+v_{F}^{2}q^{2}}}
    \frac{{1}}{M^{2}+q^{2}} \nonumber \\
    &= -\i \frac{\overline{g}T\mathrm{sgn}(\omega_n)}{2\pi(\omega_n + \frac{\Gamma}{2})} 
    \mathcal{S}(\frac{v_FM}{|\omega_n+\frac{\Gamma}{2}|}), \\
    &\Sigma_{T,g'}(\omega_n)
    =-\i\frac{{\mathcal{{N}}\overline{g}'\mathrm{{sgn}}(\omega_n)}}{4}T
    \ln\left(1+\frac{{\Lambda^{2}}}{M^{2}}\right).
  \end{align}
$\Lambda$ is the UV cutoff of momentum $q$. $\Sigma_{T, g}$ is similar to the one in MET with an energy shift. The quantum part $\Sigma_Q(\mathbf{q},\omega_n)$ is much complex,
\begin{align}
  &\Sigma_{Q}(\mathbf{{k}},\omega_n)=\Sigma_{Q,g}(\mathbf{{k}},\omega_n)+\Sigma_{Q,v}(\omega_n)+\Sigma_{Q,g'}(\omega_n), \\
  &\Sigma_{Q,g}(\omega_n) \nonumber \\
  &=-\i\overline{g}\int_{0}^{+\infty}\frac{{qdq}}{2\pi}\int_{-\infty}^{+\infty}\frac{{d\Omega_n}}{2\pi}
  \frac{{\mathrm{{sgn}}(\omega_n+\Omega_n)}}{\sqrt{{(\omega_n+\Omega_n+\frac{{\Gamma}}{2})^{2}+v_{F}^{2}q^{2}}}} 
  \nonumber \\
  &\quad\times \frac{{1}}{M^{2}+q^{2}+c^{-2}\Omega_n^{2}+c_{b}\frac{{|\Omega_n|}}{\sqrt{{(\Gamma+\Omega_n)^{2}+v_{F}^{2}q^{2}}}}}, \\
  &\Sigma_{Q,v}(\omega_n) 
  =-\i\frac{{v^{2}k_{F}}}{2v_{F}}\mathrm{{sgn}}(\omega_n)
  =-\i\frac{{\Gamma}}{2}\mathrm{sgn}(\omega_n), \nonumber \\
  &\Sigma_{Q,g'}(\omega_n)
  =-\frac{{\i\mathcal{{N}}\overline{g}'\omega_n}}{6\pi}\ln\left(\frac{{e\Lambda^{3}}}{c_{b}|\omega_n|}\right).
\end{align}
Here $c_b = \mathcal{N} \overline{g}$. An interesting thing is that there is a marginal Fermi liquid term $\sim \omega_n \ln \omega_n$, which may be a source of linear-$T$ resistivity according to Ref.~\cite{patelUniversalTheoryStrange2023}. Even though the $\Sigma_{Q, g}$ is very similar to the one in MET, we can not separate out the correction of $v$ directly by using Taylor expansion due to the convergence radius is not guaranteed when $q$ is small. The detailed derivation is included in SI Appendix 8.

We use the complete form to fit our data, and $\overline{g}$ is chosen to be the one we get in MET to reduce the number of fitting parameters. The remaining unknown parameters in this theory are $g'$, $v$, and $\Lambda$. 
We find that any physical cutoff ($\Lambda > 1/a$ with $a$ the lattice constant, which is one in our simulation) will result in near constant $v$ and $g'$, as shown in SI Appendix Fig.~S14. Thus, we fix $\Lambda = 2$ and fit $g'$ and $v$. The $g'$ we get is very close to zero, which means that the MFL term is weak in this case. Fig.~\ref{fig:hsigma} shows the fitting result of our theory. When the disorder strength is weak, our theory fits well with the data obtained in DQMC. However, when the disorder strength is strong, the approximation $G^{-1} = \i\omega_n + \i\frac{\Gamma}{2}$  fails for the first Matsubara frequency (Fig.~\ref{fig:hsigma}{\bf c}), which suffers from more severe finite-size effects and thermal fluctuations. This phenomenon has also been seen in MET~\cite{chubukovFirstMatsubarafrequencyRuleFermi2012, wuSpecialRoleFirst2019,wangSuperconductivityQuantumCriticalPoint2016}. After dropping the data of the first Matsubara frequency, the data and the fitting are in good agreement even when the disorder strength is strong (Fig.~\ref{fig:hsigma}{\bf d}).

\begin{figure*}[t]
\centering
  \includegraphics[width=17cm]{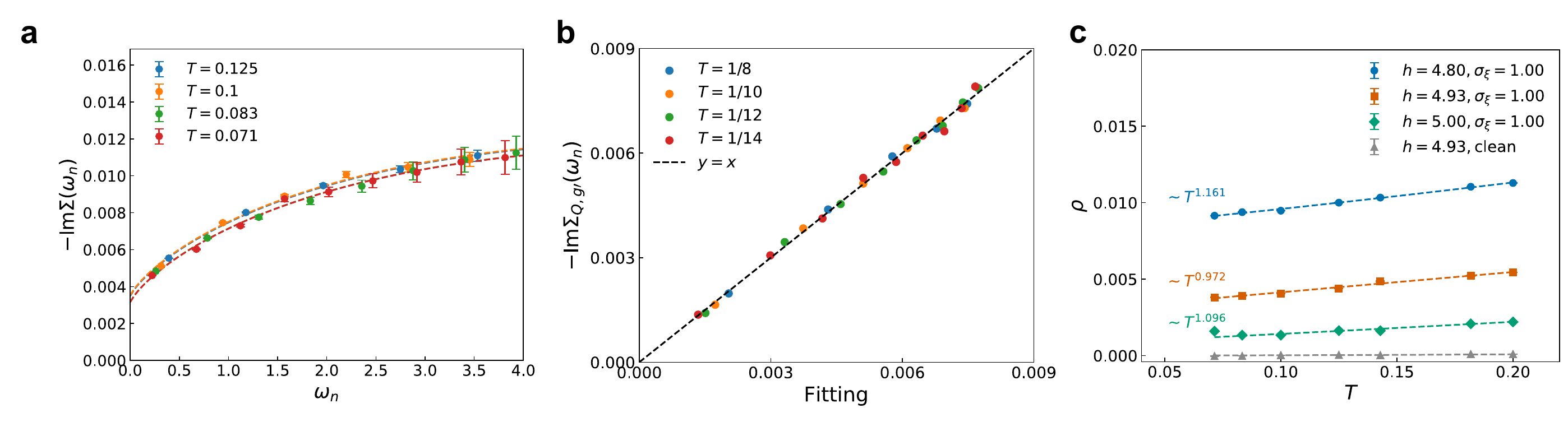}
  \centering
  \caption{\label{fig:MFL} {{Marginal Fermi liquid behavior.}
    The fermionic self-energy shows an expected MFL scaling when a completely random coupling interaction is applied. (a) The imaginary part of the total fermionic self-energy. (b) After subtracting the thermal part, the imaginary part of the remaining quantum part can be well fitted by a MFL form $a \omega_n \ln(b/\omega_n)$. The $x$-axis values denote the fitting results, and the $y$-axis values denote the Monte Carlo data; all the points are located closely on the $y=x$ line. (c) DC resistivity proxy $\rho$ extracted from the current-current correlation function.}}
\end{figure*}

{In order to observe MFL, we have to increase $g'$, and we try to use a completely random coupling interaction
\begin{equation}
    \hat{H}_{sf} = -\sum_i \xi’_i \hat{Z}_i(\hat{\sigma}^z_{i1} + \hat{\sigma}^z_{i2})
\end{equation}
with $\xi'_i$ obeys Gaussian distribution $\overline{\xi'_i \xi'_j} = \sigma_\xi^2 \delta_{ij}, \ 
\overline{\xi'_i} = 0$. At the same time, we remove the disordered field $h'$ and set $\sigma_\xi = 1$.
In this case,  the fermionic self-energy only contains one MFL term and a temperature-dependent constant caused by thermal fluctuations,
\begin{eqnarray}
    \Sigma(\mathbf{k}, \omega_n) &= \Sigma_{Q, g'}(\mathbf{k}, \omega_n) + \Sigma_{T, g'}(\mathbf{k}, \omega_n).
\end{eqnarray}
In a smeared critical region, we find the expected MFL scaling of $\mathrm{Im},\Sigma$, as shown in Fig.~\ref{fig:MFL}{\bf b} for a typical $h$ ($h=5.0$). The results for several other values of $h$ within the smeared critical region are shown in SI Appendix 9. Compared to the spatially disordered transverse-field case, the effective parameter $\bar{g}'$ in the MFL self-energy scaling form is significantly enhanced in the spatially disordered coupling case, as shown in Table S2 in SI Appendix 9. Most importantly, we measure the resistivity in this MFL region and indeed find linear-in-temperature behavior (Fig.~\ref{fig:MFL}{\bf c}). As a comparison, we also calculate the resistivity near the QCP ($h=4.93$) of the clean model and find that the resistivity is very small, which verifies that small-$q$ ferromagnetic fluctuations do not contribute to the linear-in-$T$ resistivity. Instead, the spatially disordered coupling term drives the marginal Fermi liquid behavior. The details of how to extract resistivity from current–current correlation functions~\cite{huang2019strange,ledererSuperconductivityNonFermiLiquid2017} can be found in SI Appendix 10.

\bigskip\noindent
\textbf{Bosonic properties for the spatially disordered model} \\
When spatially disordered interactions are not applied, the magnetic susceptibility is predicted to have a mean-field effective form in the HMM framework~\cite{millisEffectNonzeroTemperature1993}. However, previous studies have found that the susceptibility deviates from the mean-field prediction 
both in theory and numerics~\cite{liuItinerantQuantumCritical2018,xuNonFermiLiquid+12017,millisEffectNonzeroTemperature1993}, which means that the bosonic critical modes are strongly renormalized due to coupling to gapless fermions. We show the quantum critical scaling analysis of the boson properties of the clean model in SI Appendix 5. While in the following, we will mainly focus on the case of the spatially disordered model, and in particular the case with completely random Yukawa coupling where the MFL self-energy is found.

It has been studied that a ``quantum Griffiths phase"~\cite{vojta2010quantum} will emerge in the critical region when the spatially disordered Yukawa coupling is introduced. Localized bosonic modes may be found in this phase, which are significant for understanding the extended regime~\cite{Patel2024strange, patelLocalizationOverdampedBosonic2024}. 
To explore the bosonic localization, we demonstrate the eigenmode spectrum of the zero-frequency bosonic propagator in Fig.~\ref{fig:boson}, where the localization length $\mathcal{L}_\alpha$ is calculated by the relation to the participation ratio $\mathcal{I}_\alpha$~\cite{Patel2024strange}
\begin{eqnarray}
\begin{aligned}
 \mathcal{L}_{\alpha} = \frac{1}{\sqrt{2\mathcal{I}_\alpha}}, \quad \mathcal{I}_{\alpha} = \sum_{\mathbf{r}} |\psi_{\alpha,\mathbf{r}}|^4.
\end{aligned}
\end{eqnarray}
The bosonic eigenmodes $\psi_{\alpha,\mathbf{r}}$ in the above formula are got from diagonalization of the zero-frequency bosonic propagator,
\begin{eqnarray}
\begin{aligned}
    D(\text{i}\Omega_m=0, \mathbf{r}_1, \mathbf{r}_2) 
    &= \langle Z(\text{i}\Omega_m=0, \mathbf{r}_1) \cdot Z(\text{i}\Omega_m=0, \mathbf{r}_2) \rangle \\
    &= \sum_{\alpha = 0}^{L^2 - 1} \frac{\psi_{\alpha,\mathbf{r}_1} \psi_{\alpha,\mathbf{r}_2}}{e_{\alpha}}.
\end{aligned}
\end{eqnarray}

Clear localized bosonic modes were not observed, even down to the lowest temperature simulated ($T=1/40$), as the localization length of all modes is close to $L/2$. However, if we plot the density profiles of those modes, we indeed find some tendency of localization in the intermediate energy region (Fig.~\ref{fig:boson}{\bf f} and {\bf g}). In the disordered phase (large $h$) region, the delocalized lowest-energy mode become more and more extended  upon entering deeper into the ordered phase (small $h$), and finally becomes gapped (Fig.~\ref{fig:boson}{\bf a}) from the remaining eigenmodes and reverts to plane-wave like state (Fig.~\ref{fig:boson}{\bf d}). The presence of this energy gap aligns with previous hybrid Monte Carlo findings reported in Ref.\cite{patelLocalizationOverdampedBosonic2024}. The absence of clear localized modes can potentially be attributed to two factors: firstly, the lack of continuous symmetry of bosons in the $N_b=1$ model precludes a quantum Griffiths phase; secondly, it might stem from the finite-size and temperature limitations of our study. Importantly, this absence suggests that the self-averaging assumption remains valid, implying that the UT is likely applicable within the temperature range investigated.

We also investigate the scaling of $D^{-1}(\text{i}\Omega_m, \mathbf{q}, \mathbf{q})$ at $h = h_c$. Our findings reveal a $q^2$ dependence at all frequencies. Unlike previous studies, where zero-frequency data exhibited significant sensitivity to the specific disorder sample \cite{patelLocalizationOverdampedBosonic2024}, we observe a clear $q^2$ dependence even at zero frequency. This consistency aligns with our results from the inverse boson propagator spectrum. We also observe an obvious deviation from the $|\Omega_m|^2$ scaling of $D^{-1}(\text{i}\Omega_m, \mathbf{q}, \mathbf{q})$ and we predict a $|\Omega_m|$ scaling in the low-frequency region. The results are provided in SI Appendix 5. Those results are consistent with the UT in Ref.~\cite{patelUniversalTheoryStrange2023}.

\begin{figure*}[t]
   \includegraphics[width=17cm]{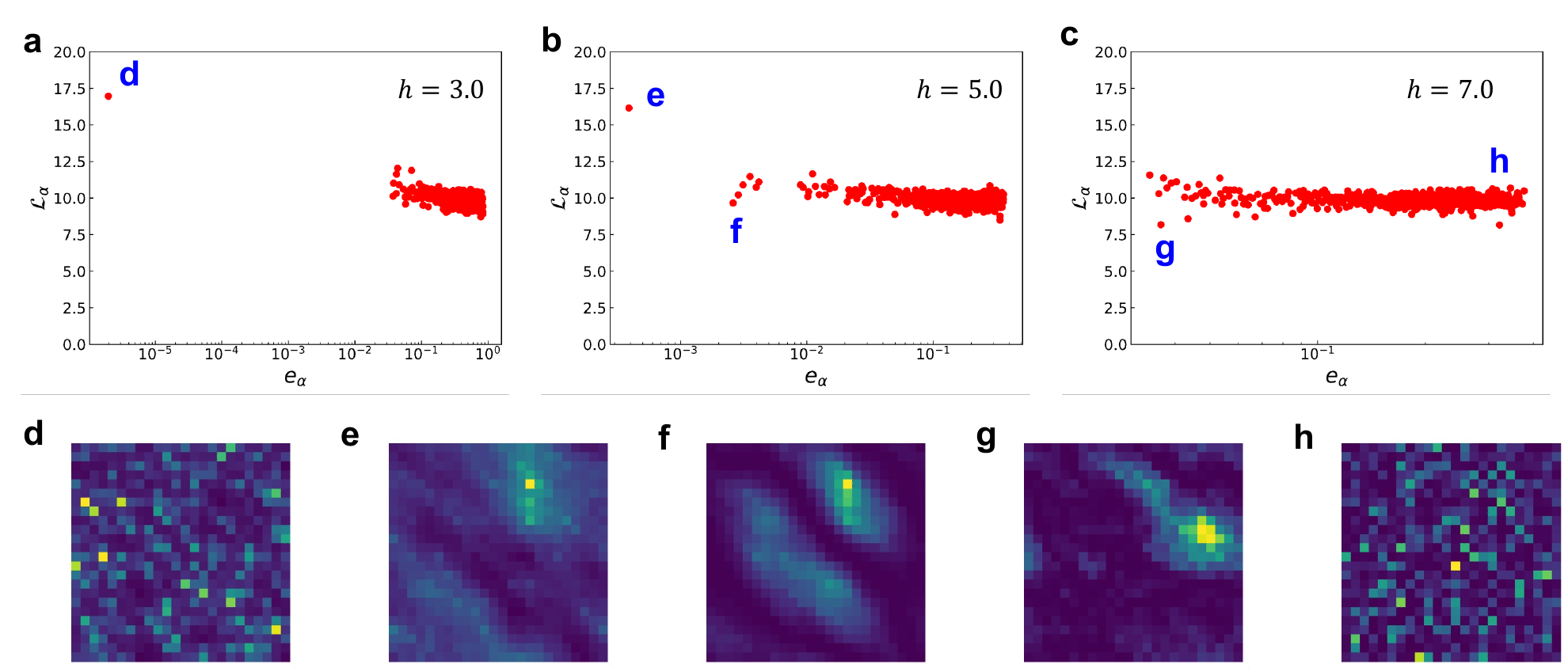}
   \centering
  \caption{\label{fig:boson} 
  {{Localization lengths $\mathcal{L}_{\alpha}$ of eigenvalues $e_{\alpha}$ and density profiles $|\psi_{\alpha,\mathbf{r}}|^2$} of one disorder realization. The results are obtained at $\beta = 40$ and $L = 24$. The wave functions in (d)-(h) have been scaled to a square for easier viewing. (d): plane-wave like mode deep in the ordered phase. (e)-(g): delocalized modes in the low and intermediate energy region. (h): plane-wave like mode in the high-energy region.}}
\end{figure*}

\bigskip \noindent {\bf Discussion}\\
In this work, we performed large-scale DQMC simulations of an FM spin-fermion bilayer model on the triangular lattice and explored the effects of spatial disorder in this model. 
When spatially disordered interactions are not applied, the fermionic self-energy data fit well with the modified Eliashberg theory (MET) which predicts a vanishing quasiparticle weight at zero temperature, a characteristic behaviors of nFL.

To deal with the case when a spatial disorder field is applied, we absorb the spirit in the universal theory (UT) of strange metals~\cite{patelUniversalTheoryStrange2023} and treat the spatial disorder in a self-average manner. This brings us a random potential and a random Yukawa coupling term in the effective action. To solve the self-consistent equations, the main approximation we take is to include explicit contributions from disorder in the Green's function, which is demonstrated to be very reasonable by our data. By applying the corrected MET we derived, the fermionic self-energy data agree well with the theory. More in detail, we considered two kinds of spatial disorder. For the spatially disordered transverse field case, the fermionic self-energy is dominated by the thermal contribution and the quantum critical part, thus no evidence of a MFL type of self-energy is found. For the spatially disordered coupling case, the MFL type of self-energy is significantly enhanced, and we found it in a smeared critical region of the phase diagram. We also observe linear in temperature scaling of resistivity in this MFL region.

{In addition, recent studies~\cite{patelLocalizationOverdampedBosonic2024} suggest that localization of the bosonic field may occur at low temperatures, which undermines the self-averaging assumption of the UT. Due to the absence of the quantum Griffith phase in our $N_b = 1$ model and the limitation of temperature, no obvious localized bosonic mode is found in our simulations, indicating that the UT is likely applicable within the temperature range investigated. However, the emergence of localized modes is itself an interesting phenomenon and is significant for understanding the extended regime in the critical region. We note that the extension of our spin-fermion model to $N_b > 1$ is straightforward, and a previous work has studied the $N_b=2$ case in the clean limit~\cite{jiangMonteCarloStudy2022}. Finding the quantum Griffith phase and localized modes in $N_b > 1$ extension of our spin-fermion model is a promising future work.

\bigskip \noindent {\bf Methods}\\
\noindent 
\textbf{DQMC simulations and data analysis.}
We implement the DQMC algorithm to simulate our system. The starting point of DQMC is the Hamiltonian in~(\ref{eq:eq1}). By integrating the fermionic quadratic form, the partition function can be written as the sum of weights corresponding to different auxiliary field configurations, and Markov chain Monte Carlo can be performed~\cite{assaadWorldlineDeterminantalQuantum2008}.
To accelerate simulations, we employ an efficient update framework for both fermionic and bosonic parts. The global update algorithms of the Ising model are very mature and can certainly be applied in our model at any point in the parameter space. 
In this work, we combine a local update, a Wolff-Cluster update~\cite{wolff1989collective} and a geometric cluster update~\cite{heringaGeometricClusterMonte1998} in one sweep. When spatial disorder is applied, we remove the geometric cluster update in our update strategy due to the broken inversion symmetry. For the update of Green's function, a general delay update and submatrix update algorithms \cite{sunDelayUpdateDeterminant2023, sun2025boosting} are proposed recently, which can effectively increase the computation efficiency, especially for large systems. Details of DQMC are given in SI Appendix 1.

\bigskip\noindent
\textbf{Modified Eliashberg theory.}
The DQMC simulations are carried out at finite temperatures due to the potential existence of a superconducting dome, and thermal fluctuations will inevitably affect our measurements. For the fermionic self-energy, this effect manifests as an increase as the temperature decreases.
Analytically, the physics of the spin-fermion bilayer model can be described by the low-energy effective theory termed the modified Eliashberg theory (MET). This theory assumes that near a QCP, a soft collective boson is a slow mode compared to a dressed fermion. This effectively decouples the fermionic and bosonic degrees of freedom and makes the problem analytically tractable. MET considers the thermal contributions from the static bosonic propagator, and we can use MET to separate this contribution from the whole. 
For details, please see Ref.~\cite{kleinNormalStateProperties2020,xuIdentificationNonFermiLiquid2020}.
We apply MET to get the effective coupling $\overline{g}$ and predict the quasiparticle weight. In SI Appendix Fig. S12, we have shown the results of the MET. All the data points fit well with the theory, and the scaling form $\Sigma \sim \omega_n^{2/3}$ will be recovered at low frequencies. Thus, we can predict the quasiparticle weight will vanish when $T \rightarrow 0$, as expected for an nFL. Due to the renormalization of high-energy fermions, the effective coupling will deviate from the bare one $\overline{g} = D_0(\xi / 2)^2$ and we obtain $\overline{g} = 0.172(8)$ along $k_x$ direction and $\overline{g} = 0.218(3)$ along $k_y$ direction.

\bigskip\noindent 
\textbf{Universal theory of strange metals.}
To explain the results when a random transverse field is applied, we introduce the universal theory of strangle metals proposed in Ref.~\cite{patelUniversalTheoryStrange2023}.
This theory treats the disorder in a self-average manner and leads to a statistical translation invariant effective action. In the large-$N$ approximation framework, the self-consistent function of self-energies can be derived by solving saddle point equations.
Following the spirit of this theory, we have corrected MET to satisfy the spatial disorder case. We find that the main effect of this random transverse field in our model is a random potential of fermions, and the random Yukawa coupling contribution is weak. Thus, we include the random potential term in Green's function, and this approximation is proven to be reasonable. The corrected MET theory fits well with the data from DQMC simulations as shown in Fig.~\ref{fig:hsigma}.
To explore the existence of marginal Fermi liquid, we remove the fixed background coupling and random field. Instead, a completely random coupling interaction is applied, and an MFL scaling self-energy is found in the critical region.
\\

\noindent\textit{Note added}: After the completion of this work, we became aware of a study that appeared on arXiv (\href{https://arxiv.org/abs/2410.05365}{arXiv:2410.05365})~\cite{Patel2024strange} last week, which investigates spatial disorder effects on an antiferromagnetic spin-fermion model with hybrid Monte Carlo.

\bigskip \noindent \textbf{Data availability}\\
The data that support the findings of this study are available from the corresponding author upon reasonable request.

\bigskip \noindent \textbf{Code availability}\\
All numerical codes in this paper are available upon request to the authors.\\


\bibliographystyle{unsrt}
\bibliography{references}

\begin{thebibliography}{10}

\bibitem{patelUniversalTheoryStrange2023}
Aavishkar~A. Patel, Haoyu Guo, Ilya Esterlis, and Subir Sachdev.
\newblock Universal theory of strange metals from spatially random interactions.
\newblock {\em Science}, 381(6659):790--793, August 2023.

\bibitem{kleinNormalStateProperties2020}
Avraham Klein, Andrey~V. Chubukov, Yoni Schattner, and Erez Berg.
\newblock Normal {{State Properties}} of {{Quantum Critical Metals}} at {{Finite Temperature}}.
\newblock {\em Physical Review X}, 10(3):031053, September 2020.

\bibitem{lohneysenFermiliquidInstabilitiesMagnetic2007a}
Hilbert~V. L{\"o}hneysen, Achim Rosch, Matthias Vojta, and Peter W{\"o}lfle.
\newblock Fermi-liquid instabilities at magnetic quantum phase transitions.
\newblock {\em Reviews of Modern Physics}, 79(3):1015--1075, August 2007.

\bibitem{leeRecentDevelopmentsNonFermi2018}
Sung-Sik Lee.
\newblock Recent {{Developments}} in {{Non-Fermi Liquid Theory}}.
\newblock {\em Annual Review of Condensed Matter Physics}, 9(1):227--244, March 2018.

\bibitem{liSuperconductivityInfinitelayerNickelate2019}
Danfeng Li, Kyuho Lee, Bai~Yang Wang, Motoki Osada, Samuel Crossley, Hye~Ryoung Lee, Yi~Cui, Yasuyuki Hikita, and Harold~Y. Hwang.
\newblock Superconductivity in an infinite-layer nickelate.
\newblock {\em Nature}, 572(7771):624--627, August 2019.

\bibitem{sunSignaturesSuperconductivity802023}
Hualei Sun, Mengwu Huo, Xunwu Hu, Jingyuan Li, Zengjia Liu, Yifeng Han, Lingyun Tang, Zhongquan Mao, Pengtao Yang, Bosen Wang, Jinguang Cheng, Dao-Xin Yao, Guang-Ming Zhang, and Meng Wang.
\newblock Signatures of superconductivity near 80 {{K}} in a nickelate under high pressure.
\newblock {\em Nature}, 621(7979):493--498, September 2023.

\bibitem{zhuSuperconductivityPressurizedTrilayer2024}
Yinghao Zhu, Di~Peng, Enkang Zhang, Bingying Pan, Xu~Chen, Lixing Chen, Huifen Ren, Feiyang Liu, Yiqing Hao, Nana Li, Zhenfang Xing, Fujun Lan, Jiyuan Han, Junjie Wang, Donghan Jia, Hongliang Wo, Yiqing Gu, Yimeng Gu, Li~Ji, Wenbin Wang, Huiyang Gou, Yao Shen, Tianping Ying, Xiaolong Chen, Wenge Yang, Huibo Cao, Changlin Zheng, Qiaoshi Zeng, Jian-gang Guo, and Jun Zhao.
\newblock Superconductivity in pressurized trilayer {{La4Ni3O10}}-{$\delta$} single crystals.
\newblock {\em Nature}, 631(8021):531--536, July 2024.

\bibitem{cao2018unconventional}
Yuan Cao, Valla Fatemi, Shiang Fang, Kenji Watanabe, Takashi Taniguchi, Efthimios Kaxiras, and Pablo {Jarillo-Herrero}.
\newblock Unconventional superconductivity in magic-angle graphene superlattices.
\newblock {\em Nature}, 556(7699):43--50, 2018.

\bibitem{cao2020strange}
Yuan Cao, Debanjan Chowdhury, Daniel {Rodan-Legrain}, Oriol {Rubies-Bigorda}, Kenji Watanabe, Takashi Taniguchi, T~Senthil, and Pablo {Jarillo-Herrero}.
\newblock Strange metal in magic-angle graphene with near {{Planckian}} dissipation.
\newblock {\em Physical review letters}, 124(7):076801, 2020.

\bibitem{stewart2001non}
{\relax GR}~Stewart.
\newblock Non-{{Fermi-liquid}} behavior in d-and f-electron metals.
\newblock {\em Reviews of modern Physics}, 73(4):797, 2001.

\bibitem{hartnollHolographicQuantumMatter2016}
Sean~A Hartnoll, Andrew Lucas, and Subir Sachdev.
\newblock {\em Holographic quantum matter}.
\newblock MIT press, 2018.

\bibitem{keimerQuantumMatterHightemperature2015}
B.~Keimer, S.~A. Kivelson, M.~R. Norman, S.~Uchida, and J.~Zaanen.
\newblock From quantum matter to high-temperature superconductivity in copper oxides.
\newblock {\em Nature}, 518(7538):179--186, February 2015.

\bibitem{hertzQuantumCriticalPhenomena1976}
John~A. Hertz.
\newblock Quantum critical phenomena.
\newblock {\em Physical Review B}, 14(3):1165--1184, August 1976.

\bibitem{millisEffectNonzeroTemperature1993}
A.~J. Millis.
\newblock Effect of a nonzero temperature on quantum critical points in itinerant fermion systems.
\newblock {\em Physical Review B}, 48(10):7183--7196, September 1993.

\bibitem{moriyaSpinFluctuationsItinerant1985}
T{\^o}ru Moriya.
\newblock {\em Spin {{Fluctuations}} in {{Itinerant Electron Magnetism}}}, volume~56 of {\em Springer {{Series}} in {{Solid-State Sciences}}}.
\newblock Springer, Berlin, Heidelberg, 1985.

\bibitem{polchinskiLowEnergyDynamics1994}
Joseph Polchinski.
\newblock Low {{Energy Dynamics}} of the {{Spinon-Gauge System}}.
\newblock {\em Nuclear Physics B}, 422(3):617--633, July 1994.

\bibitem{kimGaugeinvariantResponseFunctions1994}
Yong~Baek Kim, Akira Furusaki, Xiao-Gang Wen, and Patrick~A. Lee.
\newblock Gauge-invariant response functions of fermions coupled to a gauge field.
\newblock {\em Physical Review B}, 50(24):17917--17932, December 1994.

\bibitem{metlitskiQuantumPhaseTransitions2010}
Max~A. Metlitski and Subir Sachdev.
\newblock Quantum phase transitions of metals in two spatial dimensions. {{II}}. {{Spin}} density wave order.
\newblock {\em Physical Review B}, 82(7):075128, August 2010.

\bibitem{abanovQuantumcriticalTheorySpinfermion2001}
Ar~Abanov, Andrey~V Chubukov, and J{\"o}rg Schmalian.
\newblock Quantum-critical theory of the spin-fermion model and its application to cuprates: Normal state analysis.
\newblock {\em Advances in Physics}, 52(3):119--218, 2003.

\bibitem{mrossControlledExpansionCertain2010}
David~F. Mross, John McGreevy, Hong Liu, and T.~Senthil.
\newblock A controlled expansion for certain non-{{Fermi}} liquid metals.
\newblock {\em Physical Review B}, 82(4):045121, July 2010.

\bibitem{si2001locally}
Qimiao Si, Silvio Rabello, Kevin Ingersent, and J~Lleweilun Smith.
\newblock Locally critical quantum phase transitions in strongly correlated metals.
\newblock {\em Nature}, 413(6858):804--808, 2001.

\bibitem{coleman2001fermiliquids}
Piers Coleman, C~P{\'e}pin, Qimiao Si, and Revaz Ramazashvili.
\newblock How do fermiliquids get heavy and die?
\newblock {\em Journal of Physics: Condensed Matter}, 13(35):R723, 2001.

\bibitem{senthil2004weak}
T.~Senthil, Matthias Vojta, and Subir Sachdev.
\newblock Weak magnetism and non-fermi liquids near heavy-fermion critical points.
\newblock {\em Phys. Rev. B}, 69:035111, Jan 2004.

\bibitem{schroder2000onset}
A~Schr{\"o}der, G~Aeppli, R~Coldea, M~Adams, O~Stockert, Hv~L{\"o}hneysen, E~Bucher, R~Ramazashvili, and Piers Coleman.
\newblock Onset of antiferromagnetism in heavy-fermion metals.
\newblock {\em Nature}, 407(6802):351--355, 2000.

\bibitem{custers2003break}
Jeroen Custers, Philipp Gegenwart, H~Wilhelm, K~Neumaier, Y~Tokiwa, O~Trovarelli, C~Geibel, F~Steglich, C~P{\'e}pin, and Piers Coleman.
\newblock The break-up of heavy electrons at a quantum critical point.
\newblock {\em Nature}, 424(6948):524--527, 2003.

\bibitem{paschen2004hall}
Silke Paschen, Thomas L{\"u}hmann, Steffen Wirth, Philipp Gegenwart, Octavio Trovarelli, Christoph Geibel, Frank Steglich, Piers Coleman, and Qimiao Si.
\newblock Hall-effect evolution across a heavy-fermion quantum critical point.
\newblock {\em Nature}, 432(7019):881--885, 2004.

\bibitem{friedemann2010fermi}
Sven Friedemann, Niels Oeschler, Steffen Wirth, Cornelius Krellner, Christoph Geibel, Frank Steglich, Silke Paschen, Stefan Kirchner, and Qimiao Si.
\newblock Fermi-surface collapse and dynamical scaling near a quantum-critical point.
\newblock {\em Proceedings of the National Academy of Sciences}, 107(33):14547--14551, 2010.

\bibitem{bergSignproblemfreeQuantumMonte2012a}
Erez Berg, Max~A. Metlitski, and Subir Sachdev.
\newblock Sign-problem-free quantum {{Monte Carlo}} of the onset of antiferromagnetism in metals.
\newblock {\em Science}, 338(6114):1606--1609, December 2012.

\bibitem{schattnerIsingNematicQuantum2016}
Yoni Schattner, Samuel Lederer, Steven~A. Kivelson, and Erez Berg.
\newblock Ising {{Nematic Quantum Critical Point}} in a {{Metal}}: {{A Monte Carlo Study}}.
\newblock {\em Physical Review X}, 6:031028, July 2016.

\bibitem{ledererSuperconductivityNonFermiLiquid2017}
Samuel Lederer, Yoni Schattner, Erez Berg, and Steven~A. Kivelson.
\newblock Superconductivity and non-{{Fermi}} liquid behavior near a nematic quantum critical point.
\newblock {\em Proceedings of the National Academy of Sciences}, 114(19):4905--4910, May 2017.

\bibitem{grossmanSpecificHeatQuantum2021}
Ori Grossman, Johannes~S. Hofmann, Tobias Holder, and Erez Berg.
\newblock Specific {{Heat}} of a {{Quantum Critical Metal}}.
\newblock {\em Physical Review Letters}, 127:017601, July 2021.

\bibitem{xuNonFermiLiquid+12017}
Xiao~Yan Xu, Kai Sun, Yoni Schattner, Erez Berg, and Zi~Yang Meng.
\newblock Non-{{Fermi Liquid}} at (2 +1 ){{D Ferromagnetic Quantum Critical Point}}.
\newblock {\em Physical Review X}, 7:031058, July 2017.

\bibitem{jiangMonteCarloStudy2022}
Weilun Jiang, Yuzhi Liu, Avraham Klein, Yuxuan Wang, Kai Sun, Andrey~V. Chubukov, and Zi~Yang Meng.
\newblock Monte {{Carlo}} study of the pseudogap and superconductivity emerging from quantum magnetic fluctuations.
\newblock {\em Nature Communications}, 13(1):2655, May 2022.

\bibitem{schattnerCompetingOrdersNearly2016}
Yoni Schattner, Max~H. Gerlach, Simon Trebst, and Erez Berg.
\newblock Competing {{Orders}} in a {{Nearly Antiferromagnetic Metal}}.
\newblock {\em Physical Review Letters}, 117(9):097002, August 2016.

\bibitem{liNatureEffectiveInteraction2017a}
Zi-Xiang Li, Fa~Wang, Hong Yao, and Dung-Hai Lee.
\newblock Nature of the effective interaction in electron-doped cuprate superconductors: {{A}} sign-problem-free quantum {{Monte Carlo}} study.
\newblock {\em Physical Review B}, 95(21):214505, June 2017.

\bibitem{liuItinerantQuantumCritical2018}
Zi~Hong Liu, Xiao~Yan Xu, Yang Qi, Kai Sun, and Zi~Yang Meng.
\newblock Itinerant quantum critical point with frustration and a non-{{Fermi}} liquid.
\newblock {\em Physical Review B}, 98(4):045116, 2018.

\bibitem{liuItinerantQuantumCritical2019}
Zi~Hong Liu, Gaopei Pan, Xiao~Yan Xu, Kai Sun, and Zi~Yang Meng.
\newblock Itinerant {{Quantum Critical Point}} with {{Fermion Pockets}} and {{Hot Spots}}.
\newblock {\em Proceedings of the National Academy of Sciences}, 116(34):16760--16767, August 2019.

\bibitem{lunts2023non}
Peter Lunts, Michael~S Albergo, and Michael Lindsey.
\newblock Non-{{Hertz-Millis}} scaling of the antiferromagnetic quantum critical metal via scalable {{Hybrid Monte Carlo}}.
\newblock {\em Nature communications}, 14(1):2547, 2023.

\bibitem{leeLowenergyEffectiveTheory2009}
Sung-Sik Lee.
\newblock Low-energy effective theory of {{Fermi}} surface coupled with {{U}}(1) gauge field in 2 + 1 dimensions.
\newblock {\em Physical Review B}, 80(16):165102, October 2009.

\bibitem{michonReconcilingScalingOptical2023a}
B.~Michon, C.~Berthod, C.~W. Rischau, A.~Ataei, L.~Chen, S.~Komiya, S.~Ono, L.~Taillefer, D.~{van der Marel}, and A.~Georges.
\newblock Reconciling scaling of the optical conductivity of cuprate superconductors with {{Planckian}} resistivity and specific heat.
\newblock {\em Nature Communications}, 14(1):3033, May 2023.

\bibitem{cooper2009anomalous}
Robert~A Cooper, Y~Wang, Baptiste Vignolle, {\relax OJ}~Lipscombe, Stephen~M Hayden, Yoichi Tanabe, Tadashi Adachi, Yoji Koike, Minoru Nohara, Hidenori Takagi, et~al.
\newblock Anomalous criticality in the electrical resistivity of la2--x sr x cuo4.
\newblock {\em Science}, 323(5914):603--607, 2009.

\bibitem{greeneStrangeMetalState2020a}
Richard~L. Greene, Pampa~R. Mandal, Nicholas~R. Poniatowski, and Tarapada Sarkar.
\newblock The {{Strange Metal State}} of the {{Electron-Doped Cuprates}}.
\newblock {\em Annual Review of Condensed Matter Physics}, 11(1):213--229, March 2020.

\bibitem{guoLarge$N$Theory2022}
Haoyu Guo, Aavishkar~A. Patel, Ilya Esterlis, and Subir Sachdev.
\newblock Large \${{N}}\$ theory of critical {{Fermi}} surfaces {{II}}: Conductivity.
\newblock {\em Physical Review B}, 106(11):115151, September 2022.

\bibitem{esterlisLarge$N$Theory2021}
Ilya Esterlis, Haoyu Guo, Aavishkar~A. Patel, and Subir Sachdev.
\newblock Large \${{N}}\$ theory of critical {{Fermi}} surfaces.
\newblock {\em Physical Review B}, 103(23):235129, June 2021.

\bibitem{nosov2020}
P.~A. Nosov, I.~S. Burmistrov, and S.~Raghu.
\newblock Interaction-induced metallicity in a two-dimensional disordered non-fermi liquid.
\newblock {\em Phys. Rev. Lett.}, 125:256604, Dec 2020.

\bibitem{foster2022}
Tsz~Chun Wu, Yunxiang Liao, and Matthew~S. Foster.
\newblock Quantum interference of hydrodynamic modes in a dirty marginal fermi liquid.
\newblock {\em Phys. Rev. B}, 106:155108, Oct 2022.

\bibitem{sachdevGaplessSpinFluidGround1993}
Subir Sachdev and Jinwu Ye.
\newblock Gapless {{Spin-Fluid Ground State}} in a {{Random Quantum Heisenberg Magnet}}.
\newblock {\em Physical Review Letters}, 70(21):3339--3342, May 1993.

\bibitem{kitaev2015talks}
AY~Kitaev.
\newblock Talks at kitp, university of california, santa barbara.
\newblock {\em Entanglement in Strongly-Correlated Quantum Matter}, 2015.

\bibitem{maldacena2016remarks}
Juan Maldacena and Douglas Stanford.
\newblock Remarks on the sachdev-ye-kitaev model.
\newblock {\em Physical Review D}, 94(10):106002, 2016.

\bibitem{sachdevBekensteinHawkingEntropyStrange2015}
Subir Sachdev.
\newblock Bekenstein-{{Hawking Entropy}} and {{Strange Metals}}.
\newblock {\em Physical Review X}, 5(4):041025, November 2015.

\bibitem{patelLocalizationOverdampedBosonic2024}
Aavishkar~A. Patel, Peter Lunts, and Subir Sachdev.
\newblock Localization of overdamped bosonic modes and transport in strange metals.
\newblock {\em Proceedings of the National Academy of Sciences}, 121(14):e2402052121, April 2024.

\bibitem{blankenbeclerMonteCarloCalculations1981}
R.~Blankenbecler, D.~J. Scalapino, and R.~L. Sugar.
\newblock Monte {{Carlo}} calculations of coupled boson-fermion systems. {{I}}.
\newblock {\em Physical Review D}, 24(8):2278--2286, October 1981.

\bibitem{scalapinoMonteCarloCalculations1981}
D.~J. Scalapino and R.~L. Sugar.
\newblock Monte {{Carlo}} calculations of coupled boson-fermion systems. {{II}}.
\newblock {\em Physical Review B}, 24(8):4295--4308, October 1981.

\bibitem{assaadWorldlineDeterminantalQuantum2008}
F.F. Assaad and H.G. Evertz.
\newblock World-line and {{Determinantal Quantum Monte Carlo Methods}} for {{Spins}}, {{Phonons}} and {{Electrons}}.
\newblock In H.~Fehske, R.~Schneider, and A.~Wei{\ss}e, editors, {\em Computational {{Many-Particle Physics}}}, pages 277--356. Springer, Berlin, Heidelberg, 2008.

\bibitem{lohStableNumericalSimulations1992}
E.Y. Loh and J.E. Gubernatis.
\newblock Stable {{Numerical Simulations}} of {{Models}} of {{Interacting Electrons}} in {{Condensed-Matter Physics}}.
\newblock In {\em Modern {{Problems}} in {{Condensed Matter Sciences}}}, volume~32, pages 177--235. Elsevier, 1992.

\bibitem{sorella1988numerical}
S~Sorella, E~Tosatti, S~Baroni, R~Car, and M~Parrinello.
\newblock Numerical simulation of the {{1D}} and {{2D Hubbard}} models: {{Fermi}} liquid behavior and its breakdown.
\newblock In {\em Towards the Theoretical Understanding of High Temperature Superconductors-Proceedings of the Adriatico Research Conference and Workshop}, volume~14, page 457. World Scientific, 1988.

\bibitem{sugiyama1986auxiliary}
G~Sugiyama and {\relax SE}~Koonin.
\newblock Auxiliary field {{Monte-Carlo}} for quantum many-body ground states.
\newblock {\em Annals of Physics}, 168(1):1--26, 1986.

\bibitem{xuRevealingFermionicQuantum2019}
Xiao~Yan Xu, Zi~Hong~Liu, Gaopei Pan, Yang Qi, Kai Sun, and Zi~Yang Meng.
\newblock Revealing fermionic quantum criticality from new {{Monte Carlo}} techniques.
\newblock {\em Journal of Physics Condensed Matter}, 31:463001, November 2019.

\bibitem{xuIdentificationNonFermiLiquid2020}
Xiao~Yan Xu, Avraham Klein, Kai Sun, Andrey~V. Chubukov, and Zi~Yang Meng.
\newblock Identification of non-{{Fermi}} liquid fermionic self-energy from quantum {{Monte Carlo}} data.
\newblock {\em npj Quantum Materials}, 5:65, September 2020.

\bibitem{wuSufficientConditionAbsence2005}
Congjun Wu and Shou-Cheng Zhang.
\newblock Sufficient condition for absence of the sign problem in the fermionic quantum {{Monte Carlo}} algorithm.
\newblock {\em Physical Review B}, 71(15):155115, April 2005.

\bibitem{pfeuty1971ising}
P~Pfeuty and {\relax RJ}~Elliott.
\newblock The {{Ising}} model with a transverse field. {{II}}. {{Ground}} state properties.
\newblock {\em Journal of Physics C: Solid State Physics}, 4(15):2370, 1971.

\bibitem{kukreja2024space}
Shubham Kukreja, Afshin Besharat, and Sung-Sik Lee.
\newblock Projective fixed points for non-fermi liquids: A case study of the ising-nematic quantum critical metal.
\newblock {\em Physical Review B}, 110(15):155142, 2024.

\bibitem{bhattacharjee2001measure}
Somendra~M Bhattacharjee and Flavio Seno.
\newblock A measure of data collapse for scaling.
\newblock {\em Journal of Physics A: Mathematical and General}, 34(33):6375, 2001.

\bibitem{shao2016quantum}
Hui Shao, Wenan Guo, and Anders~W Sandvik.
\newblock Quantum criticality with two length scales.
\newblock {\em Science}, 352(6282):213--216, 2016.

\bibitem{qinDualityDeconfinedQuantumCritical2017}
Yan~Qi Qin, Yuan-Yao He, Yi-Zhuang You, Zhong-Yi Lu, Arnab Sen, Anders~W. Sandvik, Cenke Xu, and Zi~Yang Meng.
\newblock Duality between the {{Deconfined Quantum-Critical Point}} and the {{Bosonic Topological Transition}}.
\newblock {\em Physical Review X}, 7(3):031052, September 2017.

\bibitem{chubukovFirstMatsubarafrequencyRuleFermi2012}
Andrey~V. Chubukov and Dmitrii~L. Maslov.
\newblock First-{{Matsubara-frequency}} rule in a {{Fermi}} liquid. {{I}}. {{Fermionic}} self-energy.
\newblock {\em Physical Review B}, 86(15):155136, October 2012.

\bibitem{wuSpecialRoleFirst2019}
Yi-Ming Wu, Artem Abanov, Yuxuan Wang, and Andrey~V. Chubukov.
\newblock Special role of the first {{Matsubara}} frequency for superconductivity near a quantum critical point: {{Nonlinear}} gap equation below {{T}} c and spectral properties in real frequencies.
\newblock {\em Physical Review B}, 99(14):144512, April 2019.

\bibitem{wangSuperconductivityQuantumCriticalPoint2016}
Yuxuan Wang, Artem Abanov, Boris~L. Altshuler, Emil~A. Yuzbashyan, and Andrey~V. Chubukov.
\newblock Superconductivity near a {{Quantum-Critical Point}}: {{The Special Role}} of the {{First Matsubara Frequency}}.
\newblock {\em Physical Review Letters}, 117(15):157001, October 2016.

\bibitem{huang2019strange}
Edwin~W Huang, Ryan Sheppard, Brian Moritz, and Thomas~P Devereaux.
\newblock Strange metallicity in the doped hubbard model.
\newblock {\em Science}, 366(6468):987--990, 2019.

\bibitem{vojta2010quantum}
Thomas Vojta.
\newblock Quantum griffiths effects and smeared phase transitions in metals: theory and experiment.
\newblock {\em Journal of Low Temperature Physics}, 161:299--323, 2010.

\bibitem{Patel2024strange}
Aavishkar~A Patel, Peter Lunts, and Michael~S Albergo.
\newblock Strange metals and planckian transport in a gapless phase from spatially random interactions.
\newblock {\em arXiv preprint arXiv:2410.05365}, 2024.

\bibitem{wolff1989collective}
Ulli Wolff.
\newblock Collective {{Monte Carlo}} updating for spin systems.
\newblock {\em Physical Review Letters}, 62(4):361, 1989.

\bibitem{heringaGeometricClusterMonte1998}
J.~R. Heringa and H.~W.~J. Bl{\"o}te.
\newblock Geometric cluster {{Monte Carlo}} simulation.
\newblock {\em Physical Review E}, 57(5):4976--4978, May 1998.

\bibitem{sunDelayUpdateDeterminant2023}
Fanjie Sun and Xiao~Yan Xu.
\newblock Delay update in determinant quantum monte carlo.
\newblock {\em Phys. Rev. B}, 109:235140, Jun 2024.

\bibitem{sun2025boosting}
Fanjie Sun and Xiao~Yan Xu.
\newblock Boosting determinant quantum monte carlo with submatrix updates: Unveiling the phase diagram of the 3d hubbard model.
\newblock {\em SciPost Physics}, 18(2):055, 2025.

\bibitem{assaadDepletedKondoLattices2002}
F.~F. Assaad.
\newblock Depleted {{Kondo}} lattices: {{Quantum Monte Carlo}} and mean-field calculations.
\newblock {\em Physical Review B}, 65(11):115104, February 2002.

\bibitem{xuMonteCarloStudy2019}
Xiao~Yan Xu, Yang Qi, Long Zhang, Fakher~F. Assaad, Cenke Xu, and Zi~Yang Meng.
\newblock Monte {{Carlo Study}} of {{Lattice Compact Quantum Electrodynamics}} with {{Fermionic Matter}}: {{The Parent State}} of {{Quantum Phases}}.
\newblock {\em Physical Review X}, 9(2):021022, May 2019.

\bibitem{paiva2004critical}
Thereza Paiva, Raimundo~R Dos~Santos, {\relax RT}~Scalettar, and {\relax PJH}~Denteneer.
\newblock Critical temperature for the two-dimensional attractive {{Hubbard}} model.
\newblock {\em Physical Review B}, 69(18):184501, 2004.

\bibitem{chenLifshitzTransitionTwo2012}
Kuang-Shing Chen, Zi~Yang Meng, Thomas Pruschke, Juana Moreno, and Mark Jarrell.
\newblock Lifshitz {{Transition}} in the {{Two Dimensional Hubbard Model}}.
\newblock {\em Physical Review B}, 86(16):165136, October 2012.

\bibitem{chubukov2010hidden}
Andrey~V Chubukov.
\newblock Hidden one-dimensional physics in {{2D}} critical metals.
\newblock {\em Physics}, 3:70, 2010.

\bibitem{punkFiniteTemperatureScaling2016}
Matthias Punk.
\newblock Finite temperature scaling close to {{Ising-nematic}} quantum critical points in two-dimensional metals.
\newblock {\em Physical Review B}, 94(19):195113, November 2016.

\bibitem{wangNonFermiLiquidsFinite2017}
Huajia Wang and Gonzalo Torroba.
\newblock Non-{{Fermi}} liquids at finite temperature: Normal state and infrared singularities.
\newblock {\em Physical Review B}, 96(14):144508, October 2017.

\end{thebibliography}

\bigskip \noindent {\bf Acknowledgements}\\
The authors thank Subir Sachdev and Aavishkar A. Patel for helpful discussions.
This work was supported by the National Key R\&D Program of China (Grant No. 2022YFA1402702, No. 2021YFA1401400), the National Natural Science Foundation of China (Grants No. 12447103, No. 12274289), the Innovation Program for Quantum Science and Technology (under Grant No. 2021ZD0301902), Yangyang Development Fund, Shanghai Jiao Tong University 2030 Initiative, and startup funds from SJTU. The computations in this paper were run on the Siyuan-1 and $\pi$ 2.0 clusters supported by the Center for High Performance Computing at Shanghai Jiao Tong University.\\

\bigskip \noindent {\bf Author contributions}\\
T.H. and X.Y.X. performed research, analyzed data, and wrote the paper.

\bigskip \noindent {\bf Competing interests}\\
The authors declare no competing interests.





\appendix
\newpage
\clearpage
\onecolumngrid
\mbox{}
\begin{center}
\textbf{\large{Supplementary Information for\\
Monte Carlo Study of Critical Fermi Surface with Spatially Disordered Interactions}\\
Hong \textit{et al}.}
\end{center}

\setcounter{equation}{0}
\setcounter{figure}{0}
\setcounter{table}{0}
\setcounter{subsection}{0}

\renewcommand{\theequation}{\arabic{equation}}
\renewcommand{\thefigure}{\arabic{figure}}
\renewcommand{\bibnumfmt}[1]{[#1]}
\renewcommand{\citenumfont}[1]{#1}

\setcounter{equation}{0}
\setcounter{figure}{0}
\setcounter{table}{0}
\setcounter{section}{0}
\setcounter{page}{1}
\setcounter{subsection}{0}

\makeatletter

\textbf{Supplementary Note 1: DQMC Implementations}
\label{app:app1}

In DQMC, the quartic interactions are decomposed into fermionic bilinear operators at the cost of introducing coupling with auxiliary fields.
Then, the Markov chain Monte Carlo is implemented to sample the auxiliary fields and obtain statistical values of observables.
To efficiently evaluate the partition function, Suzuki-Trotter decomposition is applied, which splits the imaginary time into small time slices, e.g.
\begin{equation}
  Z = \mathrm{Tr} \left[ e^{-\beta \hat{H}} \right] = \mathrm{Tr} \left[ (e^{-\Delta \tau \hat{H}})^M \right],
\end{equation}
with $\beta = M \Delta \tau$. Considering the large transverse field across the FM-PM phase transition in our model, we choose $\Delta \tau = 0.02$ to ensure the convergence of Trotter error, as shown in Supplementary Figure~\ref{fig:dtau}.
In our case, there is no explicit four-body interaction, and the Ising field is the auxiliary field.

Due to the existence of the Ising field, the trace can be divided into a sum for Ising configurations and a trace for fermions,
\begin{eqnarray}
  Z = && \mathrm{Tr} [e^{-\beta \hat{H}}] \nonumber \\
  = && \mathrm{Tr} [(e^{-\Delta \tau \hat{H}})^M] \nonumber \\
  = && \sum_{Z_1 \cdots Z_N = \pm 1} 
  \mathrm{Tr}_F \langle Z_1 \cdots Z_N 
  | (e^{-\Delta \tau \hat{H}})^M | Z_1 \cdots Z_N \rangle.
\end{eqnarray}
We can use $\mathbf{Z} = (Z_1 \cdots Z_N)$ to denote the Ising configuration and insert the complete Ising basis between time slices; then
\begin{eqnarray}
  Z = && \sum_{\mathbf{Z}_1 \cdots \mathbf{Z}_M} \mathrm{Tr}_F \langle \mathbf{Z}_1 | e^{-\Delta \tau \hat{H}} | \mathbf{Z}_M \rangle 
  \langle \mathbf{Z}_M | e^{-\Delta \tau \hat{H}} | \mathbf{Z}_{M-1} \rangle \nonumber \\
  && \cdots \langle \mathbf{Z}_2 | e^{\Delta \tau \hat{H}} | \mathbf{Z}_1 \rangle.  
\end{eqnarray}

By tracing out the fermion degrees of freedom, we obtain the form of the partition function used in the DQMC framework and the corresponding configurational weight,
\begin{eqnarray}
  Z = \sum_{\mathbf{Z}_1 \cdots \mathbf{Z}_M} \omega_{\mathcal{C}}^{TI} \omega_{\mathcal{C}}^F.
\end{eqnarray}
Here $\mathcal{C}$ is short for configuration $\mathbf{Z}_1 \cdots \mathbf{Z}_M$. $\omega_{\mathcal{C}}^{TI}$ denotes the weight for the Ising part and can be transformed into a $(d + 1)$ dimension Ising model, e.g.,
\begin{equation}
  \omega_{\mathcal{C}}^{TI} 
  = \left( \prod_{\tau} \prod_{\langle i, j \rangle} 
  e^{\Delta \tau J Z_{i, \tau} Z_{j \tau}} \right)
  \left( \prod_{\langle \tau, \tau' \rangle} 
  \prod_i \Lambda e^{\gamma Z_{\tau, i} Z_{\tau', j}} 
  \right),
\end{equation}
with $\Lambda^2 = \sinh(\Delta \tau h)\cosh(\Delta \tau h)$ and $\gamma = -\frac{1}{2} \ln(\tanh(\Delta h))$. The fermion part is a determinant for each spin and layer. Due to the anti-unitary symmetry $i \tau_y \mathcal{K}$ (where $\tau_y$ is a Pauli matrix defined in the layer space and $\mathcal{K}$ is the complex conjugate operator), the determinants of different layers are conjugate to each other~\cite{bergSignproblemfreeQuantumMonte2012a,wuSufficientConditionAbsence2005,xuNonFermiLiquid+12017}, and consequently the fermion part weight $\omega_\mathcal{C}^F$ can be written as
\begin{eqnarray}
  \omega^F_{\mathcal{C}} 
  = && \prod_{\lambda = 1, 2} \prod_{\sigma = \uparrow, \downarrow} \det(\mathbf{I} + \mathbf{B}^{\lambda \sigma}_M \cdots \mathbf{B}^{\lambda \sigma}_1) \\
  = && \Big | \prod_{\sigma = \uparrow, \downarrow} \det(\mathbf{I} + \mathbf{B}^{1 \sigma}_M \cdots \mathbf{B}^{1 \sigma}_1) \Big |^2.
\end{eqnarray}
with 
\begin{eqnarray}
\mathbf{B}^{\lambda \sigma}_{\tau} = 
\exp(-\Delta \tau \mathbf{K}^{\lambda \sigma} + 
\Delta \tau \xi \mathrm{diag}(Z_1, \cdots, Z_N)).
\end{eqnarray}
Obviously, the fermion part weight is a non-negative number. Note that the Ising part weight is always positive, therefore, the total weight is non-negative, and the system is free of sign problem.  

Near the phase transition point, the local update suffers from the slowing down problem. Many global update strategies are invented to suppress this phenomenon, such as the Wolff-Cluster algorithm~\cite{wolff1989collective} and the geometric cluster algorithm~\cite{heringaGeometricClusterMonte1998}. 
These global methods can certainly be applied in our model at any point in the parameter space to improve the efficiency of the simulation. In this work, we combine a local update, a Wolff-Cluster update, and a geometric cluster update in one sweep. Note that in both local and global update, the Ising part and fermion part weights are properly considered to satisfy the detailed balance condition. However, the geometric cluster update is based on the lattice geometry symmetry of the Hamiltonian. When spatial disorder is applied, the lattice geometry symmetry is broken, and we only use local update and Wolff-Cluster update in this case.
For the update of Green's function, a delayed update algorithm~\cite{sunDelayUpdateDeterminant2023} has been proposed recently, which can effectively increase the computation efficiency, especially for large systems.
\\

\begin{figure}
  \includegraphics[width=14cm]{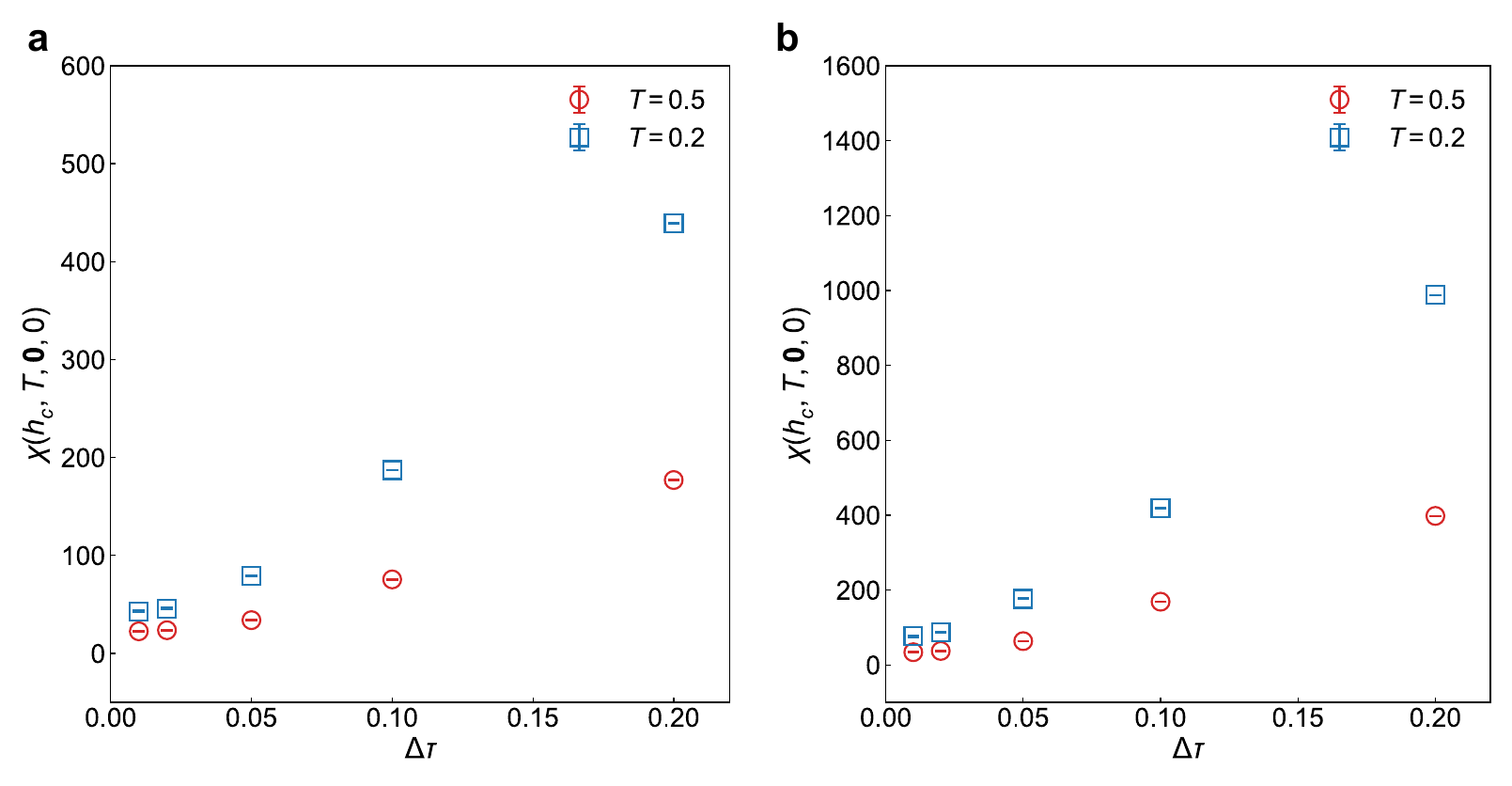}
  \centering
  \caption{{The plot of Ising spin susceptibility $\chi(h_c, T, \mathbf{0}, 0)$ 
  as it varies with Trotter step $\Delta \tau$.} 
  {(a)} $L = 12$. 
  {(b)} $L = 18$.
  We choose $\Delta \tau = 0.02$ in simulations to ensure the 
  convergence of observables.
  \label{fig:dtau} 
  }
\end{figure}

\textbf{Supplementary Note 2: Z-direction Flux}

In order to reduce the finite-size effects, we use the techniques from Ref.~\cite{assaadDepletedKondoLattices2002}. 
The main idea is to guess a site-dependent hopping parameter $t_{\vec{i}, \vec{j}}(L)$ which satisfies
\begin{equation}
\lim_{L \rightarrow \infty} t_{\vec{i}, \vec{j}}(L) = t
\label{A1}
\end{equation}
and will minimize the finite-size effects. This can be achieved by applying a uniform $z$-axis magnetic field $\vec{B}^{\lambda \sigma}$, and the hopping parameter will be multiplied by a Peierls phase factor, i.e.,
\begin{equation}
  -t \hat{c}^{\dagger}_{i \lambda \sigma} 
  \rightarrow 
  -t e^{i\phi^{\lambda \sigma}_{ij}} \hat{c}^{\dagger}_{i\lambda \sigma} \hat{c}_{j\lambda \sigma}
\end{equation}
with $\phi^{\lambda \sigma}_{ij} 
= 
[(2\pi)/\Phi_0] \int_{\vec{i}}^{\vec{j}} \vec{A}^{\lambda \sigma}(\vec{r}) \cdot d\vec{r}$ and
$\vec{B}^{\lambda \sigma}(\vec{r}) = \nabla \times \vec{A}^{\lambda \sigma}(\vec{r})$.

To maintain the invariance under anti-unitary transformation $i \tau_y \mathcal{K}$, we take 
\begin{equation}
  \phi_{ij}^{1 \uparrow} = \phi_{ij}^{1 \downarrow} = -\phi_{ij}^{2 \uparrow} = -\phi_{ij}^{2 \downarrow},
\end{equation}
which corresponds to an unreal magnetic field due to its difference in different flavors of fermions. In the following contents, we only concentrate on $\phi_{ij}^{1 \uparrow}$ and will omit the superscript. The phase for other layers and spins can be obtained by the above formula.

Since the geometry of the system is a torus, and a translation in the argument of the vector potential can be absorbed in a gauge transformation,
\begin{eqnarray}
  \vec{A}(\vec{r} + L\vec{a}) &= \vec{A}(\vec{r}) + \nabla \chi_a(\vec{r}), \\
  \vec{A}(\vec{r} + L\vec{b}) &= \vec{A}(\vec{r}) + \nabla \chi_b(\vec{r}).
\end{eqnarray}
The Landau gauge we take is $\vec{A}(\vec{r}) = -B(y, 0, 0)$ so we can choose $\chi_a(\vec{i})=-BLa_2 x$ and $\chi_b(\vec{i}) = -BLb_2 x$ for lattice with primitive cell $\vec{a}=(a_1, a_2, 0)$ and $\vec{b}=(b_1, b_2, 0)$. Then, the boundary condition is
\begin{eqnarray}
  \hat{c}^{\dagger}_{\vec{i}+L\vec{a}} &&= e^{i(2\pi/\Phi_0) \chi_a(\vec{i})} \hat{c}^{\dagger}_i \\
  \hat{c}^{\dagger}_{\vec{i}+L\vec{b}} &&= e^{i(2\pi/\Phi_0) \chi_b(\vec{i})} \hat{c}^{\dagger}_i
\end{eqnarray}
to satisfy that $[H, T_{L\vec{a}}] = [H, T_{L\vec{b}}] = 0$.

However, magnetic translation operators belong to the magnetic algebra
\begin{equation}
  T_{L \vec{a}} T_{L \vec{b}} = e^{-i 2\pi [(L^2 / \Phi_0) (\vec{a} \times \vec{b}) \cdot \vec{B}]} T_{L\vec{b}} T_{L\vec{a}}.
\end{equation}
Thus to obtain a single-valued wave function, namely, to make the above magnetic translation operators commute with each other, the condition of flux quantization has to be satisfied $\frac{L^2}{\Phi_0}(\vec{a} \times \vec{b}) \cdot \vec{B} = n$, where $n$ is an integer. Then the smallest $B$ is 
\begin{equation}
  B = \frac{\Phi_0}{L^2(a_1 b_2 - a_2 b_1)},
  \label{A9}
\end{equation}
which will vanish at the thermodynamic limit, and the Hamiltonian goes back to the original one. 

The $z$-direction flux will break the translation symmetry and make the information in the momentum space unavailable. 
When such information is needed, the $z$-direction flux is turned off, and the flux in the $x-y$ plane is applied instead, which is equivalent to the twisted boundary condition. 
In this work, we use $(0, 0)$, $\frac{\pi}{12}(1, \frac{1}{\sqrt{3}})$,
$\frac{\pi}{12}(0, \frac{2}{\sqrt{3}})$, $\frac{\pi}{12}(1, {\sqrt{3}})$
and cubic extrapolation to increase the resolution and draw the FS. 

Another important detail we want to emphasize is that the $z$-direction flux may break the Hermiticity of Hamiltonian at the boundary. 
In the original Hamiltonian, the Hermiticity is guaranteed automatically by the Hermitian conjugate part, while this may be invalid when a $z$-flux is applied due to the difference of phases.

Consider one hopping term from $n_1 \vec{a} + L \vec{b}$ to $n_2 \vec{a} + (L + 1) \vec{b}$, the phase comes from the Peierls phase factor and the boundary condition. We can calculate the total phase is
\begin{eqnarray}
\phi_1 =&& \frac{2\pi}{\Phi_0} (-\frac{B}{2}) [(n_2 - n_1) a_1 + b_1] \nonumber \\
&& \times[(n_1 + n_2)a_2 + (2L+1)b_2] \\
&& +\frac{2\pi}{\Phi_0} BL b_2(n_2 a_1 + b_1) \nonumber.
\end{eqnarray}
The inverse process is the hopping from $n_2 \vec{a} + \vec{b}$ to $n_1 \vec{a}$, and its phase is
\begin{eqnarray}
\phi_2 =&& \frac{2\pi}{\Phi_0} (\frac{B}{2}) [(n_2 - n_1) a_1 + b_1] \nonumber \\
&& \times[(n_1 + n_2)a_2 + b_2] \\ 
&& -\frac{2\pi}{\Phi_0} BL b_2(n_1 a_1 + L b_1) \nonumber.
\end{eqnarray}
Then we can get the sum of $\phi_1$ and $\phi_2$ is
\begin{eqnarray}
  \phi_1 + \phi_2 
  && =  -\frac{2\pi}{\Phi_0} B L^2 b_1 b_2 \nonumber \\
  && = - {2 n \pi} \frac{b_1 b_2}{(a_1 b_2 - a_2 b_1)}.
\end{eqnarray} 

Hermiticity requires the sum of $\phi_1$ and $\phi_2$ to be a multiple of $2\pi$. On a square lattice, there is no limitation because $b_1b_2 = 0$, while for a triangular lattice, $n$ can only be a multiple of $2$, or we should keep using bond crossing one of the boundaries, and the inverse hopping is obtained by enforcing the Hermitian conjugate.
\\

\textbf{Supplementary Note 3: Phase Transition and Crossing-point Analysis}
\begin{figure}
  \includegraphics[width=14cm]{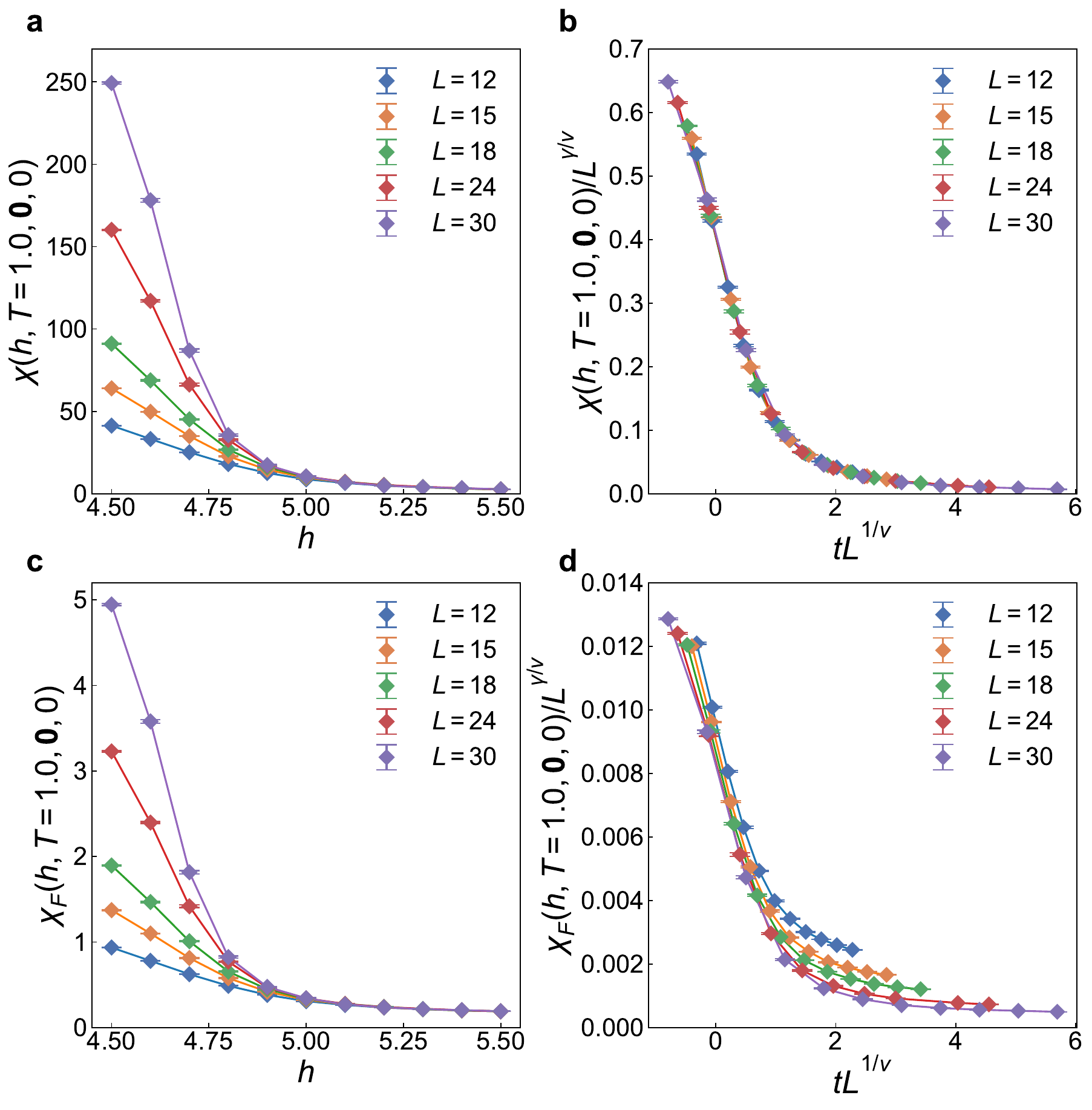}
  \centering
  \caption{
  {FM-PM phase transition at finite-temperature $T = 1.0$.} 
  {(a)} The Ising spin susceptibilities of different system sizes 
  at $\mathbf{q} = \mathbf{0}$ and $\Omega_n = 0$. 
  {(b)} The data collapse result of Ising spin susceptibility. 
  Critical exponents of 2D Ising universality 
  class $\nu = 1$ and $\gamma = 7/4$ are used. The only fitting parameter 
  is the critical field $h_N \approx 4.63$. Curves of all system sizes fall on 
  the same line.
  {(c)} The fermion spin susceptibilities of different system sizes 
  at $\mathbf{q} = \mathbf{0}$ and $\Omega_n = 0$.
  {(d)} The 
  data collapse result of fermion spin susceptibility. 
  The $h_N$ used is the one obtained from Ising susceptibility.
  Curves of all system sizes don't fall on the same line.
  \label{fig:S1} 
  }
\end{figure}

In the study, the first step is determining the phase boundary and quantum critical point (QCP), and the observable quantity we study to characterize the phase transition is the susceptibility of spins.
\begin{eqnarray}
  \chi(h, T, \mathbf{q}, \Omega_n) = \frac{1}{L^2} \sum_{ij} \int_{0}^{\beta} d\tau e^{i\Omega_n \tau - i \mathbf{q}\cdot\mathbf{r}_{ij}} 
  \langle Z_i(\tau) Z_j(0) \rangle.
\end{eqnarray}
When considering the zero-frequency ($\Omega_n = 0$) and zero-momentum ($\mathbf{q}=\mathbf{0}$) case near finite-temperature critical transverse field $h_N$, the susceptibility satisfies
\begin{eqnarray}
  \chi(h, T, \mathbf{0}, 0) = L^{\gamma/\nu} f((h - h_N) L^{1/\nu}),
\end{eqnarray}
where $f$ is a universal function and obeys the scaling law of Ising universality, so we take $\nu = 1$ and $\gamma = 7/4$. 
Obviously, once $h_N$ is determined, data $\chi L^{-\gamma/\nu}$ of all size $L$ will overlap on the curve described by $f$.
The data collapse method in Ref.~\cite{bhattacharjee2001measure} can help us obtain the value $h_N$.
We show the behavior of $\chi(h, T, \mathbf{0}, 0)$ and the date collapse results in Supplementary Figure~\ref{fig:S1}{\bf a} and {\bf b}. We fix a temperature $T = 1.0$ here and use the transverse field as a variable. The critical field given by data collapse is $h_N \approx 4.63$. 

Correspondingly, we can define a similar quantity for fermion spins,
\begin{eqnarray}
  \chi_F(h, T, \mathbf{q}, \Omega_n) = \frac{1}{L^2} \int_{0}^{\beta} d\tau 
  \sum_{ij\lambda\lambda'} e^{i\Omega_n \tau - i \mathbf{q}\cdot\mathbf{r}_{ij}} 
  \langle \hat{\sigma}^z_{i\lambda}(\tau) \hat{\sigma}^z_{j\lambda}(0) \rangle, \nonumber
\end{eqnarray}
Naturally, because of the coupling between fermion spins and Ising spins, these two quantities should reveal the same transition and give out the same $h_N$. 
However, if we take $h_N \approx 4.63$, the fermion susceptibility data can not fall on the same curve, as shown in Supplementary Figure~\ref{fig:S1}{\bf d}.

It's a result of finite-size effects and has been observed in various systems~\cite{qinDualityDeconfinedQuantumCritical2017, xuMonteCarloStudy2019, sun2025boosting}. To see the phenomenon more clearly we plot $\chi L^{-\gamma/\nu}$ versus $h$ in Supplementary Figure~\ref{fig:S2}. We can find that fermion susceptibility lines of different system sizes do not intersect at one point, and the intersection point shifts with the system size, while for Ising spin susceptibility, this phenomenon is much weaker, and the lines almost intersect at one point. 
\begin{figure}
  \includegraphics[width=14cm]{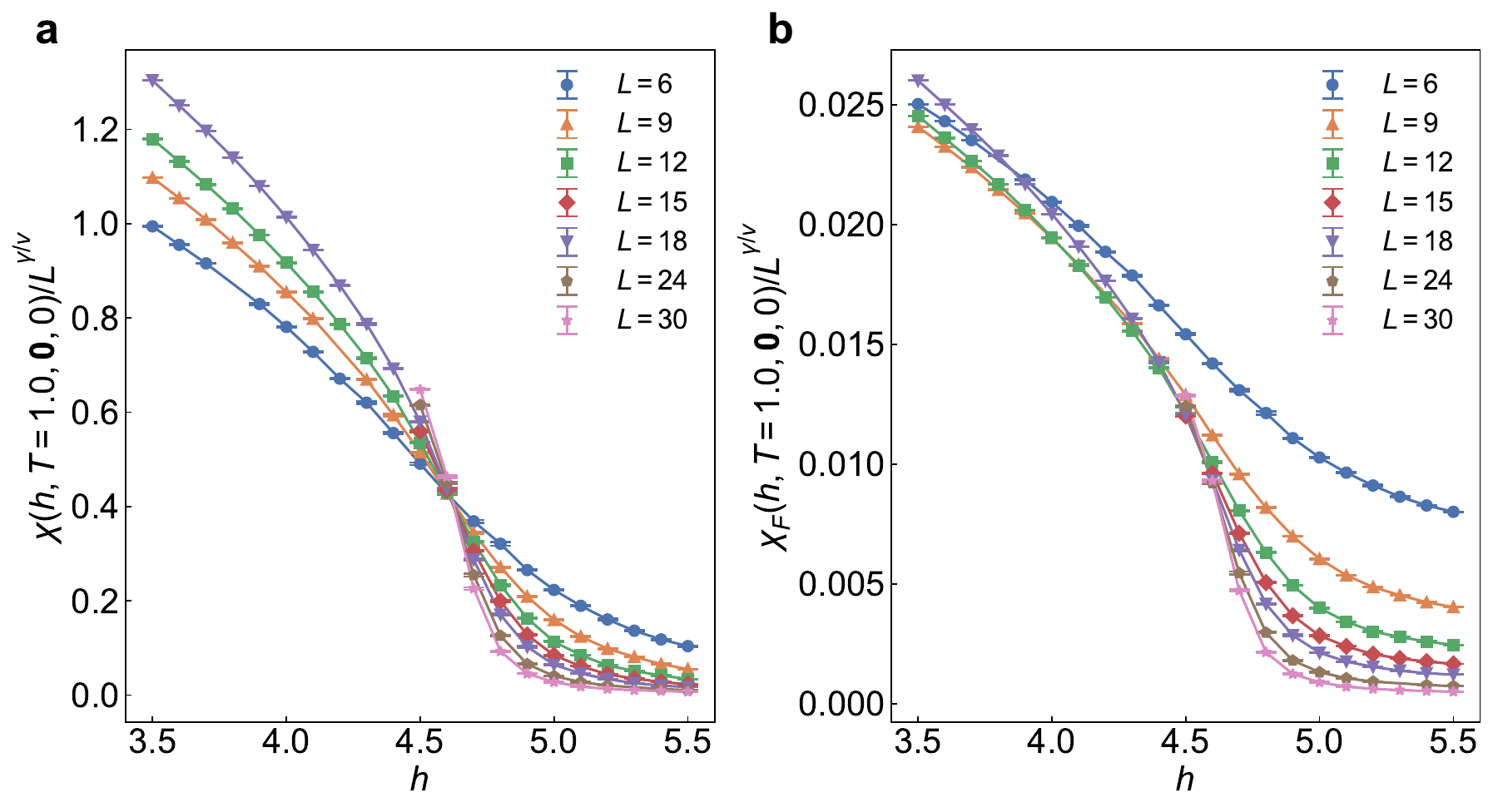}
  \centering
  \caption{
    {Ising spin and fermion spin susceptibilities are subject to finite size effects to different degrees.}
    {(a)} The rescaled susceptibility of the Ising spin. 
    Curves of different system sizes
    almost intersect at one point.
    {(b)} The rescaled susceptibility of fermion spin.
    The intersection points of small-size systems differ significantly from those of large-size systems.
  }
  \label{fig:S2}
\end{figure}
To obtain the unbiased value of the critical field, we can use the crossing-point analysis method~\cite{shao2016quantum,qinDualityDeconfinedQuantumCritical2017}. 
\begin{figure}
  \includegraphics[scale=0.50]{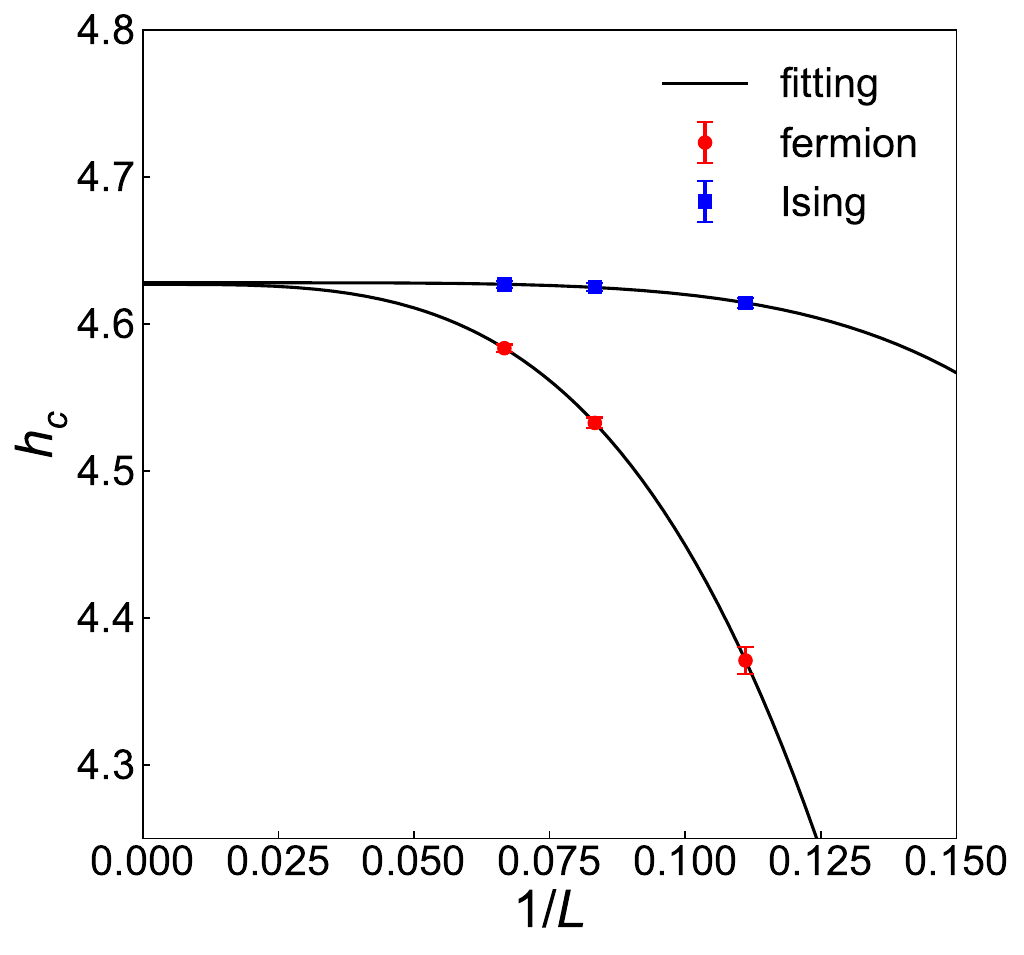}
  \centering
  \caption{
  {Crossing-point analysis of Ising susceptibility and fermion susceptibility.}
  Data points (red for fermion and blue for Ising spin) show the size dependence of the critical field defined as the crossing point for system sizes $L$ and $2L$. Fitting by the power-law form given by finite-size scaling Eq.~(\ref{eq:S29}), a common critical field $h_N \approx 4.63$ is given, which is same with the one calculated by data collapse of Ising susceptibility data.
  }
  \label{fig:S3}
\end{figure}
Consider only one irrelevant field $\lambda$ and the corresponding subleading exponent $\omega$, the standard finite-size scaling form of an observable quantity is
\begin{eqnarray}
  O(\delta, L) = L^{-\kappa/\nu} f(\delta L^{1/\nu}, \lambda L^{-\omega})
\end{eqnarray}
Here, $\delta$ denotes the distance to the critical point. When in the vicinity of critical point ($\delta, \lambda \ll 1$), we expand the scaling function $f$ to the first order,
\begin{eqnarray}
  O(\delta, L) = L^{-\kappa/\nu} ( a_0 + a_1 \delta L^{1/\nu} + b_1 \lambda L^{-\omega} + \cdots). 
\end{eqnarray}
If we have two systems of size $L$ and size $rL$ ($r > 1$), we can obtain the crossing point $\delta^*(L)$ by combining $O(\delta, rL)$ and $O(\delta, L)$,
\begin{eqnarray}
  \delta^*(L) 
  =&& \frac{a_0}{a_1} \frac{1 - r^{-\kappa/\nu}}{r^{(1-\kappa)/\nu - 1}} L^{-1/\nu} \\
  &&+ \frac{b_1}{a_1} \frac{1-r^{-(\kappa/\nu + \omega)}}{r^{(1-\kappa)/\nu-1}} L^{-(1/\nu + \omega)}.
\end{eqnarray}
When the observable quantity is dimensionless, e.g. $\kappa = 0$, the relation between $\delta^*(L)$ and $L$ is
\begin{eqnarray}
  \delta^*(L) = q_c^*(L) - q_c \propto L^{-(1/\nu + \omega)}.
\end{eqnarray}
Due to the quantity we used is the rescaled susceptibility $\chi(T, h, \mathbf{0}, 0) L^{-\gamma/\nu}$, which is also dimensionless, so we can use the above formula to get the critical field $h_N$ at the thermodynamic limit.

Considering the high computational complexity of DQMC, we take $r = 2$, and the difference of critical field between the finite-size system and the thermodynamic limit can be written as
\begin{eqnarray}
  \delta(L) = h_N(L) - h_N \propto L^{-(1/\nu + \omega)},
  \label{eq:S29}
\end{eqnarray}
with $h_N(L)$ the crossing point of size-$L$ system and size-$2L$ system while $h_N$ the critical field at thermodynamic limit. $\omega$ is the exponent of the subleading term that reflects the severity of the finite-size effect. We can fit our data using a formula of the form above to get $h_N$. Supplementary Figure~\ref{fig:S3} is the result of the crossing-point analysis. The curves of Ising spins and fermion spins tend to the same critical value $h_N \approx 4.63$ in the thermodynamic limit, which is the same as the data collapse result of Ising spin susceptibility, so we can safely use the results of Ising spins given by data collapse to get the phase diagram.
\\

\textbf{Supplementary Note 4: Superconductivity}

Various studies suggest that a superconductivity dome may form near the QCP in nFL and mask the quantum critical region. The underlying nature is the instability triggered by the critical fluctuations. In order to avoid the interference of new phases with the measurement of physical quantities, we need to measure the superfluid density $\rho_c$ and $\rho_s$ to test whether there are possible superconductivity instabilities close to the QCP. 
The definition of $\rho_{c,s}$ along $x$-axis is
\begin{eqnarray}
  \rho^{xx}_{c}&&
  =\lim_{q_y \rightarrow 0}
  \lim_{L\rightarrow\infty}
  K^{xx}_{c}(q_{x}=0,q_{y}),\\
  \rho^{xx}_{s}&&
  =\lim_{q_y \rightarrow 0}
  \lim_{L\rightarrow\infty}
  K^{xx}_{s}(q_{x}=0,q_{y}),
\end{eqnarray}
with
\begin{eqnarray}
K^{xx}_c(\mathbf{q})&&=\frac{1}{4}
[\delta\Lambda_{\uparrow\uparrow}(\mathbf{q})
+\delta\Lambda_{\downarrow\downarrow}(\mathbf{q})
+2\delta\Lambda_{\uparrow\downarrow}(\mathbf{q})], 
\nonumber \\
K^{xx}_s(\mathbf{q})&&=\frac{1}{4}
[\delta\Lambda_{\uparrow\uparrow}(\mathbf{q})
+\delta\Lambda_{\downarrow\downarrow}(\mathbf{q})
-2\delta\Lambda_{\uparrow\downarrow}(\mathbf{q})],
\end{eqnarray}
and
\begin{eqnarray}
  \delta\Lambda_{\sigma,\sigma^{\prime}}(\mathbf{q})
  =
  \Lambda_{\sigma,\sigma^{\prime}}^{xx}
  \left(q_{x}\rightarrow 0,q_{y}=0\right) - 
  \Lambda_{\sigma,\sigma^{\prime}}^{xx}
  \left(\mathbf{q}\right),
  \label{eq:S31}
\end{eqnarray}
\begin{equation}
\Lambda_{\sigma,\sigma'}^{xx}(\mathbf{q}) = \frac{1}{L^2}\int d\tau \sum_{ij\lambda\lambda'}e^{\i\mathbf{q}\cdot(\mathbf{r}_i-\mathbf{r}_j)}\langle \mathcal{J}^x_{i\lambda\sigma}(\tau)\mathcal{J}^x_{j\lambda'\sigma'}(0)\rangle.
\end{equation}
Here $\mathcal{J}^x_{i\lambda\sigma} =  \i \sum_{\delta_x > 0}\delta_{x}t e^{\i\phi^{\lambda\sigma}_{i,i+\delta}} c_{i\lambda\sigma}^{\dagger}c_{i+\delta,\lambda\sigma}+\text{H.c.}$ is the $x$-direction current density of fermions on site $i$, layer $\lambda$ and spin $\sigma$. 
We note that due to the momenta being discrete in $k$-space, we use the smallest nonzero momentum on the $x$-axis or $y$-axis to approximate the limitation $q_x \rightarrow 0$ or $q_y \rightarrow 0$.

As studied in previous work~\cite{xuNonFermiLiquid+12017}, many exotic phases are allowed near the FM QCP. The correlation of the superconducting order parameter depends on the superfluid density$\rho_{c, s}$. The correspondence between the correlation of the superconducting order parameter and superfluid density is as follows:

  Charge-4e superconductor: The order parameter is 
  $\Delta_{\uparrow} \Delta_{\downarrow}$ and the correlations go as
  $\langle (\Delta^{\dagger}_{\uparrow} \Delta^{\dagger}_{\downarrow})(\mathbf{r}) 
  (\Delta^{\dagger}_{\uparrow} \Delta^{\dagger}_{\downarrow})(\mathbf{0})
  \rangle \propto r^{-[4 / (2\pi \beta \rho_c)]}$.

  Spin-nematic: The order parameter is 
  $\Delta^{\dagger}_{\uparrow} \Delta_{\downarrow}$ and the correlations go as
  $\langle (\Delta^{\dagger}_{\uparrow} \Delta_{\downarrow})(\mathbf{r}) 
  (\Delta^{\dagger}_{\downarrow} \Delta_{\uparrow})(\mathbf{0})
  \rangle \propto r^{-[4 / (2\pi \beta \rho_s)]}$.

  Triplet superconductor: The order parameter is 
  $\Delta_{\uparrow}$ and the correlations go as
  $\langle \Delta^{\dagger}_{\uparrow}(\mathbf{r}) 
  \Delta_{\uparrow}(\mathbf{0})
  \rangle \propto r^{-\{[1 / (2\pi \beta \rho_c) + 1 / (2\pi \beta \rho_s)] \}}$.

In Supplementary Figure~\ref{fig:S4}, we show the result of $K_{c, s}$ with respect to momentum. When momentum approaches zero, the superfluid density approaches a negative number. This indicates the absence of quasi-long-range superconducting order down to the temperature available.
It is studied that the superfluid density often overestimates the exact transition temperature~\cite{paiva2004critical}.
Therefore, we can confirm that our simulations are not affected by superconductivity within our temperature range.
\\

\begin{figure}
  \includegraphics[scale=0.25]{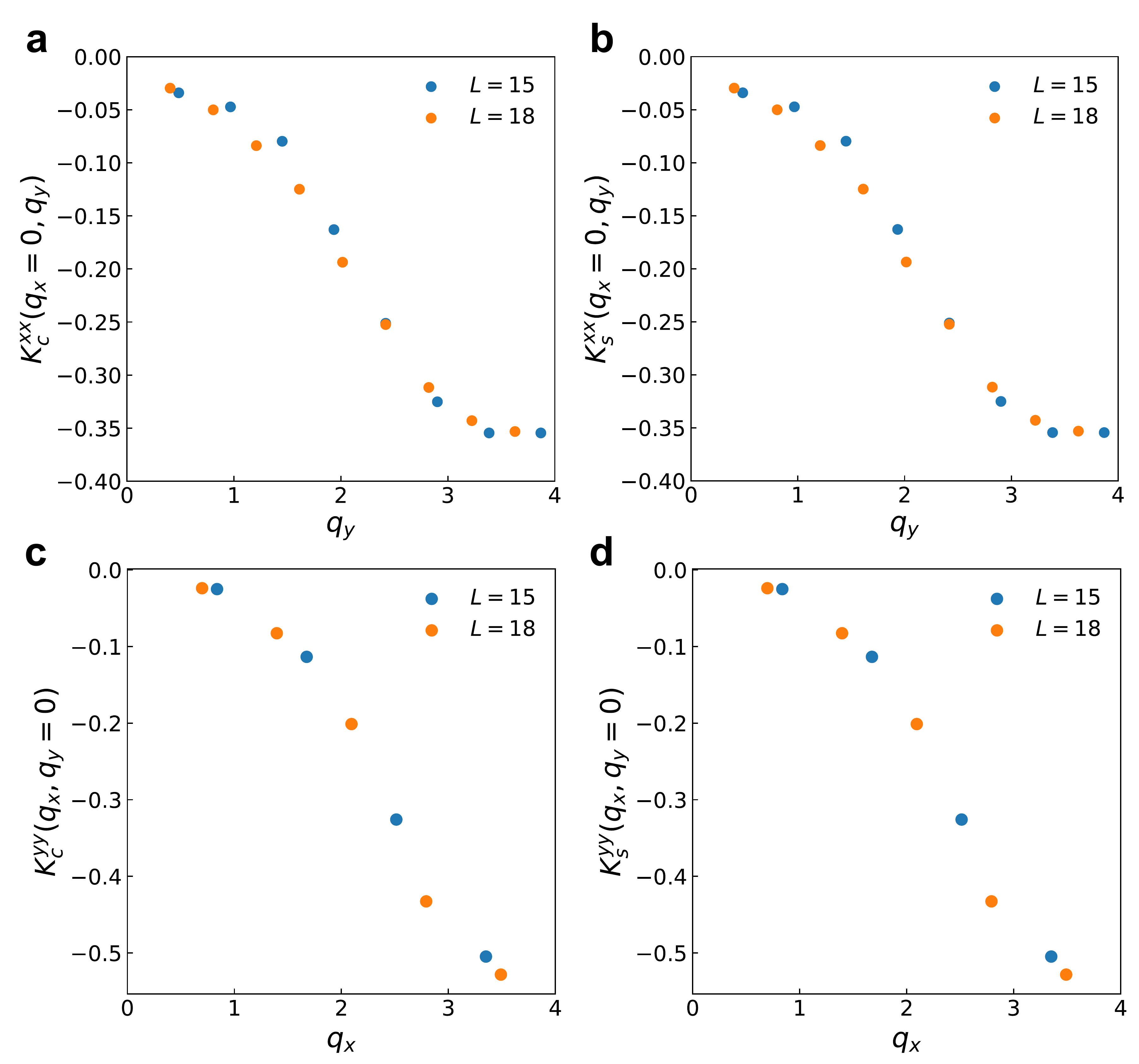}
  \centering
  \caption{
    {Graphs of $K^{xx}_{c,s}(q_x=0, q_y)$-$q_y$ 
    and $K^{yy}_{c,s}(q_x, q_y=0)$-$q_x$ for $T = 1/40$,
    $h = h_c$.}
    When momentum approaches zero, the superfluid density approaches a negative number, meaning that there are no strong superconductivity fluctuations. Curves of different sizes overlap, indicating the thermodynamic limit.
  }
  \label{fig:S4}
\end{figure}

\textbf{Supplementary Note 5: Quantum Critical Scaling Analysis of Boson} 

In this section, we try to find the scaling form of magnetic susceptibility, which can reflect the nature of QCP and serve as the bosonic propagator in MET. 

Due to coupling to gapless fermions, the bosonic critical modes are strongly renormalized and exhibit behaviors that are different from the $(2+1)$D Ising universality class. 
In theory, the HMM framework is based on the random phase approximation (RPA) and gives rise to a mean-field effective form for the bosonic critical modes~\cite{millisEffectNonzeroTemperature1993}. However, previous studies have found that the susceptibility deviates from the mean-field prediction 
both in theory and numerics~\cite{liuItinerantQuantumCritical2018,xuNonFermiLiquid+12017,millisEffectNonzeroTemperature1993}.
Here we use the modified form suggested in Ref.~\cite{xuNonFermiLiquid+12017} to fit our data, e.g.,
\begin{eqnarray}
  \chi(h, T, \mathbf{q}, \Omega_n) \nonumber = \frac{1}{c_t T^{a_t} + c_h |h - h_c|^{\gamma} + (c_q q^2
  + c_{\Omega} \Omega_n^2)^{a_q / 2} + \Delta(\mathbf{q}, \Omega_n)}, \nonumber
  \label{eq:scaling}
\end{eqnarray}
with constants $c_t, c_h, c_q, c_{\Omega}$ and non-mean-field exponents $a_t, \gamma, a_q$. $\Delta(\mathbf{q}, \Omega_n)$ term is the contribution of the fermionic fluctuations derived by RPA in HMM theory
\begin{eqnarray}
  \Delta(\mathbf{q}, \Omega_n) = c_{\mathrm{HM}} \frac{|\Omega_n|}{\sqrt{\Omega_n^2 + (v_f q)^2}}.
  \label{eq:delta}
\end{eqnarray}
Eq.~(\ref{eq:delta}) holds for an isotropic 2D Fermi fluid in the low-energy 
($\Omega \rightarrow 0$) and long-wavelength ($q \rightarrow 0$)
case.  
Here $c_{\mathrm{HM}}$ is a constant and $v_f$ is the Fermi velocity.
The behavior of $\Delta(\mathbf{q}, \Omega_n)$ is 
different in the low-frequency limit and long-wavelength limit, 
differing by a constant $c_{\mathrm{HM}}$. Thus we can get
\begin{eqnarray}
  \chi^{-1}(h=h_c, T=0, \mathbf{q}, \Omega_n=0) &&= c_q^{a_q/2} q^{a_q}, \\
  \chi^{-1}(h=h_c, T=0, \mathbf{q} = \mathbf{0}, \Omega_n) &&= c_{\Omega}^{a_q/2} \Omega_n^{a_q} + c_{\mathrm{HM}}.
\end{eqnarray}

We show the dynamic behaviors of spin susceptibility 
$\chi(h, T, \mathbf{q}, \Omega_n)$ in Supplementary Figure~\ref{fig:scaling}. 
We fit $\log \left[ \chi^{-1}(\mathbf{q}, 0) - \chi^{-1}(\mathbf{0}, 0) \right] 
= 0.5 a_q \log(c_q) + a_q \log(q)$ versus $\log(q)$ by linear function in subplot Supplementary Figure~\ref{fig:scaling}{\bf a}.
While we use the curve function to fit 
$\chi^{-1}(\mathbf{q}, \Omega_n) - \chi^{-1}(\mathbf{0}, 0) = c_{\mathrm{HM}} 
+ c_{\Omega}^{a_q/2} \Omega_n^{a_q}$ versus $\Omega_n$ in subplot 
Supplementary Figure~\ref{fig:scaling}{\bf b} due to the existence of constant $c_{\mathrm{HM}}$. 
We obtain coefficients $c_q = 1.762(8)$, $c_{\Omega} = 0.030(2)$ and exponent $a_q = 1.664(7)$,
which means that there is an anomalous dimension $\eta = 2-a_q=0.34$.
The constant $c_{\mathrm{HM}}$ in our model is $0.180(5)$.

\begin{figure*}[t]
	\includegraphics[width=17cm]{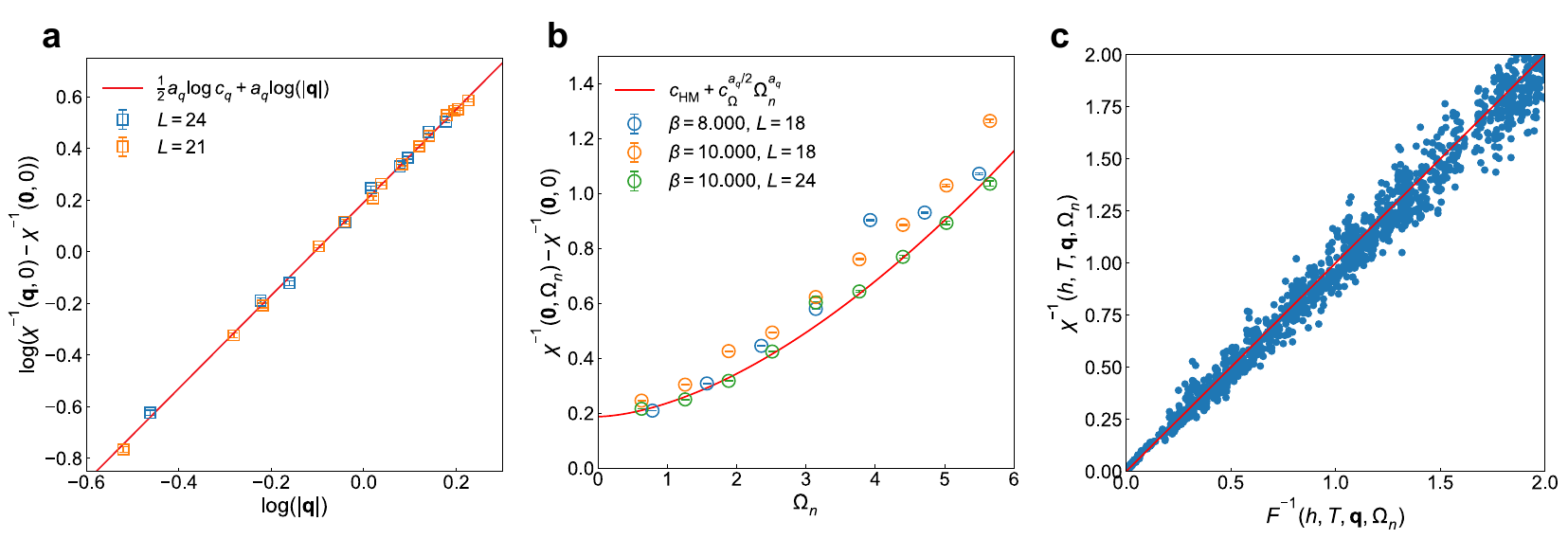}
  \caption{\label{fig:scaling} 
  {(a)} Inverse Ising spin susceptibility at $\Omega_n = 0$
  as a function of $|q|$. 
  The red linear fitting line stands for $0.5 a_q \log(c_q) + a_q \log(q)$,
  and the parameters obtained are $a_q = 1.664(7)$ and $c_q = 1.762(8)$.
  {(b)} Inverse Ising spin susceptibility at $q = 0$ as a function of $\Omega_n$. 
  The red curve fitting line shows $c_{\mathrm{HM}} + c_{\Omega}^{a_q/2} \Omega^{a_q}$
  and the parameters obtained are $c_{\mathrm{HM}} = 0.180(5)$ 
  and $c_{\Omega} = 0.030(2)$.
  {(c)} Data collapse for Ising spin susceptibility with
  respect to fitting function Eq.~\eqref{eq:scaling} 
  $F^{-1}(h, T, \mathbf{q}, \Omega_n) = 
  c_t T^{a_t} + c_h |h - h_c|^{\gamma} + (c_q q^2 + 
  c_{\Omega} \Omega_n^2)^{a_q / 2} + \Delta(\mathbf{q}, \Omega_n)$.
  All data points are close to the baseline $y = x$.
  We use the data of $L=18$ with
   $T = 1.0, 0.833, 0.625, 0.5, 0.4, 0.33, 0.25, 0.2, 0.1$ and
  $h = 4.9, 5.0, 5.1, 5.2, 5.3, 5.4$,
  as well as 
  data of $L=18$ with $T = 0.05, 0.071, 0.083, 0.1, 0.125$
  and $L = 21, 24$ with $T = 0.1, 0.125$ at QCP
  for frequency-dependent fitting.}
\end{figure*}

To get the coefficients that are independent of dynamics, we should set $\mathbf{q} = \mathbf{0}$ and 
$\Omega_n = 0$ simultaneously. This will vanish the contribution from fermionic fluctuations
$\Delta(\mathbf{q}, \Omega_n)$ according to Eq.~\eqref{eq:delta} in the main text.
Then substituting into Eq.~(\ref{eq:scaling}), 
$\chi(h, T, \mathbf{0}, 0)$ has the form
\begin{eqnarray}
  \chi(h, T, \mathbf{0}, 0) = \frac{1}{c_t T^{a_t} + c_h |h - h_c|^{\gamma}}.
\end{eqnarray}

\begin{figure}
  \includegraphics[scale=0.50]{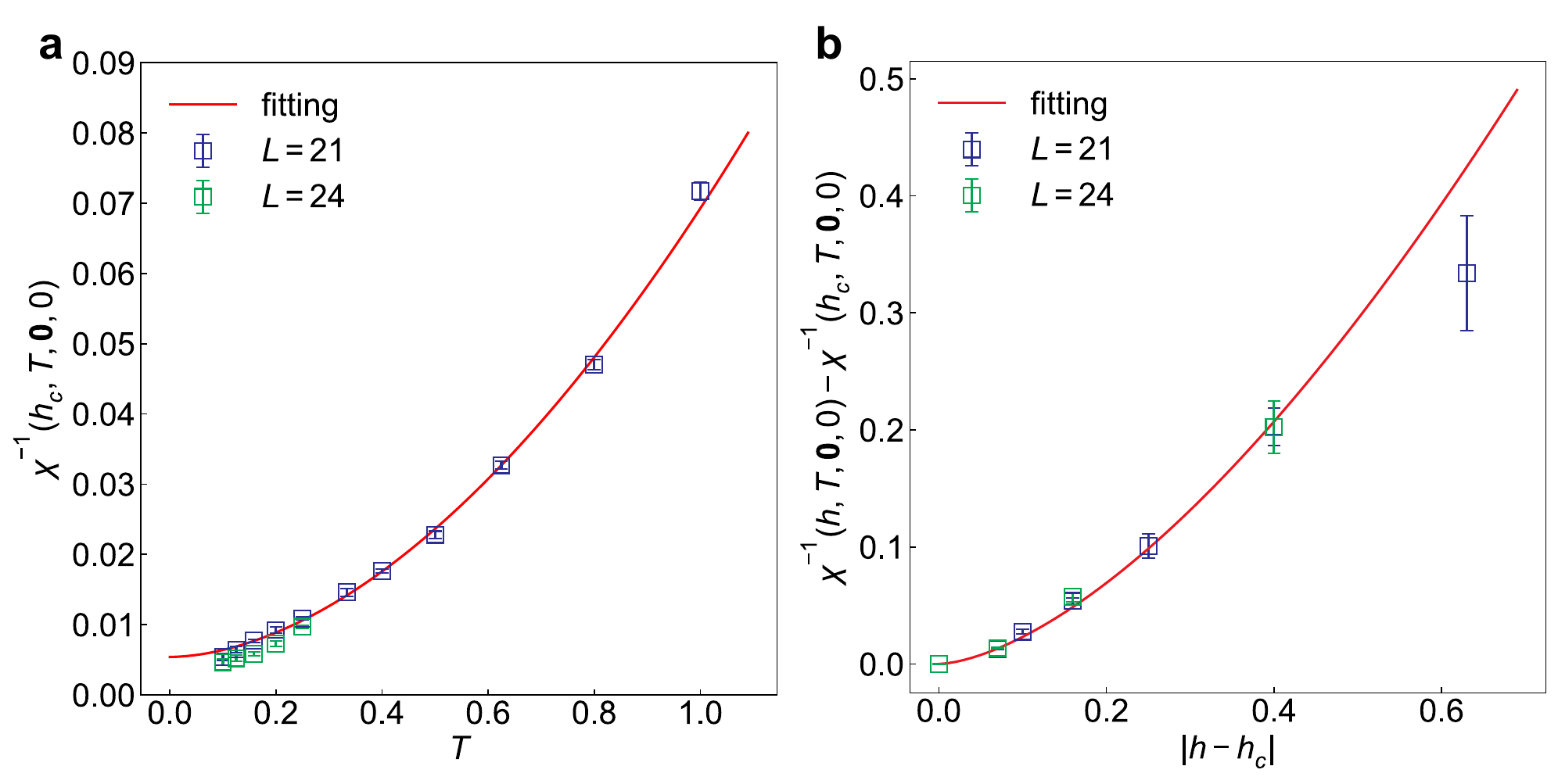}
  \centering
  \caption{
   {Inverse of Ising susceptibility at QCP as 
  a function of temperature $T$ or field $h$.}
    {(a)} As a function of temperature $T$. 
  $\chi^{-1}(h_c, T, \mathbf{0}, 0) = c_t T^{a_t}$,
  we get $c_t = 0.064(1)$ and $a_t = 1.81(6)$. 
  A small intercept characterizes the distance between the thermodynamic limit and the finite-size phase transition points.
  {(b)} As a function of field $h$.
  $\chi^{-1}(h, T, \mathbf{0}, 0) = c_t T^{a_t} + c_h |h - h_c|^{\gamma}$,
  we get $c_h = 0.9(1)$ and $\gamma = 1.58(5)$.
  }
  \label{fig:S5}
\end{figure}

Actually, due to the uncertainty about QCP $h_c$ and 
finite size effects, there may exist a constant caused by the term $c_h |h - h_c|^{\gamma}$.
Thus, we consider an additional constant and use the model function $f(x) = a x^b + c$ 
to fit $\chi^{-1}(h_c, T, \mathbf{0}, 0)$.
The coefficients we get are $c_t = 0.064(1)$ and $a_t = 1.81(6)$.
To obtain the coefficients about transverse field $h$, 
we have fitted $\chi^{-1}(h, T, \mathbf{0}, 0)
 - \chi^{-1}(h_c, T, \mathbf{0}, 0) = c_h |h - h_c|^{\gamma}$. 
The results we get are $c_h = 0.9(1)$ and $\gamma = 1.58(5)$. 
We have shown the scaling results in Supplementary Figure~\ref{fig:S5}.

As all coefficients and exponents have been determined,
we can, in principle, fit the data of any point in the parameter space 
by using Eq.~\ref{eq:scaling}. 
The results are shown in Supplementary Figure~\ref{fig:scaling}{\bf c}.
All the data points are near the line $\chi^{-1} = F^{-1}$ 
where $F^{-1} = c_t T^{a_t} + c_h |h - h_c|^\gamma + (c_q q^2 + c_{\Omega} \Omega_n^2)^{a_q/2} 
+ \Delta(\mathbf{q}, \Omega_n)$, 
especially for the data points located in the 
left bottom corner of the Supplementary Figure~\ref{fig:scaling} {\bf c}, which are from 
small momentum $q$ and small energy $\Omega_n$ case
in the critical sector (low $T$ and $h \sim h_c$).



\begin{figure}
  \includegraphics[width=18cm]{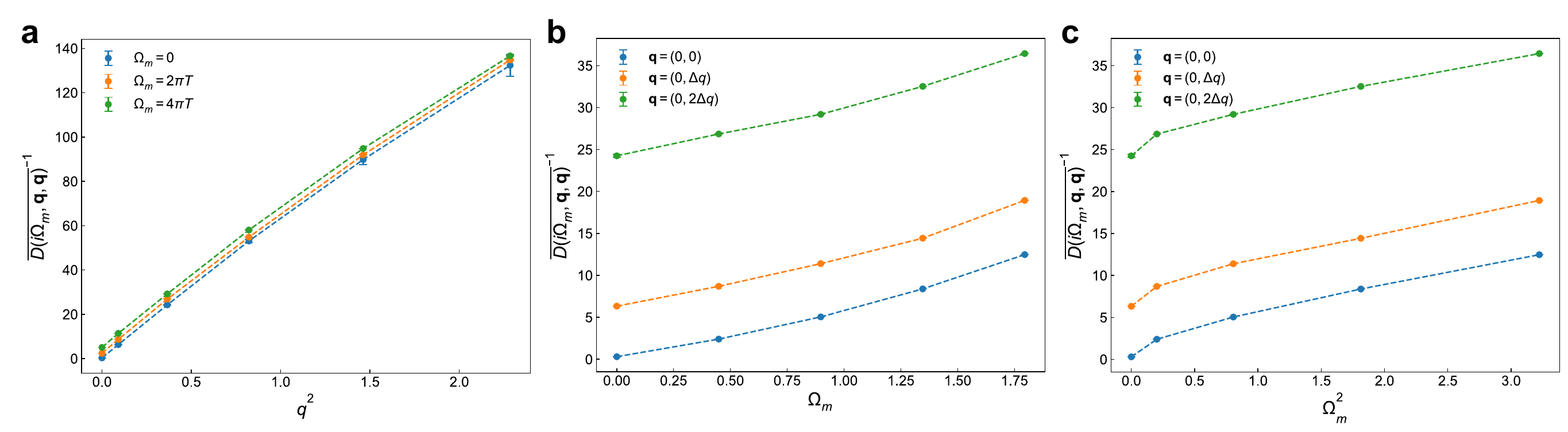}
  \centering
  \caption{  Momentum and frequency dependence of $D(\text{i} \Omega_m, \mathbf{q},\mathbf{q})$ at $h=4.93$. The results are obtained at $L = 24$ and combine 192 disorder realizations.
  (a) Momentum dependence of $D(\text{i} \Omega_m, \mathbf{q},\mathbf{q})$ at the quantum critical point $h_c$. A $q^2$ dependence is found at all frequencies.
  (b), (c) Frequency dependence of $D(\text{i} \Omega_m, \mathbf{q},\mathbf{q})$ at the quantum critical point $h_c$, where $\Delta q = \frac{4\pi}{\sqrt{3}L}$.
  An obvious derivation of $|\Omega_m|$ dependence is found in the low-frequency region.}
  \label{fig:D(w,q)} 
\end{figure}
We also investigate the scaling of $\overline{D(\text{i}\Omega_m, \mathbf{q}, \mathbf{q})}^{-1}$ at $h = 4.93$ in the presence of spatial disorder, in particular, the completely random Yukawa coupling case. Here the overline denotes average over spatial disorder, and $D(\text{i}\Omega_m,\mathbf{q},\mathbf{q}')$ is the bosonic propagator after Fourier transform and its real space form is defined in the main text. Here we consider $\mathbf{q}=\mathbf{q'}$ case. Our findings reveal a $q^2$ dependence at all frequencies, as illustrated in Supplementary Figure~\ref{fig:D(w,q)}{\bf a}. Unlike previous studies, where zero-frequency data exhibited significant sensitivity to the specific disorder sample\cite{patelLocalizationOverdampedBosonic2024}, we observe a clear $q^2$ dependence even at zero frequency. This result indicates that the self-averaging assumption in UT is still established and consistent with our results from the inverse boson propagator spectrum, where no obvious localization mode is found. We also observe an obvious deviation from the $|\Omega_m|^2$ scaling of $D^{-1}(\text{i}\Omega_m, \mathbf{q}, \mathbf{q})$. Due to the temperature limitations, we predict a $|\Omega_m|$ scaling in the low-frequency region.
\\

\textbf{Supplementary Note 6: Modified Eliashberg Theory}
\begin{figure}[htb]
  \includegraphics[scale=0.40]{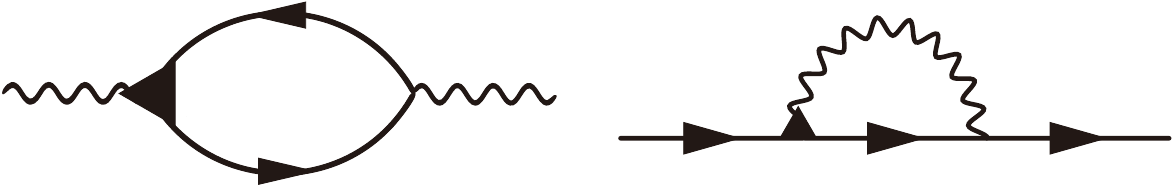}
  \centering
  \caption{\label{fig:eliashberg}{The diagrammatic representation of bosonic self-energy $\Pi(q)$ and fermionic self-energy $\Sigma(k)$ in MET.} 
  The solid lines are dressed fermion propagators $G(k)$, and the wavy lines are dressed boson propagators $D(q)$. The solid triangles are the dressed vertex corrections.}
\end{figure}

The fermionic and bosonic data in the critical region can be analyzed by using the modified Eliashberg theory (MET)~\cite{kleinNormalStateProperties2020}, which gets its name due to the similarity of Eliashberg theory in electron-phonon interaction but with both high-frequency modifications and finite-temperature corrections. MET is a low-energy effective theory, and one can obtain the self-energies of fermions and bosons by coupled one-loop equations shown diagrammatically in Supplementary Figure~\ref{fig:eliashberg}. Here, we follow the derivation in Ref.~\cite{xuIdentificationNonFermiLiquid2020} to make the supplementary note self-contained.

In the spin-fermion model we studied, though vertex corrections from low-energy fermions are logarithmic and singular at low frequencies, these corrections are rather small numerically at frequencies above the superconducting $T_c$. Other corrections from high-energy fermions can be absorbed into the renormalization of the coupling $\overline{g}$. Thus, we can neglect the corrections (solid triangle in the diagram) and get two self-consistent equations:

\begin{eqnarray}
  \Sigma(\mathbf{k},\omega_n) && = 
  N_b \overline{g} T 
  \sum_{\Omega_m} \int 
  \frac{d \mathbf{p}}{(2\pi)^2} 
  G(\mathbf{p}+\mathbf{k}, \omega_n + \Omega_m) 
  D(\mathbf{p}, \Omega_m),\label{eq:eb1}
\end{eqnarray}
\begin{eqnarray}
  \Pi(\mathbf{q}, \Omega_m) && = 
  N_f \overline{g} T 
  \sum_{\omega_n} \int 
  \frac{d \mathbf{p}}{(2\pi)^2} 
  G(\mathbf{p}+\mathbf{q}, \omega_n+\Omega_m) 
  G(\mathbf{p}, \omega_n).\label{eq:eb2}
\end{eqnarray}
Where $\Sigma$ and $\Pi$ are the self-energies of fermions and bosons. $N_f = 4$ corresponds to the $2$ layers and $2$ spins in our model. $N_b = 1$ is the flavor of bosons. In principle, the integral in Eq.~(\ref{eq:eb1}) and Eq.~(\ref{eq:eb2}) is over the entire Brillouin zone, so the contributions from high-energy fermions are still included. One of these is a static contribution to $\Pi$ and renormalizes the mass towards QCP, e.g, $M_0^2$ is shifted by
\begin{eqnarray}
  M^2 = M_0^2 - \Pi(\mathbf{q} = \mathbf{0}, \Omega_m = 0).
\end{eqnarray}
Thus, $M^2$ can be tuned to a QCP by varying $\overline{g}$. Another static contribution renormalizes $D_0$, and we can absorb it into $\overline{g}$. There are also static contributions to $\Sigma$, but they do not change the critical dynamics, so we absorb them into the fermionic dispersion. Furthermore, assuming fermionic dispersion can be linearized near the FS and ignoring the anisotropy of the lattice, we have 
$\epsilon(\mathbf{p}+\mathbf{q}) 
= \epsilon(\mathbf{p}) + v_F q \cos(\theta-\theta_q)$,
with $v_F$ the Fermi velocity and $\theta_q$, $\theta$ the angle on the FS of $\mathbf{q}$, $\mathbf{p}$ respectively.

Therefore, the self-consistent equations are decoupled, and we can first calculate the bosonic self-energy,
\begin{eqnarray}
  \begin{aligned}
    \Pi(q, \Omega_m) &= \i N_f \overline{g} T m \nu_F \int_0^{2\pi} \frac{d\theta}{2\pi}
    \frac{1}{\i\Omega_m - qv_F \cos\theta} \\
    &= \overline{g}N_f \frac{\nu_F}{2\pi} \frac{|\Omega_m|}{v_F q}
  \end{aligned}
  \label{eq:pi}
\end{eqnarray}

At finite temperature, if the condition $|\Sigma(\omega_n)| \ll \omega_n$ is satisfied, the fermionic self-energy can be split into a quantum part and a thermal part.
\begin{eqnarray}
  \Sigma(\omega_n) = \Sigma_T(\omega_n, T \neq 0) + \Sigma_Q(\omega_n, T).
\end{eqnarray}
The thermal part comes from $\omega_l = \omega_n$ piece of the sum in Eq.~(\ref{eq:eb2}) and has the form,
\begin{eqnarray}
  \Sigma_T(\omega_n) && \approx -\i \frac{\overline{g} T}{2\pi\omega_n} 
  \mathcal{S}(\frac{v_F M}{|\omega_n|}) \\
  && =
  \begin{cases}
    -\i\frac{\bar{g}T}{2\pi\omega_n}\log\left(\frac{2|\omega_n|}{\upsilon_FM}\right),
    &\upsilon_FM\ll|\omega_n|\\
    -\i\frac{\bar{g}T\sigma(\omega_n)}{4\upsilon_FM},
    &\upsilon_FM\gg|\omega_n|
  \end{cases}
\end{eqnarray}
with 
\begin{eqnarray}
  \mathcal{S}(x) = \frac{\cosh^{-1}(1/x)}{\sqrt{1 - x^2}}
  \approx 
  \begin{cases}
    \log(2/x),
    &x \ll 1\\
    \frac{\pi}{2x},
    &x \gg 1 
  \end{cases}
  \nonumber
\end{eqnarray}
In the frequency region we studied, $\omega_n \Sigma_T(\omega_n)$ is nearly a constant  $\alpha(T)$ up to a logarithmical correction. Thus providing us with a pathway to extract it from the total self-energy. 

The quantum part contains all other terms in the sum of Matsubara frequency except $\omega_l = \omega_n$. This sum can be approximated by integral $\int \frac{d \omega_l}{2 \pi T}$ and recovers the form of fermionic self-energy at $T = 0$,
\begin{eqnarray}
\Sigma_{Q}(\omega_n)\approx
-\i\overline{g}\sigma(\omega_n)\left(\frac{\omega_n}{\omega_b}\right)^{2/3}
\mathcal{U}\left(\frac{\omega_n}{\omega_b}\right),
\end{eqnarray}
with
\begin{eqnarray}
\mathcal{U}(z)
= && \int_0^\infty\frac{\mathrm{d}x\:\mathrm{d}y}{4\pi^2}
\frac{y}{y^3+\left(\upsilon_F/c\right)^2x^2yz^{4/3}+x} \nonumber \\
&& \times \left[ \frac{\sigma(x+1)}{\sqrt{1+\left(\frac{x+1}{y}\right)^2z^{4/3}}}
-\frac{\sigma(x-1)}{\sqrt{1+\left(\frac{x-1}{y}\right)^2z^{4/3}}} \right]. \nonumber
\end{eqnarray}
In this step, we assume that we are at QCP ($M=0$) and remove the angular dependence of $\upsilon_F$. The scaling function $\mathcal{U}(z)$ has the following asymptotics,
\begin{eqnarray}
\left.\mathcal{U}(z)
  =\left\{\begin{array}{ll}\frac{1}{2\pi\sqrt{3}}
  & z\ll1,\\\frac{1}{24z^{2/3}}
  &1\ll z\ll z_c,\cdot\\\frac{u_0}{z^{2/3}}
  &z_c\ll z,\end{array}\right.\right.
  \nonumber
\end{eqnarray}
In the low-frequency case, $\mathcal{U}(z)$ tends to a constant, and $\Sigma_Q$ reveals the $\omega_n^{2/3}$ scaling form suggested by one-loop calculation. At higher frequencies, this form is broken down and modified. This corresponds to 2D scattering that is not limited to being parallel to the FS.
In this work, we use the full form of fermionic self-energy to fit our data, and an additional constant $\alpha'(T)$ is added according to~\cite{xuIdentificationNonFermiLiquid2020}. This constant includes the contributions from the finite-size effects. To predict the quasiparticle weight in the entire temperature region, a linear fitting is used, as shown in Supplementary Figure~\ref{fig:S6}. \\
\begin{figure}
  \includegraphics[scale=0.60]{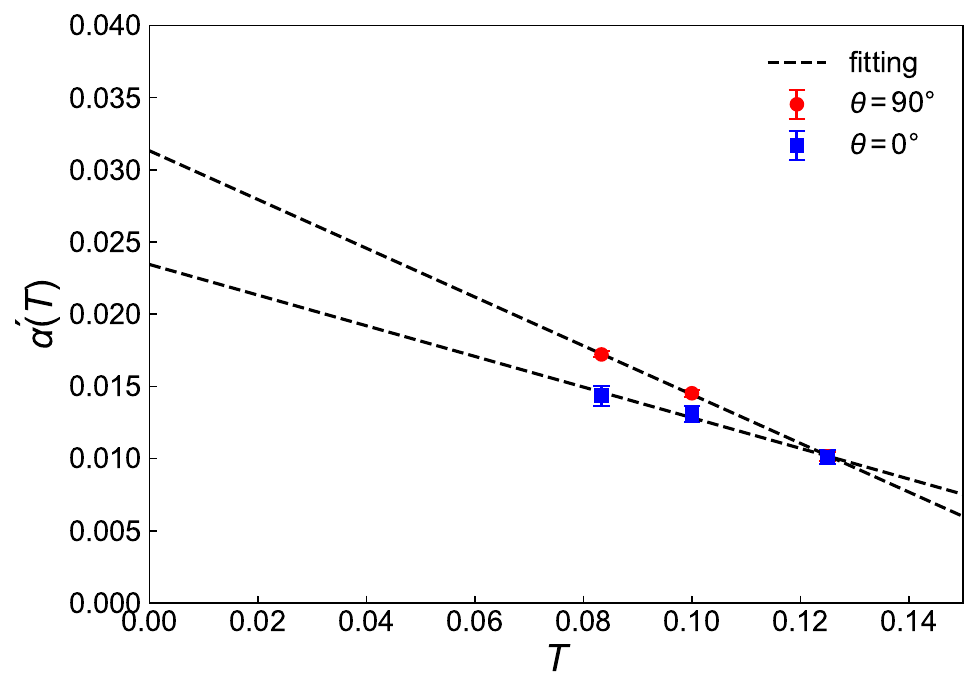}
  \centering
  \caption{
    {The extracted gap contribution $\alpha'(T)$ to 
    $-\omega_n \mathrm{Im} \Sigma(\omega_n)$.} 
    Data points of different colors represent 
    different directions 
    (blue for $k_x$ direction and 
    red for $k_y$ direction).
    The $\alpha'(T)$ versus temperature $T$ 
    is nearly linear.
  }
  \label{fig:S6}
\end{figure}

\begin{figure*}[t]
	\includegraphics[width=17cm]{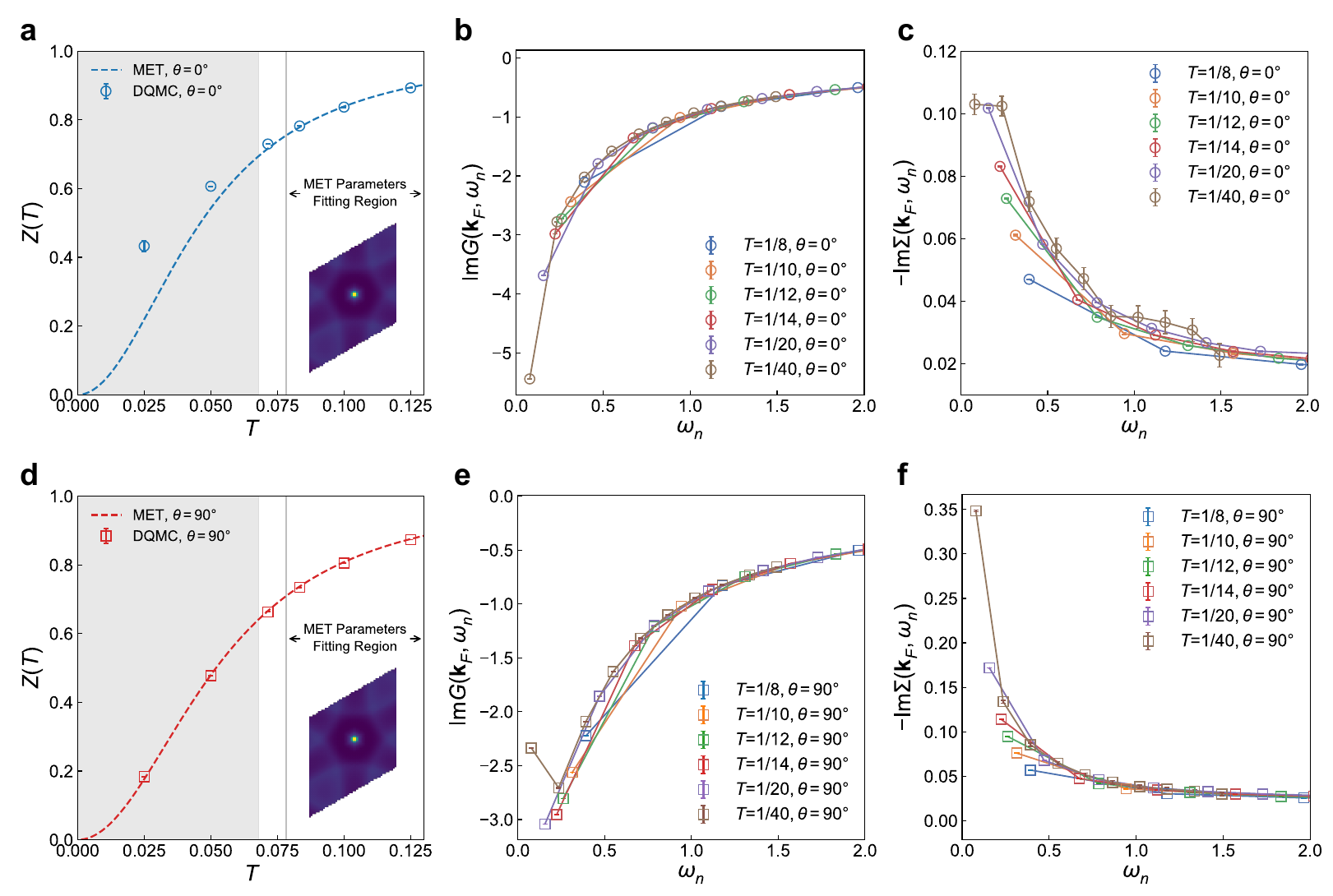}
   \centering
  \caption{\label{fig:Non-Fermi Liquid} 
  {Non-Fermi liquid behaviors along different directions.}
  The circles represent the data along the $k_x$-direction, and the squares represent the data along the $k_y$-direction. All DQMC data in this figure are obtained with a lattice size $L=24$.  
  {(a), (d)} Quasiparticle weight $Z(T)$ from DQMC and MET. Data points are obtained from DQMC, and the curves are predictions from MET only based on data with $T\ge 1/12$. The shadowed region denotes the temperature region where the relation $v_F \Delta_k \lesssim T$ along the $k_x$ direction is strongly violated ($v_F \Delta_k > 2T$). Quasiparticle weight along the $k_y$ direction decays a bit faster than the one along the $k_x$ direction, and both directions decay towards zero at QCP. The inset is the fermionic susceptibility $\chi_F$ at QCP, and no sign of a soft mode is found.
  {(b), (e)} Imaginary part $\mathrm{Im} G$ of Green's function at QCP. No evidence of a gap is found due to $\mathrm{Im} G$ being non-zero.
  {(c), (f)} Results of $-\mathrm{Im}\Sigma$ at QCP. $-\mathrm{Im} \Sigma$ increases as the temperature lowers, which is caused by thermal fluctuations and can be described by MET.}
\end{figure*}

\textbf{Supplementary Note 7: Non-Fermi Liquid Behaviors in the Clean Model}

One of the most significant characteristics of nFL in theory is the zero quasiparticle weight at the FS, while for the Fermi liquid, the quasiparticle weight is a finite value. We use the method in Ref.~\cite{chenLifshitzTransitionTwo2012} to calculate quasiparticle weight from fermionic self-energy $\Sigma(\mathbf{k}, \omega_n)$ we get in DQMC,
\begin{eqnarray}
  Z_0({\mathbf{k}_F}) = \frac{1}{1 - \frac{\mathrm{Im} \Sigma(\mathbf{k}_F, \omega_0)}{\omega_0}},
  \label{eq:Z0}
\end{eqnarray}
where $\mathbf{k}_F$ is the momentum at the FS and $\omega_0 = \pi T$ is the first fermionic Matsubara frequency. In the limit $T \rightarrow 0$, $Z_0(\mathbf{k})$ will converge to the quasiparticle weight $Z(\mathbf{k})$.

We show the behaviors of quasiparticle weight in Supplementary Figure~\ref{fig:Non-Fermi Liquid}{\bf a}. Taking into account the slight anisotropy of the FS and the $C_6$ symmetry, we plot curves of two points along different directions, i.e., along $k_x$ direction ($\theta = 0^{\circ}$, circle) and along the $k_y$ direction ($\theta = 90^{\circ}$, square). We find the quasiparticle weight $Z_0$ decreases a bit faster along the $k_y$-direction, and suffers from larger anisotropy in the lower temperature region. To search for the underlying reason for the anisotropy, we plot the fermionic susceptibility $\chi_F(T=0.1, h_c, \mathbf{q}, 0)$, which is shown in the inset of Supplementary Figure~\ref{fig:Non-Fermi Liquid}. Unlike the case on the square lattice, no soft mode is found near the FS. 
However, the Fermi velocity in the $k_y$ direction is slightly smaller than that in the $k_x$ direction, {and more importantly, the distance ($\Delta_k$) of the closest momentum points along the $k_y$ direction is much smaller than that in the $k_x$ direction. This means the particles experience larger interaction effects along the $k_y$ direction and this results in a smaller quasiparticle weight. The anisotropy becomes larger at lower temperatures as it cannot be smeared out by temperature anymore. That is to say, in the low temperature region the relation $v_F \Delta_k \lesssim T$ along the $k_x$ direction is strongly violated for $L=24$, the largest system size we considered.} We display the information of the fermiology in Table~\ref{tab:tab1}.

\begin{table}[ht]
  \centering
  \caption{{\bf Parameters of the fermiology along $k_y$
  and $k_x$ directions.} $k_F$ is the length of the Fermi vector, $v_F$ is the Fermi velocity, $E_F$ is the Fermi energy, $\nu_F = k_F/v_F$ denotes the density of states, and $\Delta_k$ is the distance of the closest momentum points along each direction to the FS for the $L=24$ lattice.}
  \begin{tabular}{lrrrrrr}
      \text{direction} & \textbf{($k_x$, $k_y$)} & \textbf{$k_F$} & \textbf{$v_F$} & \textbf{$\nu_F$} & \textbf{$E_F$} & \textbf{$\Delta_k$} \\
    \hline
      $\theta=90^\circ$ & (0, 2.723) & 2.723 & 2.444 & 1.114 & 3.328 & 0.003 \\
      $\theta=0^\circ$ & (2.666, 0) & 2.666 & 2.860 & 0.932 & 3.812 & 0.048 \\
    \hline
  \end{tabular}
  \label{tab:tab1}
\end{table}

The imaginary part of Green's function is shown in Supplementary Figure~\ref{fig:Non-Fermi Liquid}{\bf b}. As $-\text{Im}G(\mathbf{k}_F,\omega_n \rightarrow 0)$ relates to the spectral weight $A(\mathbf{k}_F,\omega \rightarrow 0)$, the $-\mathrm{Im} G(\mathbf{k}_F,\omega_n)$ remains far from zero with decreasing $\omega_n$, indicating that no evidence of opening a gap at the QCP. This is consistent with the conclusion we get in SI Appendix 4, e.g., no signature of a superconducting phase is found until the lowest temperature we can access.

In Supplementary Figure~\ref{fig:Non-Fermi Liquid}{\bf c}, the behaviors of fermionic self-energy are displayed. Fermionic self-energy is calculated as $\Sigma(\omega_n) = G_0^{-1} - G^{-1}$, where $G_0$ is the Green's function of free fermion and $G^{-1}$ is the Green's function of our model obtained in DQMC. The self-energy in different directions differs only slightly, indicating that the anisotropy is relatively weak. For self-energy in the critical region, $- \mathrm{Im} \Sigma$ increases with decreasing $\omega_n$ which deviates from the $\sim \omega_n^{2/3}$ form predicted by second-order perturbative calculations~\cite{chubukov2010hidden}. 
Due to the simulations being conducted at a finite temperature, this phenomenon can be interpreted as an effect of thermal fluctuations and has been mentioned qualitatively in previous work~\cite{xuNonFermiLiquid+12017}. A more rigorous explanation can be obtained from the modified Eliashberg theory (MET)~\cite{kleinNormalStateProperties2020}. 

\begin{figure}
  \includegraphics[width=10cm]{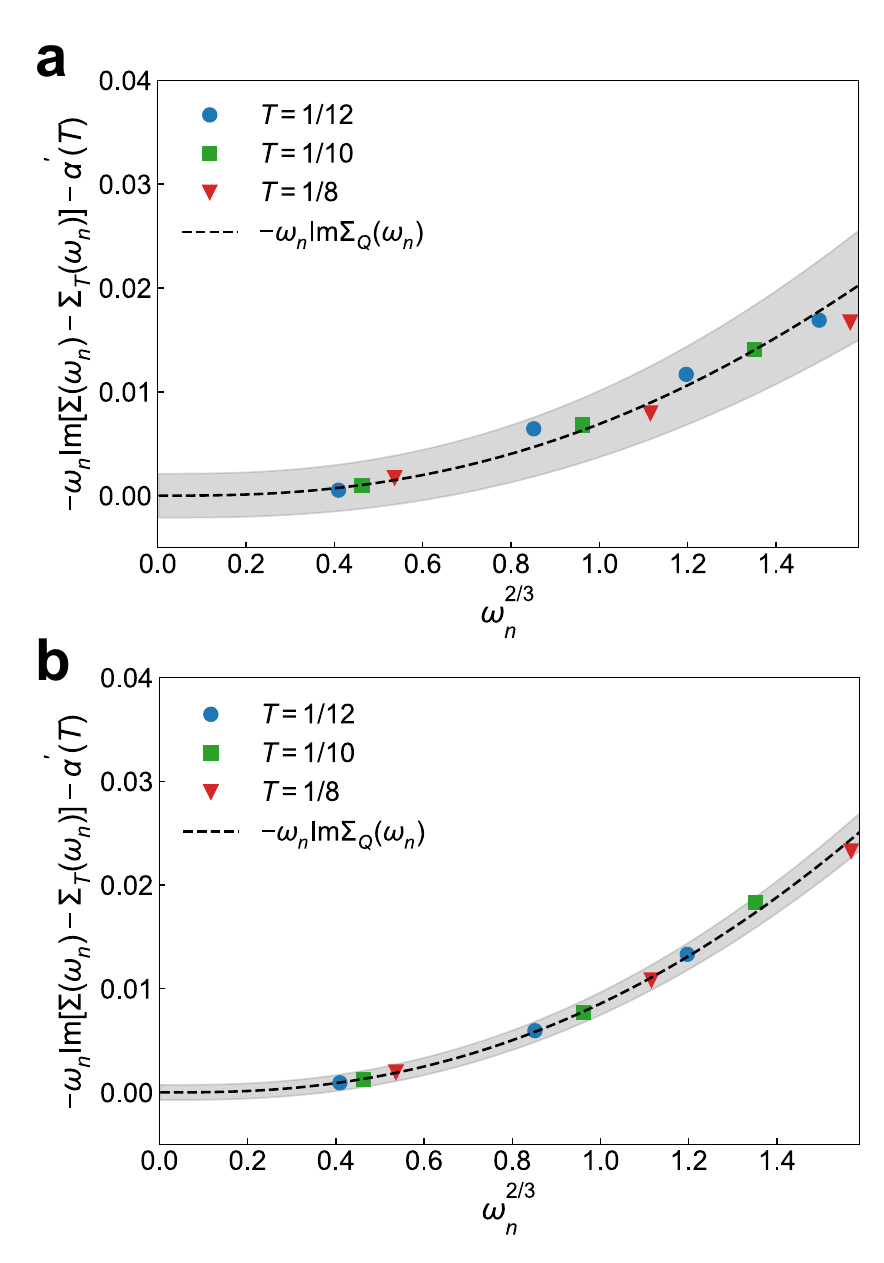}
  \centering
  \caption{\label{fig:et} {Extraction of $T=0$ self-energy by MET.}
  {(a)} Along $k_x$ direction. 
  {(b)} Along $k_y$ direction.
  The dashed line is the fitting of $-\omega_n \text{Im} \Sigma_Q(\omega_n)$, while the data dots are $-\omega_n \text{Im} \Sigma(\omega_n)$ of different $T$ after deducing the thermal part. All data points fall into the $95\%$ confidence interval (the grey region).}
\end{figure}

In MET, the bosonic and fermionic self-energy can be calculated by the coupled one-loop self-consistent equations shown diagrammatically in Supplementary Figure~\ref{fig:eliashberg}. It has been studied in several works that the main effect of finite temperature is to split the fermionic self-energy into two parts~\cite{punkFiniteTemperatureScaling2016, abanovQuantumcriticalTheorySpinfermion2001, wangNonFermiLiquidsFinite2017}, 
\begin{eqnarray}
 \Sigma(\omega_n) = 
 \Sigma_T(\omega_n) + 
 \Sigma_Q(\omega_n).
\end{eqnarray}
Here $\Sigma_T(\omega_n)$ is the thermal part that comes from static bosonic thermal fluctuations, and $\Sigma_Q(\omega_n)$ is the quantum part that comes from dynamic fluctuations. By studying the behaviors of each part, a pathway to extract the quantum part of self-energy is provided~\cite{kleinNormalStateProperties2020, xuIdentificationNonFermiLiquid2020}. A more rigorous mathematical derivation is provided in SI Appendix 6, and we just show the results here.
\begin{eqnarray}
  \begin{aligned}
    \Sigma(\omega_n) &= -\i \frac{\alpha'(T)}{\omega_n} + \Sigma_T(\omega_n) + \Sigma_Q(\omega_n), \\
    \Sigma_T(\omega_n) &= -\i\frac{\overline{g}T}{2\pi\omega_n} \mathcal{S}(\frac{v_F M}{|\omega_n|}), \\
    \Sigma_Q(\omega_n) &= -\i\overline{g} \sigma(\omega_n) \left( \frac{\omega_n}{\omega_b} \right)^{2/3}
    \mathcal{U}\left( \frac{\omega_n}{\omega_b} \right),
  \end{aligned}
\end{eqnarray}
with
\begin{eqnarray}
    \mathcal{S}(x) &= \frac{\cosh^{-1}(1/x)}{\sqrt{1 - x^2}} \nonumber
\end{eqnarray}
\begin{eqnarray}
\mathcal{U}(z) & = \int_0^\infty\frac{\mathrm{d}x\:\mathrm{d}y}{4\pi^2}
\frac{y}{y^3+\left(\upsilon_F/c\right)^2x^2yz^{4/3}+x} \nonumber \\
& \times \left[ \frac{\sigma(x+1)}{\sqrt{1+\left(\frac{x+1}{y}\right)^2z^{4/3}}}
-\frac{\sigma(x-1)}{\sqrt{1+\left(\frac{x-1}{y}\right)^2z^{4/3}}} \right].
\nonumber
\end{eqnarray}
Here, $\sigma(x)$ is a sign function, $\omega_b=\sqrt{\bar{g}N_f k_F v_F/\pi}$ is the frequency scale introduced by the boson self-energy, and $N_f$ is the number of fermion flavors. In our model, $N_f=4$. A temperature-dependent constant $\alpha'(T)$ is added, which contains the contributions from finite-size effects~\cite{xuIdentificationNonFermiLiquid2020}. We show the results of MET in Supplementary Figure~\ref{fig:et}. The fitting parameters are only the constants $\alpha'(T)$ and $\overline{g}$. The agreement is good, and all data points fall into the $95\%$ confidence interval. The results indicate that the quantum part of self-energy recovers the $\sim \omega_n^{2/3}$ scaling in the low frequency limit. 

{Along the $k_y$ direction, we get $\overline{g} = 0.218(3)$, which is approximately $30\%$ larger than the bare value  $\overline{g} = D_0 (\frac{\xi}{2})^2 \approx 0.16$. While along the $k_x$ direction, we get $\overline{g} = 0.172(8)$, which is very close to the bare value. Those results are consistent with the situation that fermions along the $k_y$ direction experience larger interaction effects. By applying MET, we plot the predicted quasiparticle weight in Supplementary Figure~\ref{fig:Non-Fermi Liquid}{\bf a}. Note that although the MET parameters fitting region is $T \ge 1/12$,  the $Z(T)$ given by MET fits well with the data given by DQMC even in the lower temperature region especially along the $k_y$ direction and predicts a vanishing quasiparticle weight at $T=0$, which supports the existence of nFL. We note that MET works better along the $k_y$ direction due to less finite size effect. Based on the above results, our subsequent research will primarily focus on the $k_y$ direction.}\\

\textbf{Supplementary Note 8: Universal Theory of Strange Metals}
\label{app:app7}

In this part, we will introduce the framework of the universal theory of strange metals in~\cite{patelUniversalTheoryStrange2023}. We absorbed the main spirit of this framework and fixed MET to satisfy the situation in our model when the spatial disorder is applied.
For our case, the number of fermion flavors is $N_f=4$, and the number of boson flavors is $N_b=1$. The effective action of our model without spatial disorder has the following form
\begin{align}
S= & \int d\tau\sum_{\mathbf{k}}\sum_{i=1}^{N_{f}}\psi_{i\mathbf{k}}(\tau)\left[\partial_{\tau}+\varepsilon_{\mathbf{k}}-\mu\right]\psi_{i\mathbf{k}}(\tau)\nonumber \\
 & +\frac{1}{2}\int d\tau\sum_{\mathbf{q}}\phi_{\mathbf{q}}(\tau)\left[-c^{-2}\partial_{\tau}^{2}+\omega_{\mathbf{q}}^{2}+M^{2}\right]\phi_{-\mathbf{q}}(\tau)\nonumber \\
 & +g\int d\tau d^{2}r\sum_{i=1}^{N_{f}}\psi_{i}^{\dagger}(\mathbf{r},\tau)\psi_{i}(\mathbf{r},\tau)\phi(\mathbf{r},\tau)
 \label{eq:pureyukawa}
\end{align}

After performing second-order perturbation, we get the following equations for fermion and boson self-energy and Green's function.
\begin{eqnarray}
  \begin{aligned}
    &\Sigma(\mathbf{r},\tau)= g^{2}D(\mathbf{r},\tau)
    G(\mathbf{r},\tau),\quad\Pi(\mathbf{r},\tau)
    =-D_0N_fg^{2}G(-\mathbf{r},-\tau)G(\mathbf{r},\tau),\\
    &G(\mathbf{k},\omega_{n})=
    \frac{1}{\i\omega_{n}-\varepsilon_\mathbf{k}
    +\mu-\Sigma(\mathbf{k},\omega_{n})},\\
    &D(\mathbf{q},\Omega_{m})=\frac{D_0}{c^{-2}\Omega_{m}^{2}
    +q^2+M^2-\Pi(\mathbf{q},\Omega_{m})},
  \end{aligned}
\end{eqnarray}
These are actually the self-consistent equations in MET. 

Now, let's apply a random transverse field to the action~\eqref{eq:pureyukawa}. In the spirit of the universal theory of strange metals~\cite{patelUniversalTheoryStrange2023}, it is equivalent to applying a spatially disordered potential for fermions
\begin{eqnarray}
  S_v=\int d\tau\frac{1}{\sqrt{N_f}}\sum_{\mathbf{r}}\sum_{i=1}^{N_f}
  v(\mathbf{r})\psi_i^\dagger(\mathbf{r},\tau)\psi_i(\mathbf{r},\tau),
\end{eqnarray}
and a spatial disorder Yukawa coupling
\begin{eqnarray}
  \begin{aligned}
    S_{g'}&=
    \int d\tau\frac{1}{\sqrt{N_f}}\sum_{\mathbf{r}}\sum_{i=1}^{N_f}
    g'(\mathbf{r})\psi_i^\dagger(\mathbf{r},\tau)
    \psi_i(\mathbf{r},\tau)\phi(\mathbf{r},\tau),
  \end{aligned}
\end{eqnarray}
with 
\begin{eqnarray}
  \overline{v(\mathbf{r}) v(\mathbf{r'})} = v^2 \delta^2(\mathbf{r} - \mathbf{r}'),
  \quad
  \overline{g'(\mathbf{r}) g'(\mathbf{r'})} = g'^2 \delta^2(\mathbf{r} - \mathbf{r}'),
\end{eqnarray}
After self-average over the spatial disorder and performing perturbation to the order of $g^2$, $g'^2$ and $v^2$, we reach the following self-consistent equations for fermion and boson self-energy and Green's function.

\begin{align}
	\Pi(\mathbf{{r}},\tau)&=-D_0 N_f g^{2}
  G(-\mathbf{{r}},-\tau)G(\mathbf{{r}},\tau)
  -D_0 g'^{2}G(\mathbf{{r}}=0,-\tau)
  G(\mathbf{{r}}=0,\tau)\delta^{2}(\mathbf{{r}}), \\
	\text{\ensuremath{\Sigma(\mathbf{{r}},\tau)}}&
  =g^{2}D(\mathbf{{r}},\tau)G(\mathbf{{r}},\tau)
  +v^{2}G(\mathbf{{r}}=\mathbf{{0}},\tau)\delta^{2}(\mathbf{{r}})
  +g'^{2}G(\mathbf{{r}}=\mathbf{{0}},\tau)
  D(\mathbf{{r}}=\mathbf{{0}},\tau)\delta^{2}(\mathbf{{r}}), \label{eq:sigma_disorder} \\
	G(\mathbf{{k}},\omega_n)&=\frac{{1}}{\i\omega_n-\varepsilon_{\mathbf{k}}
  +\mu-\Sigma(\mathbf{{k}}, \omega_n)}, \\
	D(\mathbf{{q}},\Omega_m)&=\frac{{D_0}}{c^{-2}\Omega_m^{2}+q^{2}+M^{2}-\Pi(\mathbf{{q}},\Omega_m)}.
\end{align}

In MET of Ref.~\cite{xuIdentificationNonFermiLiquid2020}, we ignore the fermionic self-energy $\Sigma(\mathbf{k},\omega_n)$ in Green's function. Here, in order to explore the corrections from disorder, we include the leading explicit contribution of $\Sigma(\mathbf{k},\omega_n)$, which is $\i \frac{\Gamma}{2}\mathrm{sgn}(\omega_n)$ with $\Gamma = \frac{v^2 k_F}{v_F}$. Following the same programs in MET, we will first calculate the bosonic self-energy 
$\Pi(\mathbf{q},\Omega_m)$,
\begin{eqnarray}
\begin{aligned}
  \Pi_{g}(\mathbf{{q}},\Omega_m)
  &=-D_0N_fg^{2}T\sum_{\omega_n}\int\frac{{d\mathbf{{k}}}}{(2\pi)^{2}}
  G(\mathbf{{k}},\omega_n)G(\mathbf{k}+\mathbf{q},\omega_n+\Omega_m)\\
  &=-\i D_0N_f\frac{{g^{2}k_{F}}}{2v_{F}}\int_{-\infty}^{+\infty}
  \frac{{d\omega}}{2\pi}\int_{-\pi}^{+\pi}
  \frac{{d\theta}}{2\pi}
  \left[\mathrm{{sgn}(\omega)}-\mathrm{{sgn}}(\omega+\Omega_m)\right]
  \left(\frac{{1}}{\i\frac{{\Gamma}}{2}\mathrm{{sgn}}(\omega+\Omega_m)
  -\i\frac{{\Gamma}}{2}\mathrm{{sgn}(\omega)}+\i\Omega_m-v_{F}q\cos\theta}\right)\\
  &=-\i D_0N_f\frac{{g^{2}k_{F}}}{2\pi v_{F}}\int_{-\pi}^{+\pi}
  \frac{{d\theta}}{2\pi}
  \frac{{\Omega_m}}{\i\Gamma\mathrm{{sgn}}(\Omega_m)+\i\Omega_m-v_{F}q\cos\theta}\\
  &=-\mathcal{N}\overline{g} 
  \frac{{|\Omega_m|}}{\sqrt{{(\Gamma+\Omega_m)^{2}+v_{F}^{2}q^{2}}}}.
\end{aligned} 
\end{eqnarray}
\begin{eqnarray}
\begin{aligned}
  \Pi_{g'}(\Omega_m)&=-D_0g'^{2}k_{F}^{2}
  \int_{-\infty}^{+\infty}
  \frac{{d\omega}}{2\pi}
  \int_{-\infty}^{+\infty}
  \frac{{dk}}{2\pi}
  \int_{-\infty}^{+\infty}\frac{{dk'}}{2\pi}
  \frac{{1}}{\i\omega+\i\frac{{\Gamma}}{2}\mathrm{{sgn}(\omega)}-v_{F}k}
  \frac{{1}}{\i(\omega+\Omega_m)+\i\frac{{\Gamma}}{2}\mathrm{{sgn}}(\omega+\Omega_m)-v_{F}k'}\\
  &=-\frac{{\pi}}{2}\mathcal{\mathcal{{N}}}^{2}\overline{g}'|\Omega_m|. 
\end{aligned}
\end{eqnarray}
\begin{eqnarray}
  \begin{aligned}
  \Pi(\mathbf{{q}}, \Omega_m)
  &=\Pi_{g}(\mathbf{{q}},\Omega_m)+\Pi_{g'}(\Omega_m)\\
  &=-\text{\ensuremath{\mathcal{{N}}\overline{g}}}\frac{{|\Omega_m|}}{\sqrt{{(\Gamma+\Omega_m)^{2}+v_{F}^{2}q^{2}}}}
  -\frac{{\pi}}{2}\mathcal{\mathcal{{N}}}^{2}\overline{g}'|\Omega_m|.
  \end{aligned}
  \label{eq:pi_disorder}
\end{eqnarray}
Here $\mathcal{N} = \frac{k_F}{2\pi v_F}$ and $\overline{g}' = D_0 g'^2$, $\overline{g} = D_0 g^2$. Clearly, Eq.~(\ref{eq:pi}) will be recovered when the disorder strength is zero. Considering $\mathcal{N}$ and $\overline{g}'$ are small, we can neglect the second term when calculating the fermionic self-energy. The thermal part of the fermionic self-energy comes from the static boson. Thus, only the first and third terms in Eq.~(\ref{eq:sigma_disorder}) make a contribution,
\begin{eqnarray}
  \begin{aligned}
    \Sigma_{T}(\mathbf{{k}},\omega_n)
    &=\Sigma_{T,g}(\mathbf{{k}},\omega_n)+\Sigma_{T,g'}(\omega_n).\\
  \end{aligned}
\end{eqnarray}
\begin{eqnarray}
  \begin{aligned}
  \Sigma_{T,g'}(\omega_n)
  &=\overline{g}'Tk_{F}\int_{-\infty}^{+\infty}\frac{{dk}}{2\pi}\int_{0}^{+\infty}\frac{{qdq}}{2\pi}G(\mathbf{{k}},\omega_n)D(\mathbf{{q}},\Omega_m=0)\\
  &=\overline{g}'Tk_{F}\int_{-\infty}^{+\infty}\frac{{dk}}{2\pi}\int_{0}^{+\infty}\frac{{qdq}}{2\pi}\frac{{1}}{\i\omega_n+\i\frac{{\Gamma}}{2}\mathrm{{sgn}(\omega_n)}-v_{F}k}\frac{{1}}{M^{2}+q^{2}}\\
  &=-\i\overline{g}'T\frac{{k_{F}}}{2v_{F}}\mathrm{{sgn}}(\omega_n+\frac{{\Gamma}}{2}\mathrm{{sgn}}(\omega_n))\int_{0}^{+\infty}\frac{{qdq}}{2\pi}\frac{{1}}{M^{2}+q^{2}}\\
  &=-\i\overline{g}'T\frac{{k_{F}}}{2v_{F}}\mathrm{{sgn}}(\omega_n)\frac{{1}}{4\pi}\ln\left(1+\frac{{\Lambda^{2}}}{M^{2}}\right)\\
  &=-\i\frac{{\mathcal{{N}}\overline{g}'\mathrm{{sgn}}(\omega_n)}}{4}T\ln\left(1+\frac{{\Lambda^{2}}}{M^{2}}\right).
  \end{aligned}
\end{eqnarray}
\begin{eqnarray}
  \begin{aligned}
    \Sigma_{T,g}(\mathbf{{k}},\omega_n)
    &=\overline{g}T\int\frac{{d\mathbf{{q}}}}{(2\pi)^{2}}
    G(\mathbf{{q+k}},\omega_n)D(\mathbf{{q}},\Omega_m=0)\\
    &=\overline{g}T\int_{-\pi}^{+\pi}\frac{{d\theta}}{2\pi}
    \int_{0}^{+\infty}\frac{{qdq}}{2\pi}
    \frac{{1}}{\i\omega_n+\i\frac{{\Gamma}}{2}\mathrm{{sgn}}(\omega_n)-v_{F}q\cos\theta}
    \frac{{1}}{M^{2}+q^{2}}\\
    &=-\i\overline{g}T\mathrm{{sgn}}(\omega_n)
    \int_{0}^{+\infty}\frac{{qdq}}{2\pi}
    \frac{{1}}{\sqrt{(\omega_n+\frac{{\Gamma}}{2})^{2}+v_{F}^{2}q^{2}}}
    \frac{{1}}{M^{2}+q^{2}}.
  \end{aligned}
\end{eqnarray}
Here $\Lambda$ is the UV-cutoff. We can find that $\Sigma_{T, g}$ has the same form as that in MET, and the only difference is a shift of frequency $\omega$. However, we can not separate the correction of $\Gamma/2$ directly by using Taylor's expansion because the convergence radius is not guaranteed when $q$ is small.

The remaining is the quantum part of the fermionic self-energy. 
\begin{eqnarray}
  \Sigma_{Q}(\mathbf{{k}},\omega_n)=
  \Sigma_{Q,g}(\mathbf{{k}},\omega_n)
  +\Sigma_{Q,v}(\omega_n)+\Sigma_{Q,g'}(\omega_n).
\end{eqnarray}
\begin{eqnarray}
  \begin{aligned}
  \Sigma_{Q,v}(\omega_n)
  &=v^{2}\int\frac{{d\mathbf{{k}}}}{(2\pi)^{2}}
  G(\mathbf{{k}},\omega_n)\\
  &=v^{2}k_{F}\int_{-\infty}^{+\infty}\frac{{dk}}{2\pi}
  \frac{1}{\i\omega_n+\i\frac{{\Gamma}}{2}\mathrm{{sgn}}(\omega_n)-v_{F}k}\\
  &=-\i\frac{{v^{2}k_{F}}}{2v_{F}}\mathrm{{sgn}}(\omega_n)\\
  &=-\i\frac{{\Gamma}}{2}\mathrm{sgn}(\omega_n).
  \end{aligned}
\end{eqnarray}
\begin{eqnarray}
  \begin{aligned}
  \Sigma_{Q,g'}(\omega_n)
  &=\overline{g}'k_{F}\int_{-\infty}^{+\infty}
  \frac{{dk}}{2\pi}\int_{0}^{+\infty}
  \frac{{qdq}}{2\pi}\int_{-\infty}^{+\infty}
  \frac{{d\Omega}}{2\pi}
  \frac{{G(\mathbf{{k}},\omega_n+\Omega)}}{q^{2}+\mathcal{{N}}g^{2}\frac{{|\Omega|}}{v_{F}q}}\\
  &=-\i\overline{g}'\pi\mathcal{{N}}\int_{0}^{+\infty}\frac{{qdq}}{2\pi}\int_{-\infty}^{+\infty}\frac{{d\Omega}}{2\pi}
  \frac{{\mathrm{{sgn}}(\omega_n+\Omega)}}{q^{2}+c_{b}\frac{{|\Omega|}}{q}}\\
  &=-\frac{{\i\mathcal{{N}}\overline{g}'\omega_n}}{6\pi}\ln\left(\frac{{e\Lambda^{3}}}{c_{b}|\omega_n|}\right).
  \end{aligned}
\end{eqnarray}
\begin{eqnarray}
  \begin{aligned}
    \Sigma_{Q,g}(\mathbf{{k}},\omega_n)
    &=\overline{g}\int\frac{{d\mathbf{{q}}}}{(2\pi)^{2}}
    \int_{-\infty}^{+\infty}
    \frac{{d\Omega}}{2\pi}
    G(\mathbf{\mathbf{{k+q}}},\omega_n+\Omega)
    D(\mathbf{{q}},\Omega)\\
    &=\overline{g}\int_{-\pi}^{+\pi}\frac{{d\theta}}{2\pi}
    \int_{0}^{+\infty}
    \frac{{qdq}}{2\pi}\int_{-\infty}^{+\infty}
    \frac{{d\Omega}}{2\pi}\frac{{1}}{\i(\omega_n+\Omega)+\i\frac{{\Gamma}}{2}\mathrm{{sgn}}(\omega_n+\Omega)-v_{F}q\cos\theta}
    \frac{{1}}{M^{2}+q^{2}+\Omega^{2}+c_{b}\frac{{|\Omega|}}{\sqrt{{(\Gamma+\Omega)^{2}+v_{F}^{2}q^{2}}}}}\\
    &=\overline{g}\int_{0}^{+\infty}\frac{{qdq}}{2\pi}
    \int_{-\infty}^{+\infty}
    \frac{{d\Omega}}{2\pi}
    \frac{{-\i\mathrm{{sgn}}(\omega_n+\Omega)}}{\sqrt{{(\omega_n+\Omega+\frac{{\Gamma}}{2})^{2}+v_{F}^{2}q^{2}}}}
    \frac{{1}}{M^{2}+q^{2}+\Omega^{2}+c_{b}\frac{{|\Omega|}}{\sqrt{{(\Gamma+\Omega)^{2}+v_{F}^{2}q^{2}}}}}\\
    &=-\i\overline{g}\int_{0}^{+\infty}\frac{{qdq}}{2\pi}\int_{-\infty}^{+\infty}\frac{{d\Omega}}{2\pi}
    \frac{{\mathrm{{sgn}}(\omega_n+\Omega)}}{\sqrt{{(\omega_n+\Omega+\frac{{\Gamma}}{2})^{2}+v_{F}^{2}q^{2}}}}\frac{{1}}{M^{2}+q^{2}+\Omega^{2}+c_{b}\frac{{|\Omega|}}{\sqrt{{(\Gamma+\Omega)^{2}+v_{F}^{2}q^{2}}}}}\\
  \end{aligned}
\end{eqnarray}
The second term $\Sigma_{Q, v}$ is the reason we add a $-\i\frac{\Gamma}{2}\mathrm{sgn}(\omega_n)$ in the Green's function. The main part $\Sigma_{Q, g}$ also has a similar but more complex form than the previous one in MET, and we still cannot separate the contribution from disorder directly because of the same reason as in $\Sigma_{T, g}$. 

In the main text, we use the full form of $\Sigma(\mathbf{k},\omega_n)$ to fit our data. $\overline{g}$ obtained in MET is used to reduce the number of fitting parameters. We choose the UV cut-off $\Lambda$ by hand and the remaining fitting parameters are $\overline{g}'$ and $\Gamma$ ($g'$ and $v$). It's found that any physical $\Lambda$ will result in a $\overline{g}' \approx 0$ and an almost invariant $v$.
\\
\begin{figure}
  \includegraphics[width=10cm]{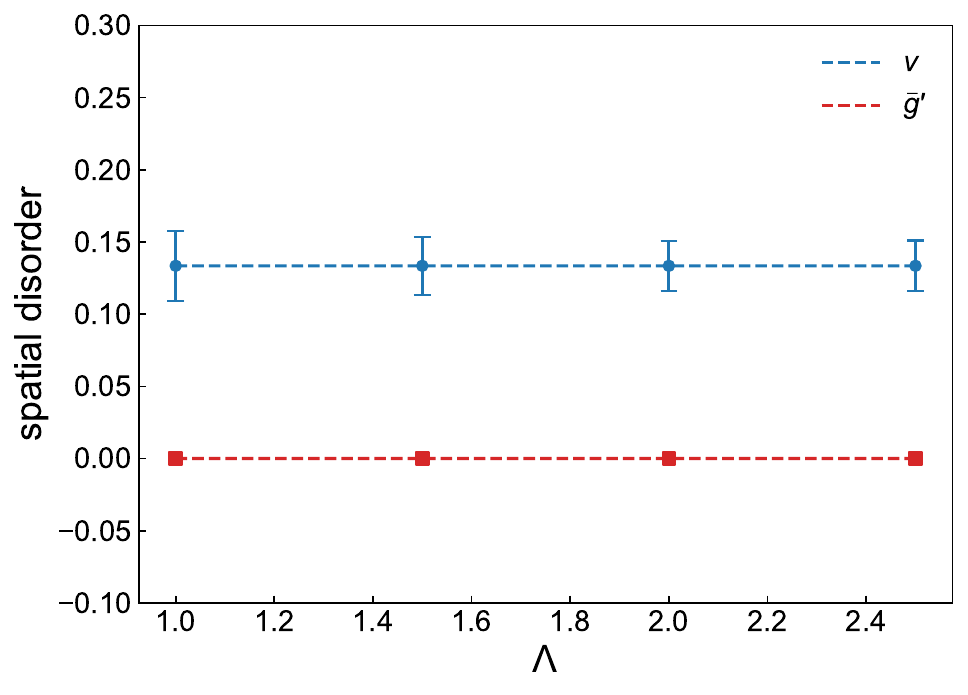}
  \centering
  \caption{
   {$v$ and $g'$ change with UV cutoff $\Lambda$.}
   We choose the UV cutoff $\Lambda$ by hand in the fitting. 
   We find that any physical $\Lambda$ will result in an almost 
   invariant $v$ and $g'$.
  }
  \label{fig:S7}
\end{figure}

\textbf{Supplementary Note 9: Additional Results of Fermionic Self-Energy for the Case with Completely Random Yukawa Coupling} 

In this part, we provide some additional results of the fermionic self-energy of the case with completely random Yukawa coupling. When the field is small, the self-energy deviates from the MFL scaling, as shown in Supplementary Figure~\ref{fig:Sigma_order}.  When approaching the quantum critical point ($h_c \approx 4.93$), an MFL scaling self-energy can be observed in a wide region. We fit the data for various transverse fields $h$ and temperatures $T$ by using the form $f(\omega_n) = a \omega_n \ln(b/\omega_n) + c$ functions, and the fitted parameters are shown in Table~\ref{tab:tab1}. \\

\begin{figure}
  \includegraphics[width=16cm]{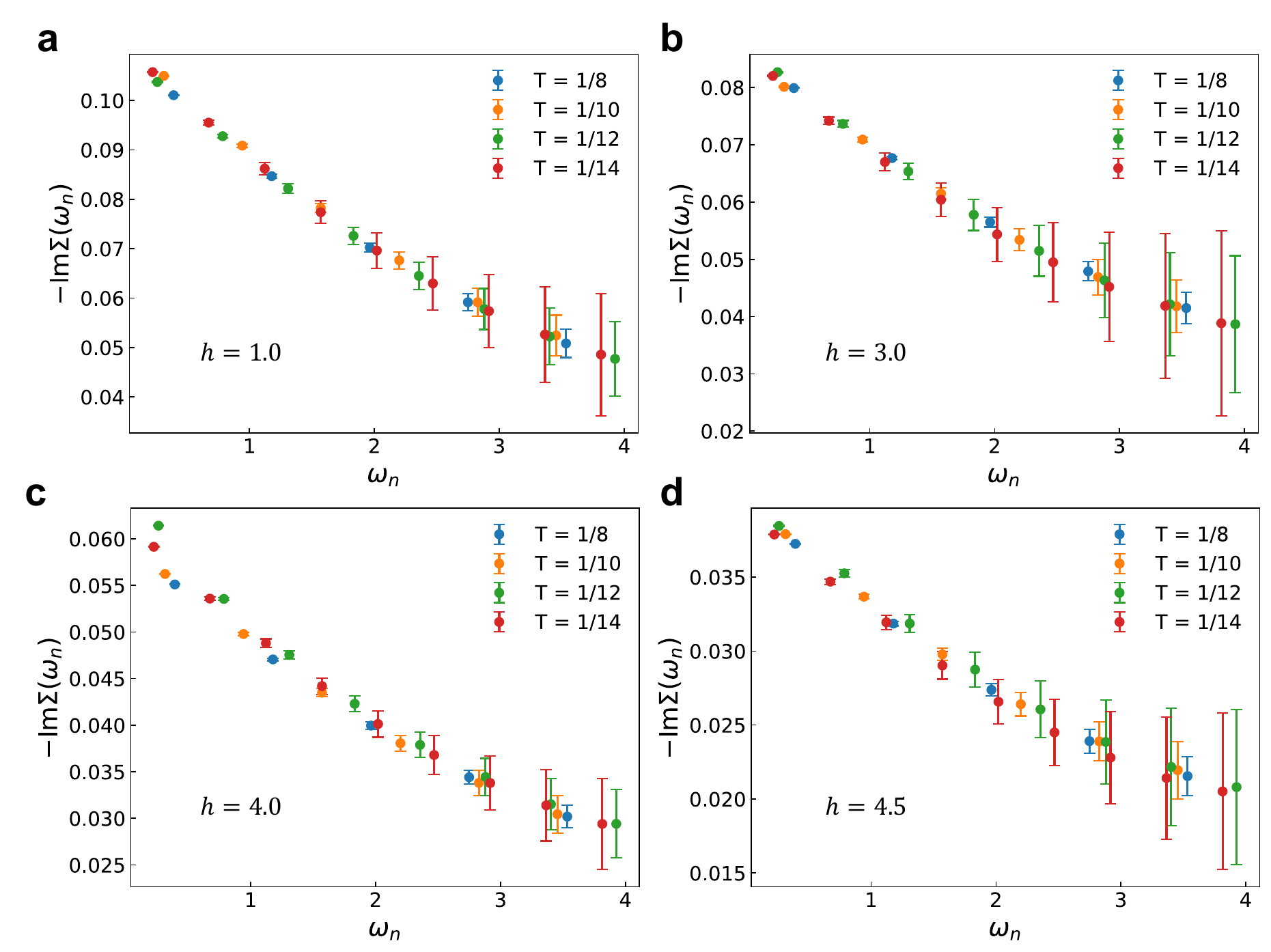}
  \centering
  \caption{
   {Fermionic self-energy at small fields when a completely random coupling is applied.}
   {(a)} $h = 1.0$.
   {(b)} $h = 3.0$.
   {(c)} $h = 4.0$.
   {(d)} $h = 4.5$.
  }
  \label{fig:Sigma_order}
\end{figure}

\begin{figure}
  \includegraphics[width=16cm]{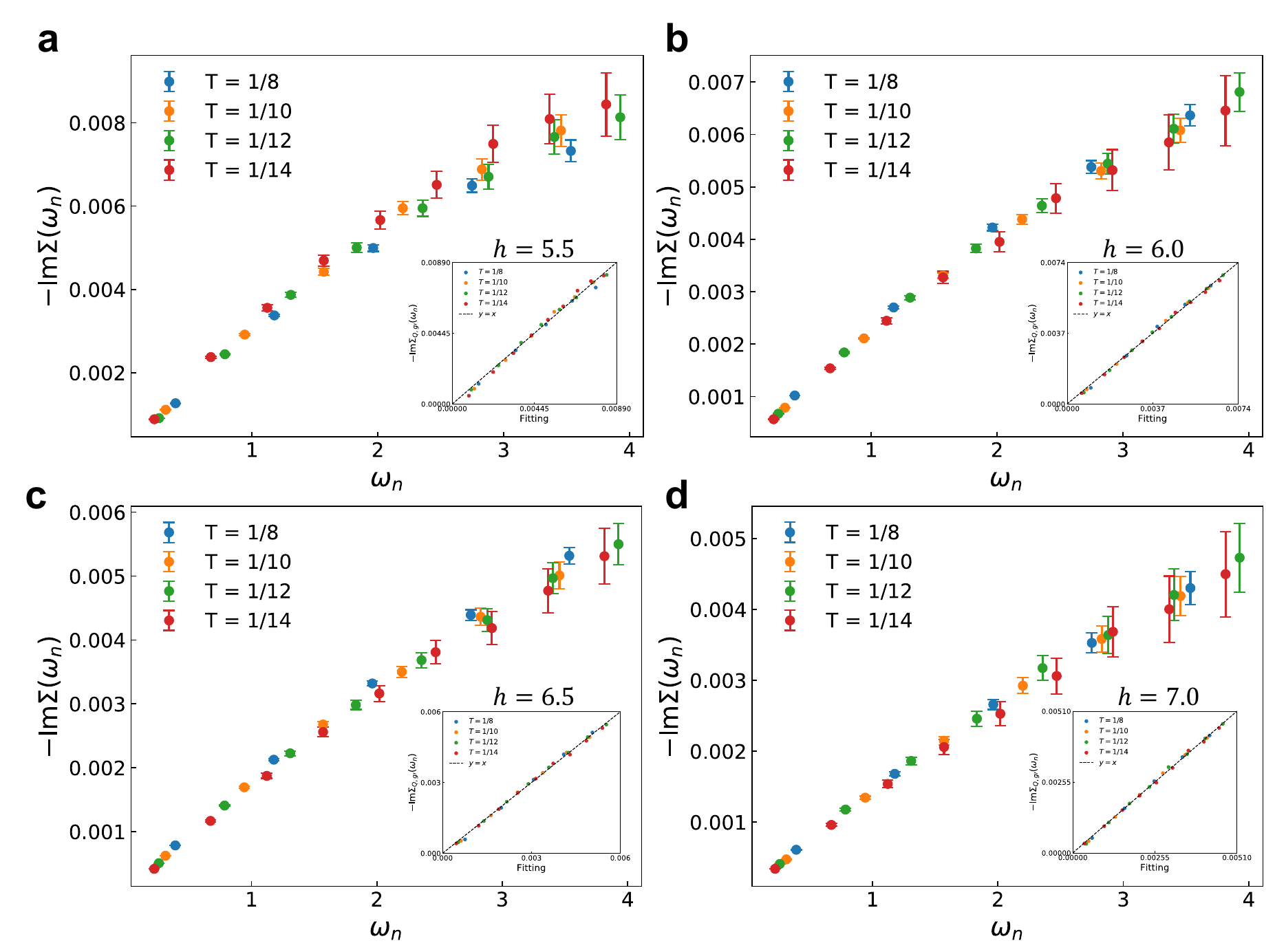}
  \centering
  \caption{
   {Fermionic self-energy at the critical region when a completely random coupling is applied. We fit the data with the MFL form and show the results in the insets.}
   {(a)} $h = 5.5$.
   {(b)} $h = 6.0$.
   {(c)} $h = 6.5$.
   {(d)} $h = 7.0$. After subtracting the thermal part, the imaginary part of the remaining quantum part can be well fitted by a MFL form $a \omega_n \ln(b/\omega_n)$. In the inset of each figure, the $x$-axis values denote the fitting results, and the $y$-axis values denote the Monte Carlo data, all the points are located closely on the $y=x$ line.
  }
  \label{fig:Sigma_critical}
\end{figure}

\begin{table}[ht]
  \centering
  \caption{{\bf Fitted MFL parameters at different transverse fields $h$ when $T = 1/14$.}}
  \begin{tabular}{cccc}
      \text{parameter}  & \textbf{$a$} & \textbf{$b$} & \textbf{$\bar{g}'$} \\
    \hline
      $h=5.0$ & $1.6(2)\times10^{-3}$ & $12.2(18)$ & $0.17(2)$ \\
      $h=5.5$ & $9.9(12)\times10^{-4}$ & $4.1(10) \times 10^{1}$ & $0.104(12)$ \\
      $h=6.0$ & $5.2(6)\times10^{-4}$ & $1.1(4) \times10^{2}$ & $0.055(6)$ \\
      $h=6.5$ & $3.2(5)\times10^{-4}$ & $3(2) \times 10^{2}$ & $0.034(5)$ \\
      $h=7.0$ & $1.6(5)\times10^{-4}$ & $6.8(15)\times10^{3}$ & $0.017(6)$ \\
    \hline
  \end{tabular}
  \label{tab:tab1}
\end{table}

\begin{figure}
  \includegraphics[width=14cm]{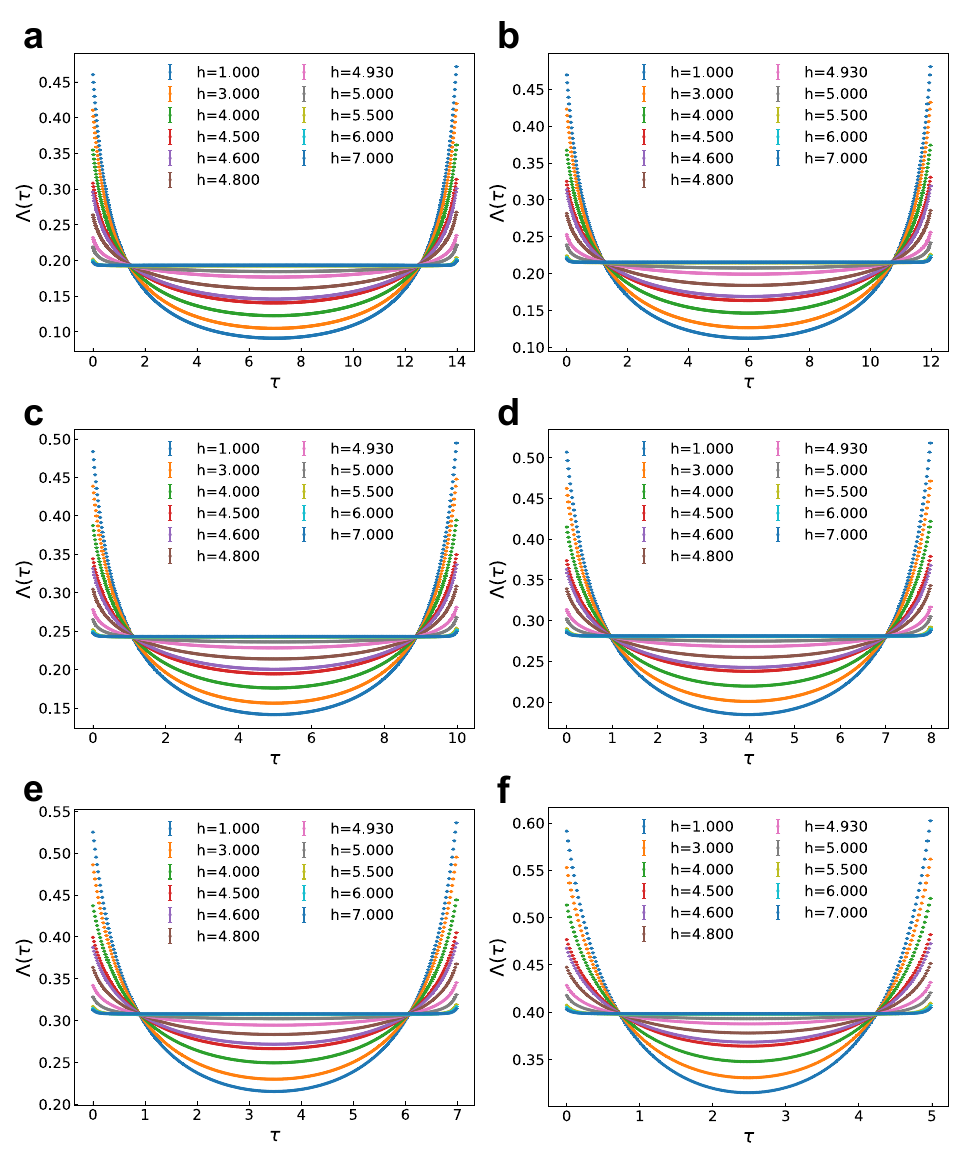}
  \centering
  \caption{
   {Current-current correlation function $\Lambda(\tau)$ corresponding to different transverse fields $h$ and temperatures $T$. }
   {(a)} $T = 1/14$.
   {(b)} $T = 1/12$.
   {(c)} $T = 1/10$.
   {(d)} $T = 1/8$.
   {(e)} $T = 1/7$.
   {(f)} $T = 1/5$.
  }
  \label{fig:chi_tau}
\end{figure}

\begin{figure}
  \includegraphics[width=10cm]{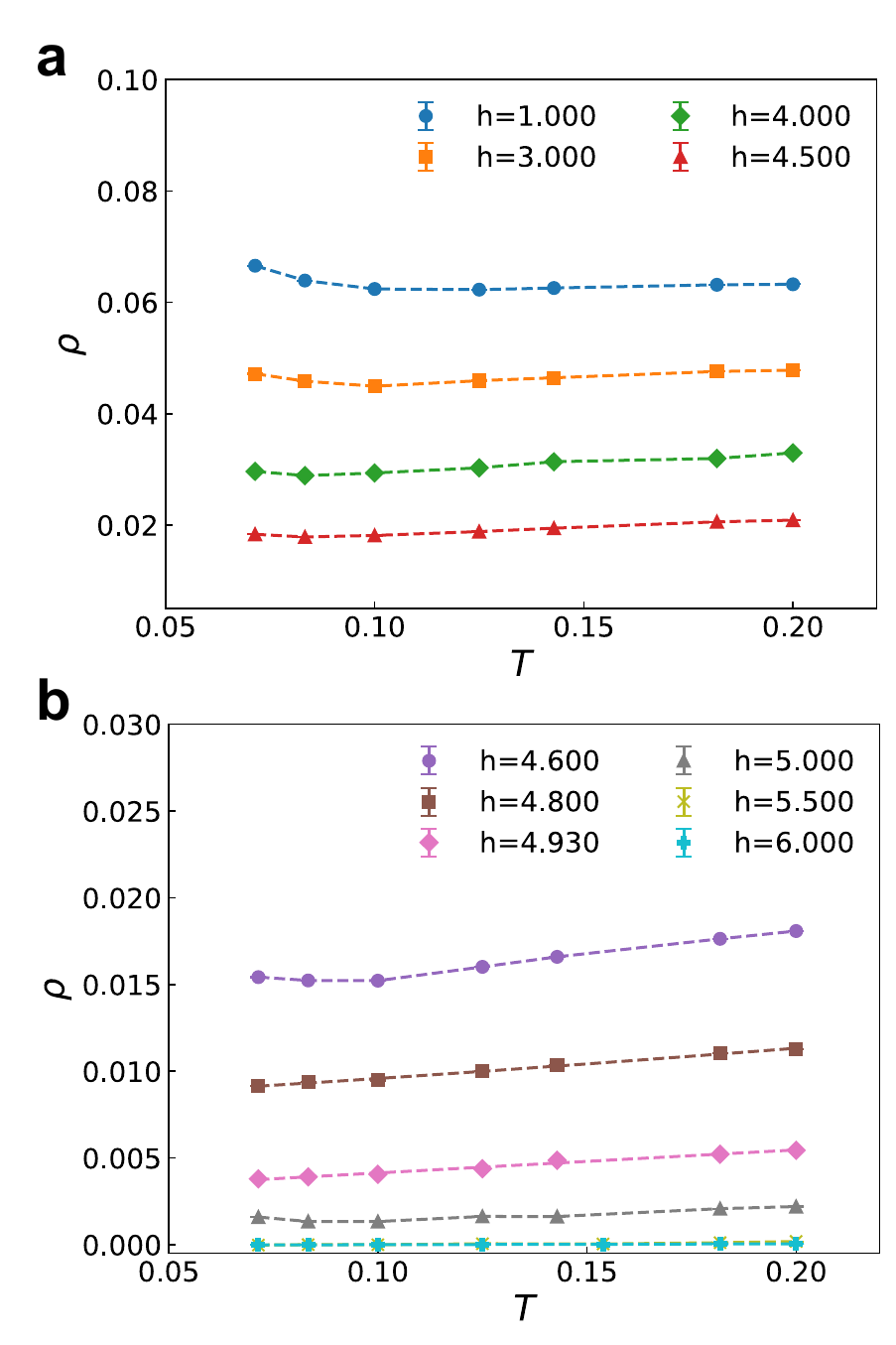}
  \centering
  \caption{
   {DC resistivity proxy $\rho$ extracted from the current-current correlation function $\Lambda(\tau)$.}
   }
  \label{fig:rho}
\end{figure}

\textbf{Supplementary Note 10: Resistivity Proxy from Current-Current Correlation Function} 

The current-current correlations are related to the conductivity by
\begin{equation}
    \Lambda(\i\omega_n) = \int \frac{d\omega}{\pi} \frac{\omega^2 \sigma'(\omega)}{\omega^2 + \omega_n^2}.
\end{equation}
\begin{equation}
    \Lambda(\tau) = \int \frac{d\omega}{2\pi} \sigma'(\omega) \frac{\omega \cosh{\left[(\frac{\beta}{2} - \tau)\omega\right]}}{\sinh\left(\frac{\beta\omega}{2}\right)}.
\end{equation}

Extracting the DC resistivity from the imaginary-time current-current correlator, $\Lambda(\tau)$, obtained via DQMC simulations, is a notoriously ill-conditioned inverse problem that requires analytic continuation. In light of the numerical instability of this procedure, various approximation schemes have been developed to serve as a proxy for resistivity. Prominent examples include extrapolation from imaginary-time self-energy~\cite{Patel2024strange}, simple functional fits to $\Lambda(\i\omega_n)$~\cite{ledererSuperconductivityNonFermiLiquid2017}, and leveraging the correlator's curvature at $\tau=\beta/2$~\cite{huang2019strange, ledererSuperconductivityNonFermiLiquid2017}. In this work, our simulations reveal that $\Lambda(\tau)$ is rather flat over the imaginary-time axis, especially for small $h$ region, as shown in Fig~\ref{fig:chi_tau}. This flatness results in a broad and sharply peaked $\Lambda(\i\omega_n)$ in Matsubara space. Under these conditions, we conclude that the correlator's value at $\tau=\beta/2$ provides the most robust and reliable DC resistivity proxy $\rho = \Lambda''(\beta / 2)/(2\pi \Lambda(\beta/2)^2)$ at the temperatures we investigated~\cite{huang2019strange, ledererSuperconductivityNonFermiLiquid2017}. The DC resistivity proxy of different transverse field $h$ is shown in Supplementary Figure~\ref{fig:rho}. We observe linear in temperature resistivity in the smearing region where MFL shows up. The small upturn of resistivity in low temperature and small $h$ region is possible due to strong finite size effect and slow dynamics of Ising spins. This is also manifest in the behavior of fermionic self-energy, the magnitude of their imaginary part increase when reducing the Matsubara frequency (Supplementary Figure \ref{fig:Sigma_order}), indicating strong scattering rate in this region.

\end{document}